\documentclass[ACMTOCHI]{acmtrans2m}
\usepackage{graphicx}
\usepackage{rotating}
\usepackage{multirow}
\usepackage{booktabs}

\newenvironment{myindentpar}[1]%
{\begin{list}{}%
         {\setlength{\leftmargin}{#1}}%
         \item[]%
}
{\end{list}}

\acmVolume{0}
\acmNumber{0}
\acmYear{09}
\acmMonth{12}

\newcommand{\BibTeX}{{\rm B\kern-.05em{\sc i\kern-.025em b}\kern-.08em
    T\kern-.1667em\lower.7ex\hbox{E}\kern-.125emX}}

\title{Reconstructing Experiences through Sketching}
\author{EVANGELOS KARAPANOS\\Eindhoven University of Technology\\
JEAN-BERNARD MARTENS\\Eindhoven University of Technology \and
MARC HASSENZAHL\\Folkwang University}

\begin{abstract}
We present iScale, a survey tool for the retrospective elicitation of longitudinal user experience data. iScale employs sketching in imposing a process in the reconstruction of one's experiences with the aim to minimize retrospection bias. Two versions, the \emph{Constructive} and the \emph{Value-Account} iScale, were motivated by two distinct theories on how people reconstruct emotional experiences from memory. These two versions were tested in two separate studies. Study 1 aimed at providing qualitative insight into the use of iScale and compared its performance to that of free-hand sketching. Study 2 compared the two versions of iScale to free recall, a control condition that does not influence the reconstruction process. Significant differences between iScale and free recall were found. Overall, iScale resulted in an increase in the amount, the richness, and the test-retest reliability of recalled information. These results provide support for the viability of retrospective techniques as a cost-effective alternative to longitudinal studies.
\end{abstract}


\begin{document}

\setcounter{page}{1}

\begin{bottomstuff}
Author's address: E. Karapanos, Eindhoven University of Technology,
E, P.O. Box 513, 5600 MB, Eindhoven, Netherlands.
\end{bottomstuff}
\maketitle

\noindent A video demonstration of the tool may be found at http://ekarapanos.com/iscale. This paper has been submitted to ACM Transactions on Computer-Human Interaction. It is available as a pre-print at http://arxiv.org/abs/0912.5343v1. Cite this paper as:

\begin{itemize}
  \item Karapanos, E., Martens, J.-B., Hassenzahl, M. (2009) Reconstructing Experiences through Sketching. CoRR abs/0912.5343.
\end{itemize}

\section{Introduction}
Understanding the use and acceptance of interactive products beyond initial interactions has always been a key interest in the HCI community \cite{Erickson1996,prumper1992}. A number of recent trends however are highlighting the importance of longitudinal studies in the HCI domain \cite{Karapanos2009UXOT}. First, legislation and competition within the consumer electronics industry has resulted in an increase in the time-span of product warranties, resulting in an alarmingly increasing number of products being returned on the basis of failing to satisfy users' true needs \cite{Ouden2006}. Secondly, products are increasingly becoming service-centered. Often, products are being sold for lower prices and revenues are mainly coming from the supported service \cite{Karapanos2009UXOT}. Thus, the overall acceptance of a product shifts from the initial purchase to establishing prolonged use. This increasing interest in understanding prolonged use of interactive products is also reflected in the HCI community \cite{Gerken2007longitudinal,Barendregt2006,FenkoSchifferstein2009,Karapanos2008UXOT,Wilamowitz06,Courage2009Jain,vaughan2008longitudinal,Kjeldskov2008longitudinal}.

From a methodological perspective, one could distinguish between three dominant approaches in understanding the development of users' behavior and experience over time \cite{Wilamowitz06}. \emph{Cross-sectional} approaches are the most popular in the HCI domain \cite{prumper1992,Bednarik2005}. Such studies distinguish user groups of different levels of expertise, e.g. novice and expert users. Differences between the user groups are then attributed to the manipulated variable, e.g. expertise. Such approaches are limited as one may fail to control for external variation and may falsely attribute variation across the different user groups to the manipulated variable. \citeN{prumper1992} already highlighted this problem, by showing that different definitions of novice and expert users lead to varying results.

Beyond the cross-sectional, one may further distinguish \emph{pre-post} and \emph{longitudinal} approaches in repeated sampling designs. Pre-post designs study the same participants at two points in time. For instance, \citeN{Kjeldskov2008longitudinal} studied the same 7 nurses, using a healthcare system, right after the system was introduced to the organization and 15 months later, while \citeN{Karapanos2008UXOT} studied how 10 individuals formed overall evaluative judgments of a novel pointing device, during the first week of use as well as after four weeks of using the product. While these approaches study the same participants over an extended period of time, they cannot inquire much into the exact form of change, due to the limited number of measurements. Longitudinal designs take more than two measurements. Because of their laborious nature, however, they are only rarely used in practice and research. \citeN{Wilamowitz06} distinguished different "resolutions" in those studies: a micro perspective (e.g. an hour), a meso perspective (e.g. 5 weeks) and a macro perspective, with a scope of years of use and the idea to map the whole product lifecycle. Studies with a micro-perspective assess how users' experience changes through increased exposure over the course of a single session. For instance,  \citeN{Minge2008} elicited judgments of perceived usability, innovativeness and the overall attractiveness of computer-based simulations of a digital audio player at three distinct points: a) after participants had seen but not interacted with the product, b) after 2 minutes of interaction and c) after 15 minutes of interaction. An example of a study with a meso-perspective is \citeN{Karapanos2009UXOT}. They followed 6 individuals after the purchase of a single product over the course of 5 weeks. One week before the purchase of the product, participants started reporting their expectations. After product purchase, during each day, participants were asked to narrate the three most impactful experiences of the day using a retrospective diary method, the Day Reconstruction Method \cite{Kahneman2004}. Studies with a macro-perspective are "nearly non-existent" \cite{Wilamowitz06}.

A third approach is the retrospective recall of personally meaningful experiences from memory. Variants of the critical incident technique, popular in the fields of marketing and service management research \cite{Edvardsson2001CRitical,Flanagan1954critical}, ask participants to report critical incidents over periods of weeks, months or the complete time-span of the use of a product or service. In a survey study, \citeN{FenkoSchifferstein2009}, for example, asked participants to recall their single most pleasant and unpleasant experience with different types of products and to assess the most important sensory modality (i.e. vision, audition, touch, smell and taste) at different points in time, i.e. when choosing the product in the shop, during the first week, after the first month, and after the first year of usage. Von Wilamowitz-Moellendorff et al. \citeyear{Wilamowitz06,wilamowitz2007} proposed a structured interview technique named CORPUS (Change Oriented analysis of the Relation between Product and User) for the retrospective assessment of the dynamics in users' perceptions of product quality. CORPUS starts by asking participants to compare their current opinion on a given product quality (e.g. ease-of-use) to the one they had right after purchasing the product. If change has occurred, participants are asked to assess the direction and shape of change (e.g., accelerated improvement, steady deterioration). Finally, participants are asked to elaborate on the reasons that induced these changes in the form of short reports, the so-called "change incidents".

One may wonder about the degree to which these recalls are biased or incomplete. However, we argue that the veridicality of one's remembered experience is of minimal importance, as these memories (1) will guide future behavior of the individual and (2) will be communicated to others. In other words, it may not matter how good a product is objectively, its quality must also be "experienced" subjectively to have impact \cite{hassenzahl2006Law}. See also \citeN{Norman2009Memory}.

Although the validity of remembered experiences may not be crucial, their reliability is. It seems at least desirable that participants would report their experiences consistently over multiple trials. If recall is random in the sense that different experiences are perceived to be important at different recalls, then the importance of such elicited reports may be questioned. In other words, what we remember might be different from what we experienced; however, as long as these memories are consistent over multiple recalls, they provide valuable information. In the area of critical incident research, interviewing techniques have been developed with the aim of helping the participant in cueing more contextual information surrounding an experienced critical incident \cite{Edvardsson2001CRitical}. Interviews may however elicit only a limited number of reports. Self-reporting approaches, for instance through online surveys, have far more impact because one can survey large samples and, thus, also inquire into rare experiences. Such approaches, however, are less controlled than face-to-face interviews. Thus, the question at hand is: \emph{How can a survey procedure support a participant in recalling her experiences with a product in a reliable way?}

This paper presents iScale, a survey tool that was designed with the aim of increasing participants' effectiveness and reliability in recalling their experiences with a product. iScale uses sketching to impose specific guiding procedures, assumed to improve the participant's ability to recall experiences from memory. In the following we will describe the theoretical motivations for the development of iScale and present the results of two studies. Study 1 aimed at acquiring a qualitative understanding of the use of iScale in comparison to its analog equivalent, i.e. a free-hand sketching, and aimed at informing its redesign. Study 2 aimed at assessing how iScale compares to an experience reporting tool that provides no sketching, and, thus, can be seen as a control condition to assess the impact of iScale on participants' effectiveness and reliability in recalling.

\section{Sketching and memory}
Memory was for long understood as a faithful account of past events, which can be reproduced when trying to remember details of the past. This idea was first challenged in Barlett's \citeyear{Bartlett1932} seminal work. He suggested that remembering is an act of reconstruction that can never produce the exact past event, but instead, every attempt to recall results in a new, often altered representation of the event. \citeN{Bartlett1932} asked participants to recall an unfamiliar story that they were told 20 hours before. Recalled stories differed from the original one in missing details, altering the order and importance of events, or in applying rationalizations and interpretations to the original story. Stories were further distorted through repeated reconstruction.

The notion that remembering is an act of reconstruction instead of mere reproduction has received wide support. At the heart of reconstruction lies the distinction between episodic and semantic memory \cite{Tulving2002}. While episodic memory "is specific to a particular event from the past, semantic memory is not tied to any particular event but rather consists of certain generalizations (i.e. beliefs) that are rarely updated" \cite{Robinson2002}. These two types of memory serve different needs such as learning new information quickly - a capacity of episodic memory - or developing relatively stable expectations about the world - a capacity of semantic memory \cite{Robinson2002}. Reconstruction happens through the retrieval of cues from episodic memory. In the absence of contextual cues in episodic memory, beliefs found in semantic memory may be used to reconstruct the past, resulting in distortions such as the ones found in Barlett's study. Thus, overall, the accuracy of one's remembered events lies in the degree to which contextual cues are still present in the person's episodic memory.

But, how do we reconstruct emotional experiences that contain not only contextual details of the experienced event, but also value-charged information such as emotions or overall evaluative judgments on the event? One can distinguish between two distinct approaches to the reconstruction of value-charged experiences. The first one, the \emph{Constructive} approach, assumes that felt emotion cannot be stored in memory but is instead reconstructed from recalled contextual cues. The second approach, the \emph{Value-Account} approach, proposes the existence of a memory structure that is able to store the frequency and intensity of one's responses to a stimulus. This information may in turn be used to cue the recall of contextual details of one's experiences. In the next sections we describe the two approaches in more detail.

\subsection{Two approaches to experience reconstruction}

\subsubsection{The Constructive Approach}

The constructive approach assumes that reconstruction happens in a forward temporal order \cite{Anderson1993,Barsalou1988,Means1989}. \citeN{Barsalou1988} asked people to recall their experiences during the summer. Most participants started in the beginning of the summer and proceeded in a chronological order. Often, the recall of an event cues the reconstruction of more events and contextual information surrounding the event \cite{Anderson1993} - like a string of pearls.

\citeN{Robinson2002} further argued that "emotional experience can neither be stored nor retrieved" (p. 935), but can only be reconstructed on the basis of recalled contextual cues. They propose an accessibility model that distinguishes between four types of knowledge used to construct an emotion. First, \emph{experiential knowledge} is used when an emotion is constructed online, i.e. as the experience takes place. When experiential knowledge is inaccessible, people will resort to \emph{episodic information}, i.e. recall contextual cues from episodic memory in reconstructing the emotional experience. When episodic memories become inaccessible, people will shift to semantic memory. People will first access \emph{situation-specific beliefs}, i.e. "a belief about the emotions that are likely to be elicited in a particular type of situation". If event-specific beliefs are inaccessible, e.g. due to rarity of the event, people will access \emph{identity-related beliefs}, i.e. "beliefs about their emotions in general".

Motivated by the accessibility model of \citeN{Robinson2002}, Daniel Kahneman and colleagues \citeyear{Kahneman2004,Schwarz2008global} developed the Day Reconstruction Method (DRM), an offline diary method that attempts to minimize retrospection biases when recalling emotional experiences. DRM starts by asking participants to mentally reconstruct their daily experiences as a continuous series of episodes, writing a brief name for each one. This aims at eliciting contextual cues within each experiential episode but also the temporal relations between the episodes. As a result, participants reconstruct the emotional experience on the basis of sufficient episodic information, thus avoiding retrospective biases that most offline methods suffer from as participants draw on semantic information to reconstruct the emotional experience. \citeN{Kahneman2004} demonstrated that DRM may achieve an accuracy close enough to that of online reporting as in the case of the  Experience Sampling Method \cite{Hektner2007}.

\subsubsection{The Value-Account Approach}
The Value-Account approach assumes that reconstruction happens in a top-down fashion. It assumes that people may recall an overall emotional assessment of an experience without recalling the exact details of the experienced event. \citeN{Betsch2001} proposed the existence of a new memory structure called Value-Account, that is able to store the frequency and intensity of positive or negative responses to stimuli. Since Value-Account is assumed to be more easily accessible than concrete details from episodic memory, it may cue episodic information in reconstructing the experienced event, or inform the construction of an overall evaluation even in the absence of episodic information \cite{Koriat2000,Neisser1981}.

While some studies have shown that value-account may be falsely recalled even in the presence of accurate episodic information, it is generally accepted that value-account information is better retained over time than episodic information \cite{Koriat2000} and may be used in cuing episodic information. In a related field of memory research, that of autobiographical memories, researchers distinguish between three levels of specificity in memory: lifetime periods, general events, and event-specific knowledge. Reconstruction has been found to take place in a top-down fashion where knowledge stored at the level of a lifetime period may cue information at the two lower levels \cite{Conway2000}.

Both approaches, constructive and value-account, suggest specific processes of retrieving emotional experiences from memory. While the constructive approach suggests a \emph{chronological order} in recalling episodic information that subsequently cues the reconstruction of the experienced emotion, the value-account approach suggests a \emph{top-down progression} where the affective information stored in value account is used to cue the recall of episodic information. In the following section, we will illustrate how these two processes were operationalized in two distinct versions of the iScale tool.

\subsection{Can sketching affect the reconstruction process?}

Imagine being asked to "sketch" how the perception of the usability of your mobile phone changed over time; you are given a timeline that starts at the moment of purchase and ends in the present. How would you start? One may go back to the past, right after the purchase of the product, and try to recall her first experience with the product. What was it about? Was it positive or negative? What else happened after that? Reconstruction is assumed to take place in a chronological order and the sketch, the overall evaluation of one's experiences, is constructed from the recalled details of the experiences. Another person may start the sketching exercise by thinking of the overall experience, the change over time. Did my opinion about the product's ease-of-use increase overall? If so, was it mainly in the beginning or at the end? This might then cue the recall of experienced events that caused these changes \cite{Wilamowitz06,wilamowitz2007}.

Sketching, in the above scenario, provides what \citeN{Goldschmidt1991dialectics} calls \emph{interactive imagery}, i.e. "the simultaneous or almost simultaneous production of a display and the generation of an image that it triggers". This imagery is an incomplete, reconstructed representation of the experienced past. It consists of two sources of information: a) contextual details of experienced events such as the temporal, factual, and social context of an experience, and b) value-charged information such as emotions and evaluations of the experienced event. Product evaluations are here seen as carriers of affective information, that is, affect that is attributed to the product \cite{Hassenzahl2007}.

The veridicality of this reconstructed representation, i.e. the convergence between the representation and the past, is likely to be influenced by the process that the participant follows in reconstructing it from memory. Sketching may, thus, impose a certain process on the reconstruction of the past and by that crucially influence the way experiences are remembered. In the remainder of this section we describe iScale, a survey tool that elicits experiences with a product through sketching how one's opinion changed over time. We will introduce two different versions of iScale, each trying to lead to a different experience reconstruction process. For reasons of simplicity we will call them the Constructive and the Value-Account iScale.

\begin{figure*}
\centerline{\includegraphics[width=12cm]{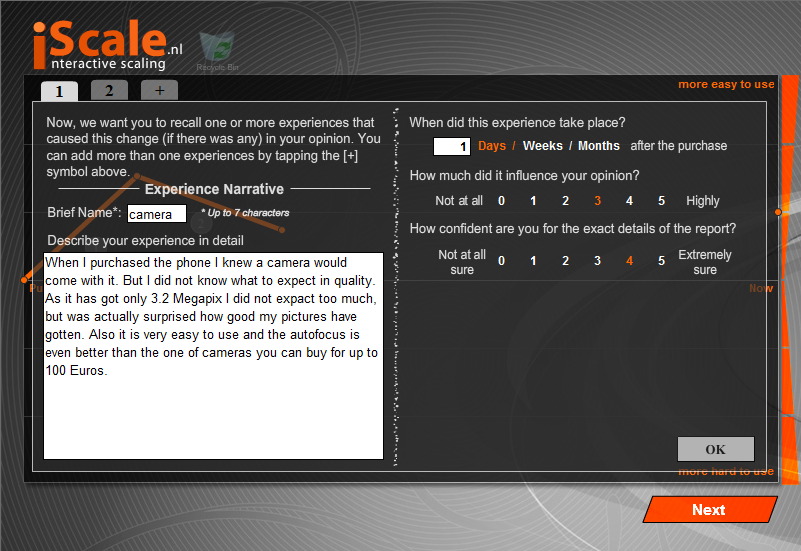}}
\centerline{}
\centerline{\includegraphics[width=12cm]{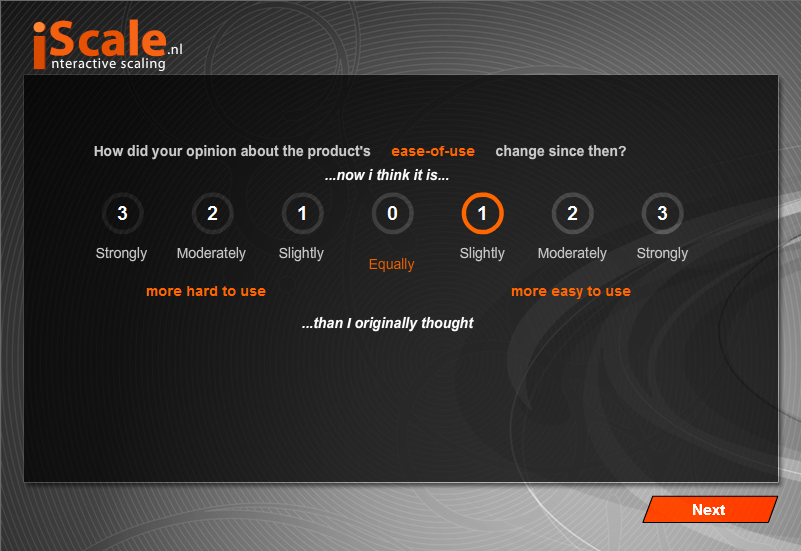}}
  \caption{(a) interface for reporting experiences, and (b) overall questions asked in the beginning of the survey}
  \label{iScaleP2}
\end{figure*}

Interacting with iScale is done in three steps. First, the participant is asked to respond to two questions (figure \ref{iScaleP2}b): a) "\emph{What was your opinion about the product's [certain quality] just before you purchased it", and b) "How did your opinion about the product's [certain quality] change since then}". While one could use participants' ratings to elicit a between-subjects estimation of how the perceived quality of a certain product develops over time, with the participants' time of ownership of the product as the independent variable, this is not the primary information that we are interested in. Instead, we assume that these questions can help the participant in positioning herself in the past and in recalling contextual details before the start of the sketching activity.

Second, the participant is presented with a timeline that starts at the moment of purchase and ends at the present time. The participant is asked to sketch how her opinion about a certain quality of the product has developed over time. The two distinct modes of sketching will be discussed below. Overall, the participant may sketch linear segments that represent an increase or decrease in her opinion over a certain period. Each period can be annotated along the time by specifying the time that has passed from moment of purchase.

\begin{figure*}
\centerline{\includegraphics[width=12cm]{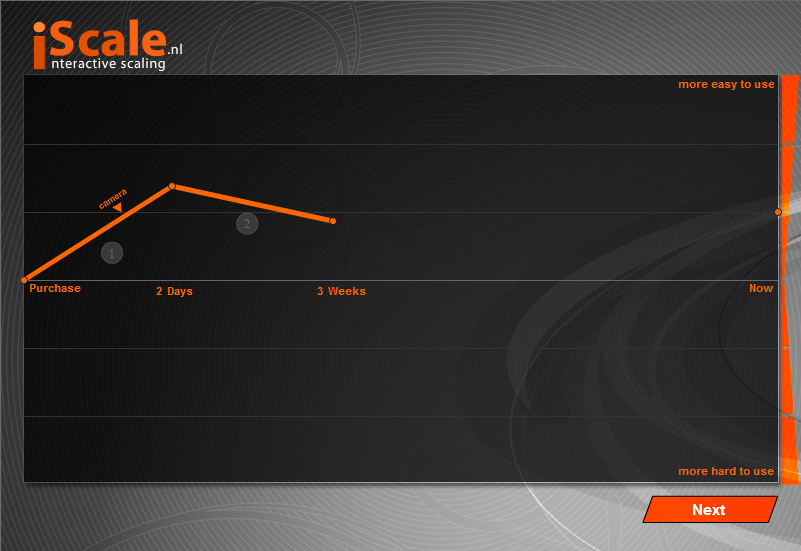}}
\centerline{}
\centerline{\includegraphics[width=12cm]{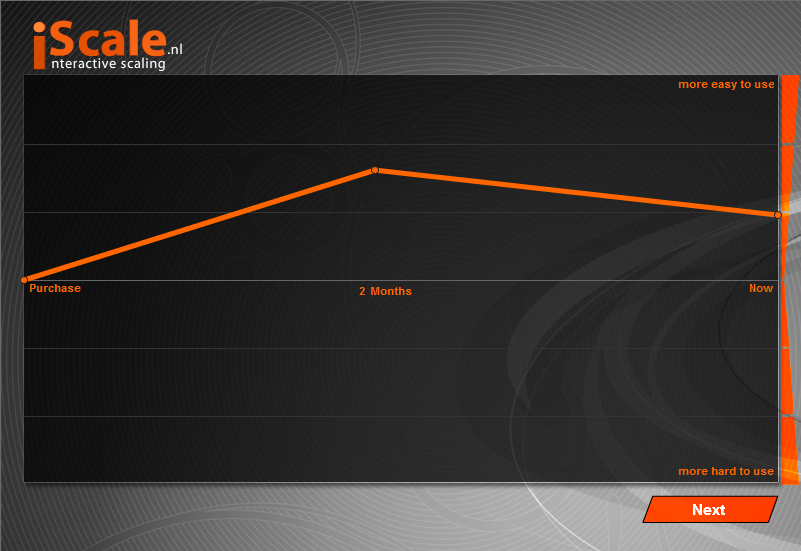}}
  \caption{(a) Constructive iScale, (b) Value-Account iScale}
  \label{iScaleP1}
\end{figure*}

Third, each line segment is associated with a line identifier that is displayed below the segment (figure \ref{iScaleP1}a). A participant may click on the segment and an interface is presented for reporting one or more experienced events that are perceived to have caused the sketched change in the participant's overall opinion (figure \ref{iScaleP2}a). For each experience report, the participant may provide a brief name (identifier), a more elaborate description of the experienced event - \emph{experience narrative}, and respond to a number of event-specific questions. For the goals of the specific study we present in this paper, we asked participants to recall a) the exact time that the event took place, b) the impact of the event on the participant's overall opinion, and c) the participant's confidence on the exact details of the narrative.
 	
\subsubsection{The Constructive and the Value-Account iScale}
The two versions of iScale differ only in the second component, that of sketching. These aim at imposing distinct modes of reconstruction of experiences from memory. More specifically: the existence or absence of concurrency between sketching and reporting, and feed-forward or top-down progression of sketching.

\begin{myindentpar}{1cm}
\textbf{Feed-forward - Top-down progression of sketching:} The constructive approach to reconstruction suggests that recalling experiences in a chronological order will cue more contextual details surrounding the experience and that this will in turn lead to a better reconstruction of the experienced emotion as well as a recall of further temporally aligned experiences. On the other hand, the value-account approach assumes that participants follow a top-down approach where a participant may first form an overall evaluation of the change over the full time of ownership, and proceed by forming subsequent judgments on the sub-parts of each change. Thus, in the constructive iScale one starts by plotting points in a serial order; in the Value-Account iScale a line connects the start with the end of the timeline using the participant's response to the question asked in the first step regarding how the overall opinion has change from the moment of purchase to the present. The participant may then proceed by splitting the full segment into parts.

\textbf{Concurrent - Non-concurrent reporting:} The constructive approach to reconstruction assumes that the affective component of a past experience, in other words the value-charged information, can only be reconstructed from recalled contextual details of the event. On the contrary, the value-account approach assumes that individuals may recall an overall emotional assessment of an experience even without being able to recall the underlying contextual details. Thus, according to the constructive approach reporting should be concurrent with sketching as reporting would increase the contextual details and thus result in richer recall. On the other hand, in the value-account approach, concurrent reporting might bias or hinder the process of recalling this value-charged information. Thus, in the constructive iScale the participant is asked to report on experiences right after a line segment is sketched. Sketching and reporting is thus proceeding concurrently in a step-by-step process. In the Value-Account iScale, this process is split into two distinct steps: the participant is urged to first sketch the desired pattern before proceeding to the next step where she may report one or more experiences for each sketched line segment. Both methods however retain flexibility, as the sketched pattern can be modified even after experiences have been reported for the existing pattern.
\end{myindentpar}

Overall, sketching is expected to provide temporal context for the recall of experienced events. This is expected to increase the amount and test-retest reliability of the information that the participants are able to recall. This assumption will be tested in study 2.

\section{Study 1}
The first study attempts a qualitative understanding of sketching as a process for supporting the reconstruction of one's experiences. First, it questions some of the core assumptions that underlie the design of iScale through the observation of users' behavior when employing free-hand-sketching (FHS), the paper version of Constructive iScale. Secondly, it compares the two iScale tools to the FHS approach and identifies the design qualities that a sketching tool should have in order to support the recall process.

\subsection{Method}

\subsubsection{Participants}
A convenience sample of 12 graduate students in HCI (7 male, median age 30 years) participated in the study. They were chosen due to the diversity in their educational backgrounds. They were: five Computer Scientists, three Industrial Engineers, two Linguists, one Psychologist and one Industrial Designer.

\subsubsection{Procedure}
The study consisted of two main parts. In the first part, each participant used the three different sketching techniques, i.e. free-hand sketching and the two iScale tools. All tasks were carried out on a Wacom Cintiq 21UX Interactive Pen Display. The order in which the tools and qualities were employed was counterbalanced across participants; FHS was always used first to avoid any bias from the iScale tools as we wished to understand users' natural behavior in free-hand sketching.

Participants were asked to sketch how their opinion on three distinct product qualities of their mobile phone developed over time (see table \ref{tab:QualityDefinitions}). Each quality was described by a brief definition and three words to support the definition \cite{Hassenzahl2004,Wilamowitz06}. Participants were instructed to think aloud; interactions and verbal data were captured on video.

\begin{myindentpar}{1cm}
\emph{"While sketching, you are asked to report \textbf{experiences and events} that induced these changes in your opinion about the product. We are interested in knowing your exact thoughts as you perform the sketching activity. What makes you sketch something? Do you remember something? Is it just a feeling? We ask you to \textbf{think aloud} while doing this task."}
\end{myindentpar}

\begin{table}
\centering
\caption{The three product qualities that participants reported on, along with definitions and word items.} \label{tab:QualityDefinitions}
\begin{tabular}{l@{\hspace{2em}}p{5cm}l} \toprule
\textbf{Name} & \textbf{Definition} & \textbf{Word items} \\ \midrule
Usefulness & The ability of a product to provide the necessary functions for given tasks. & Useful, Practical, Meaningful \\
Ease-of-use & The ability of a product to provide the functions in an easy and efficient way. &	Easy to use, Simple, Clear \\
Innovativeness & The ability of a product to excite the user through its novelty. &	Innovative, Exciting, Creative \\ \bottomrule
\end{tabular}
\end{table}

In the second part, participants were interviewed about the differences between the three sketching techniques, using a structured interview technique, called the Repertory Grid \cite{Fransella2003}. Participants were given three cards, each providing a name and a screenshot of one of the three sketching techniques. Participants were first asked to identify the three techniques. Next, they were asked to "\emph{think of a property or quality that makes two of the sketching techniques alike and discriminates them from the third}". They were instructed to feel free to make any combination of the three alternatives. Contrary to common practice with the Repertory Grid Technique, we did not probe participants in providing a bipolar construct \cite{Karapanos2008Workshop} while we instructed them to elaborate when possible.

Participants were further probed using the \emph{laddering} and \emph{pyramiding} techniques \cite{reynolds1988laddering}. Laddering seeks to understand what motivates a given statement and thus ladders up in an assumed means-ends-chain \cite{Gutman1982} towards more abstract qualities of the stimuli; in laddering we first asked the participant whether the mentioned quality is positive, and subsequently why this quality is important to him/her, e.g. "\emph{why is expressiveness important to you?}". Pyramiding, on the other hand, also known as negative laddering, seeks to understand the lower level attributes that make up for a given quality; in pyramiding we asked the participant to elaborate on what makes the given technique to be characterized with the respective attribute, e.g. "\emph{what makes free-hand-sketching more expressive?}".

\subsection{Analysis and results}
\subsubsection{Understanding free-hand sketching}
Throughout the development of iScale, FHS acted as our reference for testing assumptions and gaining confidence for the respective design decisions. Often, compromises had to be made. For instance, enabling users to create non-linear curves would increase the expressiveness in sketching but would on the other hand also increase the complexity of the task by either increasing the number of actions needed in enabling users to define all parameters of a curve, or minimizing users' control over the result by imposing restrictions to the curve in an effort to minimize the number of actions required. Thus, it is reasonable to explore the value of non-linearity in such sketches. In other terms, do users sketch non-linear curves in free-hand-sketching and if so, what does this non-linearity express?

Next, iScale assumes that sketching supports and is supported by the reconstruction of discrete experiences. It is thus assumed that users will associate changes in their opinion to one or more discrete experiences. This notion can however be questioned. First, the degree to which a sketched change is associated to one or more discrete experiences will depend on the mode of reconstruction, being it either \emph{constructive} or \emph{value-account}. In the constructive mode, users recall contextual cues about the discrete experience and reconstruct the overall value-judgment based on the recalled facts. In the value-account mode users may recall this overall evaluative judgment first and may or may not further reason to associate the recalled evaluative judgment to underlying reasons for this, e.g. one or more experiences. Thus, to what extend do users succeed in recalling discrete experiences? Second, assuming that a user recalls a discrete experience: Is it associated to a continuous change or a discontinuous one? In other terms, if a user thinks in terms of discontinuities, i.e. in terms of discrete events instead of overall opinion, does he/she relate these discontinuous recalls to a continuous graph?

These questions were explored by observing users' reported experiences and sketched patterns while employing FHS. Users' free-hand sketches were segmented in discrete units based on cues available in participants' verbalized thoughts as well as users' pauses in their sketching behavior as observed in the video recorded sessions. A new unit was coded when both conditions were observed: a semantic change in the participant's report following a pause in sketching. Often this was combined with a change in the slope of the curve, but this was not always the case.

Each unit was then coded for the type of curve and the type of verbal report.  Curves were classified under four categories: a) \emph{Constant (C)} signifying no change in participant's opinion over a certain period, b) \emph{Linear (N)}, either Increasing or Decreasing, c) \emph{Non-linear (NL)} when there were no grounds that the curve could be approximated by a linear one or when a single report was associated with two discrete linear curves of different curvature (see \ref{fig:FHS}b), and d) \emph{Discontinuous (D)} when the slope was significantly higher than on average.

\subsubsection{What kinds of curves do users sketch?}
\begin{figure*}
\centerline{\includegraphics[width=6cm]{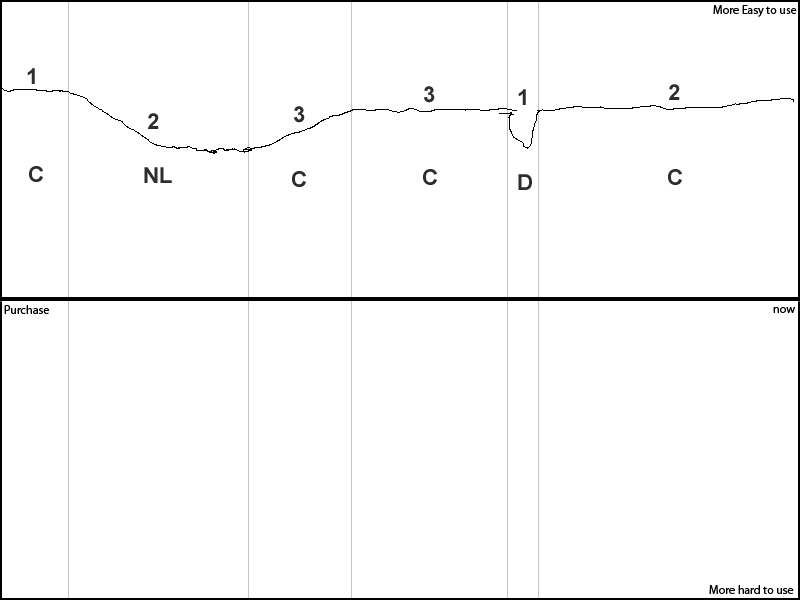} \includegraphics[width=6cm]{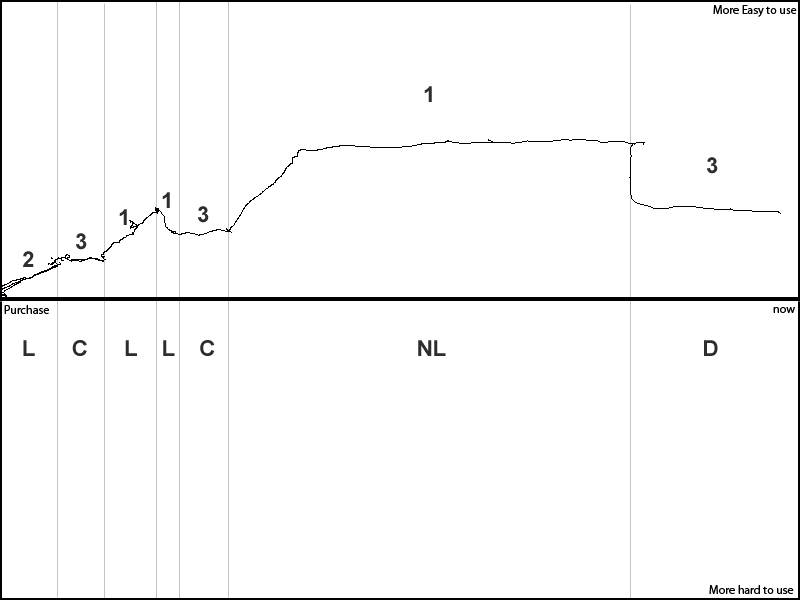}}
\centerline{}
\centerline{\includegraphics[width=6cm]{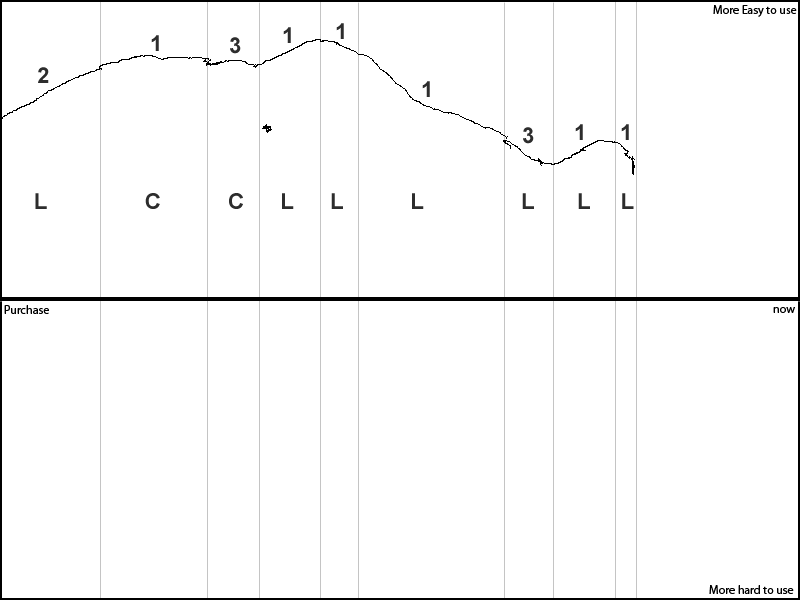} \includegraphics[width=6cm]{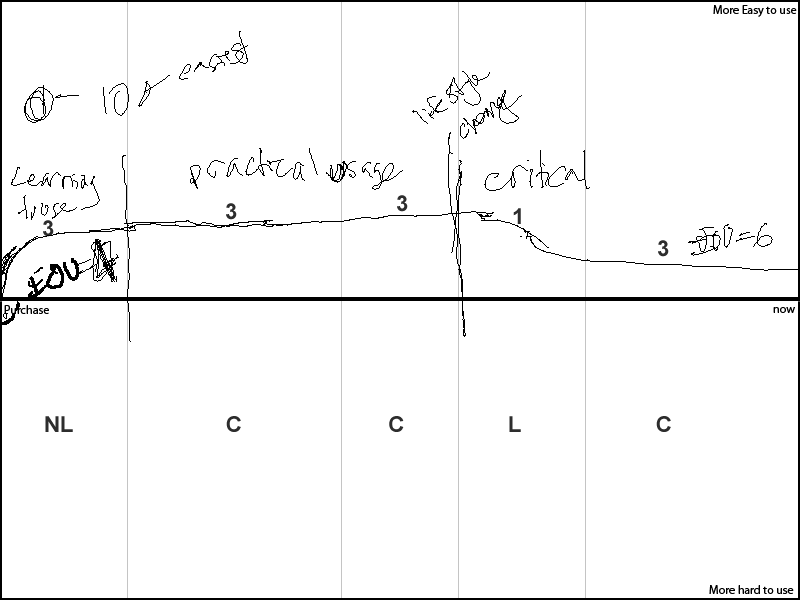}}
\centerline{}
\centerline{\includegraphics[width=6cm]{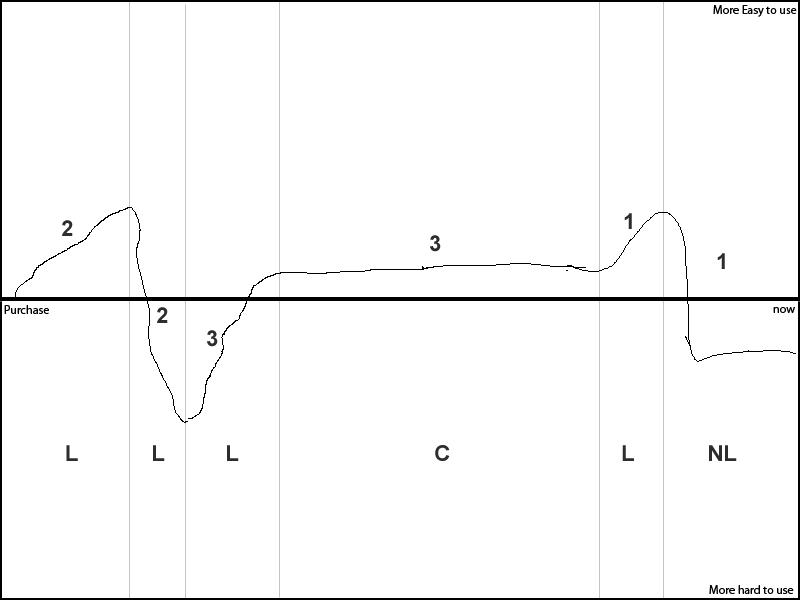} \includegraphics[width=6cm]{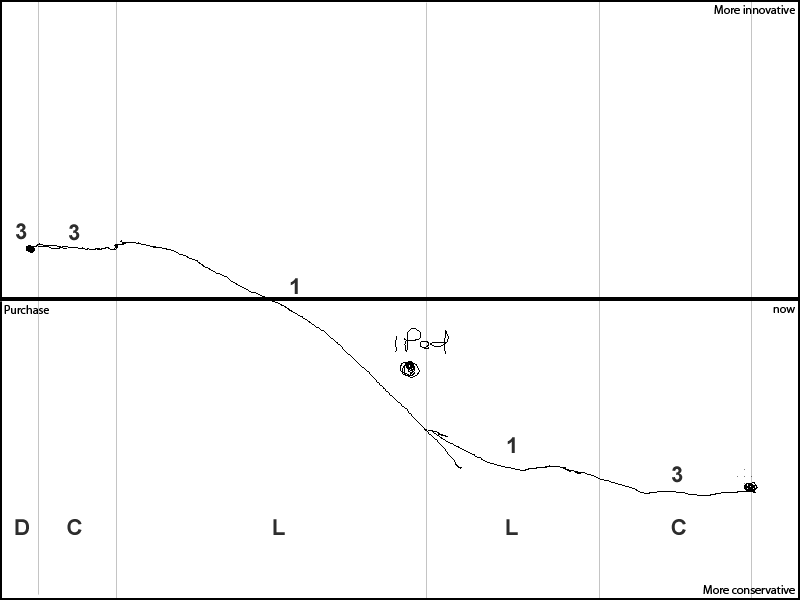}}
  \caption{Examples of free-hand sketching. Identified segments are indicated by vertical lines. Each segment is coded for the type of report (1: Reporting a discrete experience, 2: Reporting an overall evaluation, reasoning through experience, 3: Reporting an overall evaluation with no further reasoning) and type of sketch (C: Constant, L: Linear, NL: Non-Linear, D: Discontinuous).}
  \label{fig:FHS}
\end{figure*}

Table \ref{tab:SketchingReporting} illustrates the distribution of users' sketches across the four types of curves. The majority of curves (44 of 74, 60\%) were categorized as linear, signifying a change that can reasonably be approximated by a linear curve. Only 5\% (4 of 74) curves were non-linear. For these curves, a single report was associated with two or more linear curves with different slopes (cf. figure \ref{fig:FHS}a segment 2, figure \ref{fig:FHS}b segment 6, figure \ref{fig:FHS}d, segment 1). Thus while in certain cases users draw non-linear curves, the majority of curves will be linear ones, and therefore the overall value of non-linearity in a sketching tool appears to be limited.

In a similar vein, only 4 of 74 (5\%) instances of discontinuity were observed in users' sketches. One might expect that the recall of discrete events increases users' tendency to sketch discontinuous curves, as the mode of recalling, being it continuous or discontinuous, should relate to the mode of sketching. This expectation is not supported by the data.

\begin{table}
\centering
\caption{Relationship between sketching and reporting in free-hand sketching. Types of sketching: C=Constant, L=Linear, NL=Non-Linear and D=Discontinuous} \label{tab:SketchingReporting}
\begin{tabular}{p{4cm}@{\hspace{2em}}ccccc} \toprule
 &	\multicolumn{4}{c}{Type of sketch} & \\ 	
\emph{\textbf{Type of report}} & C & L & NL & D & \\ \midrule
Reporting a discrete experience & 3 & 30 & 2 & 2 & 37 (50\%) \\
Reporting an overall evaluation with no further reasoning & 17 &	4 &	1 &	2 &	25 (34\%) \\
Reporting an overall evaluation, reasoning through experience  & 2 & 10 & 1 & 0 & 12 (16\%) \\
Overall & 22 (30\%) & 44 (60\%) &	4 (5\%) & 4 (5\%) & 74 \\ \bottomrule
\end{tabular}
\end{table}

\subsubsection{How do curves relate to experiences?}
Participants' reports were classified into three broad categories. First, reports rooted in the \emph{recall of a discrete experience}. Such reports represented the constructive mode of reconstruction: recalling contextual information from a specific experience followed by the reconstruction of the value judgment from the recalled facts. Such reports provided one or more contextual cues about the past experience, such as temporal information (i.e. when the event took place), changes in the context of use (e.g. "then I went on vacation..."), information related to the participant's social environment (e.g. "a friend of mine was very positive..."), etc. They constituted the most dominant type of reporting (37 of 74, 50\%).

\begin{myindentpar}{1cm}
\emph{\textbf{Recall of a discrete experience}: "The reason I got this device was to develop applications for it. [the company] has a special program for educational institutions to which provides free licenses for development. But when we contacted them, they even questioned the existence of our institution... this should have happened around here [points to the curve]"}
\end{myindentpar}

On the contrary, other reports provide no contextual information about a recalled experience, but instead, the participant \emph{reports an overall evaluation without further reasoning}. Such reports represented the value-account mode of reconstruction: recalling an overall evaluation of a specific experience or period, while failing to recall contextual cues or facts about an exact experience.

\begin{myindentpar}{1cm}
\emph{\textbf{Recall an overall evaluation without further reasoning}: "after that, [my opinion] is about constant, it hasn't changed lately"}
\end{myindentpar}

Table \ref{tab:SketchingReporting} illustrates a strong association of discrete experiences with linear curves (10/13), while reporting merely an overall evaluative judgment takes place mostly when participants sketch a constant line (17/24), signifying no change in their perception. Reporting only evaluative judgments may thus be a side effect of sketching, rooted in the fact that participants are asked to sketch a continuous timeline.

Last, we found a third type of reporting that combines the two core types. Those reports were grounded in the recall of an overall evaluation, but participants proceeded to reason about this value-judgment through reporting discrete experiences. Most of them (10 of 13) reflected linear changes.

\begin{myindentpar}{1cm}
\emph{\textbf{Recall an overall evaluation followed by reasoning about an experience}: "[my opinion] decreased as I expected that it would be easier than that, for example, I would like to have the automatic tilting to landscape view as it has an accelerometer"}
\end{myindentpar}

\subsubsection{How does iScale compare to free-hand sketching?}

The two iScale tools were also compared to free-hand sketching. Participants' verbal reports were transcribed and analyzed using Conventional Qualitative Content Analysis \cite{Hsieh2005}. We started with \emph{open coding} \cite{Strauss1998} where we aimed at identifying an overpopulated list of design qualities that appear to influence the design space of the three sketching techniques. Subsequently we grouped the initial codes into overall categories through an iterative process. Each statement was coded for the type of quality that the participant refers to as well as to whether or not this quality affects the sketching or the recalling process. Statements were always differentiating two of the approaches from a third as this was imposed by the structure of the interview technique.

Table \ref{tab:DesignQualities} illustrates the dominant qualities that were elicited in the interview sessions. For each quality, it displays the number of participants mentioning it as present for a given sketching technique, and the number of participants mentioning the given quality as affecting the sketching or recalling process. The design qualities can be distinguished in three broad groups: \emph{expressiveness, control}, and \emph{concurrency}. Meaningful differences emerged across the three different techniques; these should only be considered as tentative qualitative insights, however, due to the small number of observations.

\begin{sidewaystable}
\centering
\caption{Design qualities elicited in the interview sessions. Number of participants using a given quality to differentiate the three sketching techniques, and number of participants mentioning the given quality as affecting the sketching or recalling process.} \label{tab:DesignQualities}
\begin{tabular}{l@{\hspace{1em}}p{7cm}ccccc} \toprule
 & & \multicolumn{3}{c}{Tool} & \multicolumn{2}{c}{Impact on} \\
Name & Description & FHS & CON & VA & Sketch & Recall \\ \midrule
\multicolumn{2}{l}{\emph{Expressiveness}} & & & & & \\
Sketching freedom & Increasing the degrees-of-freedom in sketching & 7 & & & 6 & 1 \\
Annotation & Enabling free-hand annotation on the graph & 3 & & & & 3 \\
 & & & & & & \\
\multicolumn{2}{l}{\emph{Control}} & & & & & \\
Interactivity & Providing instant visual feedback through quick actions & & 8 & 8 & 4 & 4 \\
Modifiability & Providing the ability to modify the sketch as new experiences are being recalled & & 8 & 6 & 8 & 2 \\
Temporal Overview & Providing feedback on overall opinion across full timeline & & & 7 & 3 & 4 \\
 & & & & & & \\
\multicolumn{2}{l}{\emph{Interplay sketching-recalling}} & & & & & \\
Chronological linearity & Chronological linearity in recalling & 5 & 5 & & 1 & 5 \\
Concurrency & Concurrency of time in sketching \& recalling & 5 & 5 & & 2 & 5 \\ \bottomrule
\end{tabular}
\end{sidewaystable}

\subsubsection{Expressiveness}
As expected, the majority of users perceived the free-hand sketching approach as more expressive. This was mostly due to the \emph{\textbf{freedom in sketching}} that the free-hand approach provides as opposed to the iScale tools that restrict the user in plotting points connected through line segments. As some participants commented:

\begin{myindentpar}{1cm}
\emph{"[p6] it is easier to convey my thought into paper, the other ones I could only draw a straight line and that constrained me", "I think this allows you more to relate to your opinion, because you can really sketch what you want to... you can give slight ditches on the curve"}
\end{myindentpar}

The majority of participants emphasized the effect this has on the sketching activity. While all participants expressed the perceived benefit of FHS as it enables them communicating their thoughts, only one participant mentioned that this also affects the recalling process as the sketch provides richer feedback and therefore more cues for recalling. One participant stated freedom in sketching as a positive quality for expressing impressions for which she fails to recall any supportive facts.

\begin{myindentpar}{1cm}
\emph{"[p1] free-hand sketching provides the freedom to sketch something without having to justify the exact reason for this... I might not recall something specific; still I can communicate my impression through the curve... with the electronic ones you select points, therefore you feel obliged to justify them"}
\end{myindentpar}

Next, some participants mentioned the ability to \emph{\textbf{annotate}} the graph as a positive quality that enhances the recalling process. Annotations helped in providing \emph{contextual} and \emph{temporal} cues from the past such as positioning a specific experience along the timeline, splitting the timeline in terms of periods, but also in externalizing thoughts that participants thought they might fail to recall afterwards:

\begin{myindentpar}{1cm}
\emph{"[p12] I can annotate things... it helps as I might forget the reasoning in the beginning of the graph and when I am at the end, ...like in ease of use, I looked back and it helped me remember what my reference point meant, I didn't find it very intuitive at the zero point, and I was able to put my scale on diagram", "[p6] free-hand sketching provides the opportunity to add notes, you can sketch and then split the x-axis in terms of periods.",  "[p3] in free-hand sketching you can make notes in the meanwhile, bring other thoughts you have..."}
\end{myindentpar}

\subsubsection{Control}
Most participants stated that the iScale tools provide more overall control, related to three core aspects of the interaction. Firstly, eight out of the twelve participants found the constrained interaction a positive aspect of iScale, providing better \textbf{\emph{interactivity}}, as it consumes less resources, thus providing them \emph{better control of the output} (4 participants) and \emph{enabling them to focus on recalling their experiences} (4 participants):

\begin{myindentpar}{1cm}
\emph{"[p8] the digital ones in some way restrict your freedom... I think it's a good thing because you think before you put the dot, so elaborate more before you sketch, it is a more thoughtful account", "[p5] It's difficult to just draw on a free line whatever you think... it feels like a drawing, it doesn't feel like making a realistic graph", "[p10] in free-hand sketching I didn't have the control of the output as much as with this one", "[p12] it constrains me to draw line segments, at the same time focusing my efforts on just creating the events... I wouldn't have to worry about how neat the line looked..."}
\end{myindentpar}

\begin{myindentpar}{1cm}
\emph{"the electronic ones are more playful, you add points, move them, it has a certain dynamic, with free-hand sketching you cannot play and see the result... with the other ones, you may with quick actions alter the output, they are more interactive. I think I am more fond of them as I can communicate what I have in mind"}
\end{myindentpar}

Second, participants differentiated iScale from FHS in terms of the ability to modify the sketch while new experiences are recalled. Some participants further differentiated between the two iScale tools in terms of modifiability:

\begin{myindentpar}{1cm}
\emph{"[p7] it's easier to adapt these lines in that you just move part of the line, in free-hand sketching you have to clean some parts of the line", "[p6] in the value-account tool, you had to decide on the overall slope right from the start... This constrained me in..., which is not actually true because I could select the point in between and move it, ... but in any case it was harder."}
\end{myindentpar}

Third, seven out of the twelve participants acknowledged that the value-account tool provides a better overview of the full timeline; approximately half of the participants were concerned about the accuracy of the graph, while the rest thought that this \emph{\textbf{temporal overview}} enhances their recall process.

\begin{myindentpar}{1cm}
\emph{"[p8] these ones [free-hand, constructive] surprise you at the end, you can end up somewhere you didn't expect, the value-account gives you a better overview of totality.", "[p9] it provides an explicit timeline, you are thinking 'I have to do some clustering...' it is easier for recalling, it helps in rethinking"}
\end{myindentpar}

\subsubsection{Concurrency}
At the same time some participants perceived the temporal overview as a negative quality of the value-account tool. As one participant mentions:

\begin{myindentpar}{1cm}
\emph{"[p4] while I don't know where the story ends, I can add whatever event that fits to the story, it is more open ended... with the end point, I had to structure my story more towards a coherent story"}
\end{myindentpar}

Five participants in total mentioned \emph{\textbf{temporal linearity}} as a quality that differentiated free-hand sketching and the constructive iScale from the value-account tool. Most of those participants mentioned that recalling events in a step-by-step order helped them in recalling more events, while some of them were negative towards value-account as they felt that it constrained them when recalling events due to a focus in compiling a coherent story:

\begin{myindentpar}{1cm}
\emph{"[p6] it allows you to visualize things step by step and also historically... you ascertain that the decision that you make in each step is correct... you just make one step, you don't have to think about the whole curve", "[p5] in free-hand sketching you are really forced to think about the experience from the moment of purchase till now", "[p5] the more you do the more you remember... I think with the value-account in that sense you can relate to less experiences, you first draw an overall picture and you are less inclined to relate to experiences..., in both other cases, you are more focused on the spot...", "[p4] value-account felt more limited, I had to think where I was going to go, I would disregard things that didn't go where I wanted to go"}
\end{myindentpar}

Similarly, five participants highlighted that the \emph{\textbf{concurrency}} of sketching with reporting, that was lacking from the value account tool, enhances the output of both the recall and the sketching process:

\begin{myindentpar}{1cm}
\emph{"[p7] you go step by step with what you want to report and the visualization. ...It aligns the action that you are making with the thinking process that you want to make", "[p7] [in value-account iScale], you have to really judge... I was trying to compare the purchase point to my overall impression now and this changes, there are different periods, I am trying to average over different time periods and experiences, which is not actually the purpose... the decisions I made at first were wrong decisions because I have to change it, the other tools are more accurate"}
\end{myindentpar}

\subsection{Conclusion}
Overall, the study provided support for the iScale tool. The need for sketching non-linear and discontinuous curves was limited. In addition, most non-linear curves could be approximated by two linear segments. The need for annotation was highlighted by participants in the post-use interviews and two forms of annotation were added to the tool: (a) a timeline annotation that allowed users to set the start and end date of sketched segments, thereby splitting the timeline in specified periods, and (b) a visualization of experiences along the respective line segment that they belong to, with a brief identifier for the experience (see figure \ref{iScaleP1}a). Annotation, provided users with the ability to combine an overview of the different periods as well as the experiences that defined these periods.  Annotation also promotes interactivity as users have a better overview of the sketched pattern and are therefore more likely to modify it. The interviews confirmed that free-hand sketching is more expressive than using iScale due to the increased degrees of freedom in sketching as well as due to its ability to easily annotate sketches. Participants also reported qualities that were not present in the free-hand sketching, such as the two-step interactivity and modifiability of the electronic sketches that resulted in a better interoperability between the sketching and the recalling activity. Last but not the least, participants also reported benefits for both the constructive and the Value-Account over FHS. The Value-Account provided a temporal overview which influenced both the sketching and the recalling process. The constructive approach provided benefits such as chronological order and concurrency between sketching and reporting which had a positive impact on the recall process.

\section{Study 2}
The second study aimed at comparing the two different versions of iScale, the constructive and the Value-Account version, to a control condition that entails reporting one's experiences with a product without performing any sketching. It aims at testing whether sketching impacts the number, the richness, and the test-retest reliability of the elicited experience reports.

\subsection{Method}
\subsubsection{Participants}
48 participants (17 Female), ranging from 18 to 28 years (median=23 years), took part in the experiment. They were all students at a technical university; 19 of them majored in management related disciplines, 16 in design, and 13 in natural sciences and engineering. They all owned a mobile phone for no less than four and no more than eighteen months; 16 participants owned a smart phone.

\subsubsection{Materials}
Three different versions of iScale were used in the experiment: \emph{Constructive}, \emph{Value-Account}, and \emph{Experience Report}. The Constructive and Value-Account versions employ two discrete sketching approaches aiming at inducing two distinct modes of reconstruction from memory as described earlier. Experience Report constitutes a stripped down version of iScale where the sketching interface is removed; instead the user may only provide report experiences. It uses the same interface as the two sketching tools in reporting experiences.

\subsubsection{Experimental design}
The experiment used a balanced incomplete design with three factors: \emph{presence or absence of sketching} (the two sketching tools versus Experience Report), \emph{mode of reconstruction} (i.e. Constructive or Value-Account), and \emph{product quality} being reported (i.e. ease-of-use versus innovativeness). Mode of Reconstruction was between-subjects with the other two variables being within-subject.

\subsubsection{Procedure}
Participants joined two sessions, each one lasting approximately 40 minutes. The second session took place roughly one week after the first one (minimum: 7 days, maximum: 10 days).

\begin{table}
\centering
\caption{Experimental design: manipulated variables (rows) and process (columns). Each participant joined two sessions, each consisting of two tasks. Participants in group A, for instance, which belong in the Constructive condition, in the first session, they used the sketching tool to report on ease-of-use followed by the non-sketching tool to report on innovativeness. In the second session the order was counterbalanced but the two coupling of the two variables, sketching and quality, remained the same.} \label{tab:ExperimentalDesign}
\begin{tabular}{l@{\hspace{2em}}llcccc} \toprule
\multicolumn{3}{c}{} & \multicolumn{2}{c}{Session 1}  & \multicolumn{2}{c}{Session 2}  \\ \morecmidrules \cmidrule(rl){4-5} \cmidrule(rl){6-7}
\multicolumn{3}{c}{} & Task 1 & Task 2 & Task 1 & Task 2 \\ \midrule
\multirow{4}{*}{Constructive} & \multirow{2}{*}{Sketching} & Ease-of-use & A & D & D & A \\
 &  & Innovativeness & B & C & C & B  \\
 & \multirow{2}{*}{No-sketching} & Ease-of-use & C & B & B & C \\
  &  & Innovativeness & D & A & A & D \\
  & & & & & & \\
 \multirow{4}{*}{Value-Account} & \multirow{2}{*}{Sketching} & Ease-of-use & E & H & H & E \\
 &  & Innovativeness & F & G & G & F  \\
 & \multirow{2}{*}{No-sketching} & Ease-of-use & G & F & F & G \\
  &  & Innovativeness & H & E & E & H \\ \bottomrule
\end{tabular}
\end{table}

In each session, participants used two different tools for reporting on two different qualities of their mobile phones: a tool that allowed both sketching and reporting and a tool that allowed only reporting. We did not follow a complete design as that would mean that participants would have to report on each quality twice. This however restricts the power of the experiment as only between-subjects analyses may be performed on the data.

Participants were split into two groups; the first group used the Constructive iScale while the second used the Value-Account iScale. The product qualities used in this study were \emph{ease-of-use} and \emph{innovativeness}. Both qualities were introduced to participants through the definitions used in study 1. The order of tool and product quality was counterbalanced within each group using a latin square design, and reversed across the two sessions so that each participant would use, in the second session, the same tool for reporting on the same quality but in the reverse order (see table \ref{tab:ExperimentalDesign}).

\subsubsection{Assumptions}
Despite the fact that the study was explorative in nature, a number of predictions about the differences in performance of the three versions of the tool can be made.

First, based on existing evidence that the reconstruction of events in a serial chronological order cues the recall of both a) more temporally surrounding experiences and b) more contextual cues about the experience being reported \cite{Anderson1993}, it is expected that within the constructive condition, sketching will result in an increase of a) the number of experiences being reported and b) the richness of reported experiences. For the Value-Account iScale which obstructs participants in reconstructing their experiences in a chronological order, the difference to experience reporting is expected to be smaller.

Second, as long as participants are expected to recall more contextual cues in the constructive iScale, this is expected to influence their test-retest reliability in recalling factual details of the past experiences, such as temporal information (e.g. when did the experience take place). In a similar vein, this is expected to happen only in the constructive condition, but not in the value-account condition.

Third, the consistency of the participants' sketches, i.e. value-charged information, is expected to be higher in the value-account version, where participants cue this directly through a hypothetical memory structure, compared to the constructive version, where participants are assumed to reconstruct this information from concrete contextual cues recalled from episodic memory. This is based on an assumption that repeated chronological reconstruction might cue a different set of experiences and thus lead to a different path in reconstructing the overall pattern of how one's experience with a product developed over time.

\subsection{Analysis and results}
A total of 457 experience narratives were elicited. Participants provided an average of 3.7 to 6 experience reports depending on the recall condition. 95\% of all experiences related to the first six months of use. We compare the two sketching tools to the no-sketching condition in terms of a) number of elicited experiences, b) qualitative richness of elicited experiences, c) reliability of recalled event-specific information, and d) reliability of overall value-charged information.

\subsubsection{Does sketching impact recalling?}
\subsubsection{Does sketching result in an increase in the number of reported experiences?}

\begin{figure*}
\centerline{\includegraphics[width=6.2cm]{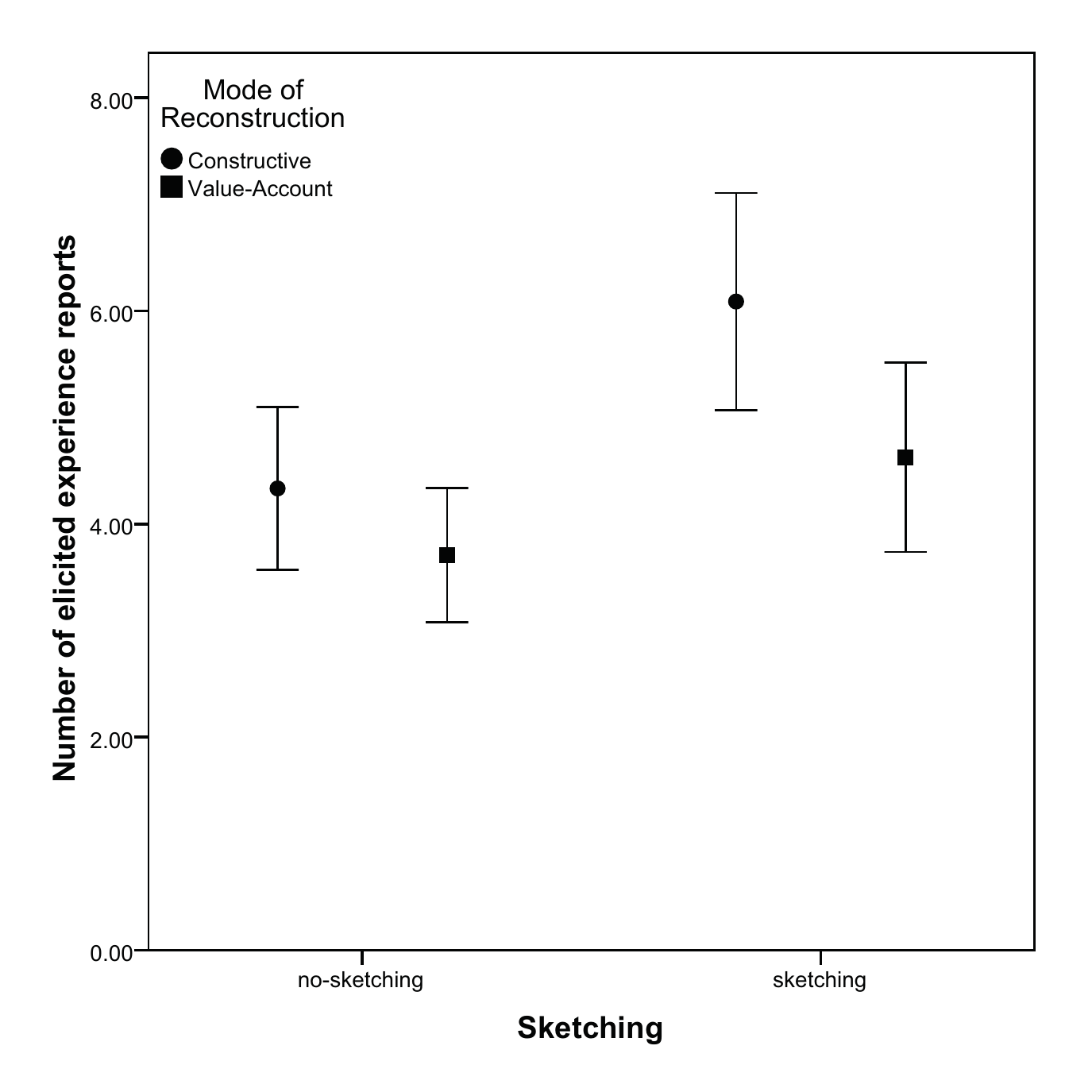} \includegraphics[width=6.2cm]{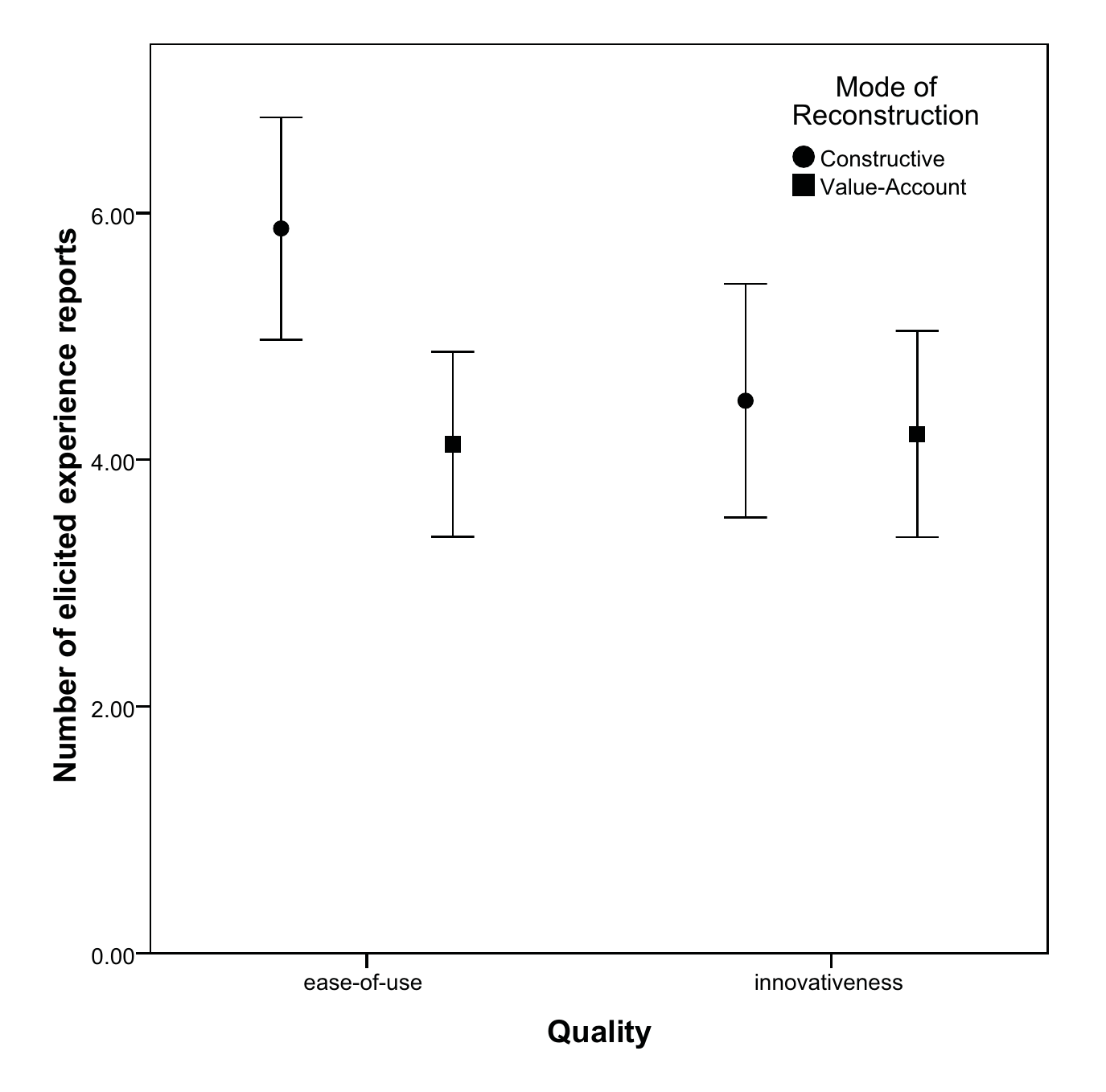}}
  \caption{Average number of experience reports elicited (a) in the presence or absence of sketching in the constructive and in the Value Account modes of reconstruction, and (b) for the two respective product qualities in relation to mode of reconstruction.}
  \label{fig:NoReports}
\end{figure*}

Figure \ref{fig:NoReports}a shows the number of reported experiences as a function of the mode of elicitation. In the constructive mode of recalling, an average of 6 experience reports was elicited when using iScale with an average of 4.4 when only reporting was used. In the value-account condition, the number was 4.6 when reporting was combined with sketching and 3.7 when participants only reported, without sketching, on their experiences.

An analysis of variance with number of experience reports as dependent variable and mode of reconstruction (constructive, value-account), sketching (no-sketching, sketching) and quality (ease-of-use, stimulation)  as independent variables displayed significant effects for mode of reconstruction, F(1,88) = 7.16, p$<$.05, and the presence of sketching, F(1,88) = 10.55, p$<$.05. As expected, sketching resulted in a higher number of elicited experience reports compared to non-sketching (sketching: $\mu$ = 5.3, no-sketching: $\mu$ = 4.0) as it provided a temporal context during recall, which cued the recall of further experiences. This did not only happen for the constructive sketching tool, which imposes a chronological order of events as existing retrospective diary methods \cite{Kahneman2004}, but also for the Value-Account sketching tool. However, the constructive sketching tool resulted overall in more experience reports than the Value-Account sketching tool (constructive: $\mu$ = 5.2, value-account: $\mu$ = 4.2).

Further, while there was no overall effect of the product quality being reported, we found a marginal interaction effect between mode of reconstruction and quality, F(1, 88) = 2.92, p=.09. In the constructive condition participants elicited a significantly higher number of experience reports when reporting on ease-of-use ($\mu$=5.9) than when reporting on innovativeness ($\mu$=4.5), t(46)=-2.14, p$<$.05. This was not observed in the Value Account condition. One possible interpretation might tap into the different nature of events that relate to ease-of-use and innovativeness. \citeN{wilamowitz2007} observed that participants often recall with greater ease contextual cues about experiences relating to ease-of-use rather than stimulation.  Ease of use is tied to concrete action, whereas stimulation cannot be allocated to specific events. Thus, the effect of chronological order in reconstruction may be more salient in case of contextually rich experiences than in case of more abstract ones. All other main effects and interactions remained insignificant.

\subsubsection{Does sketching result in richer reported experiences?}
Based on existing knowledge \cite{Barsalou1988}, one would expect that reconstructing in a chronological order would lead to more contextual cues in the elicited experience reports, thus provide richer insight into users' experiences. Such contextual information may relate to temporal (i.e. when did the event happen), factual (i.e. what happened), social (i.e. who was present) etc. To identify these different factors of richness, we submitted the experience reports to a qualitative content analysis. Open coding \cite{Strauss1998} was performed by the first author and resulted in three main types of information present in experience reports: \emph{temporal information} summarizes references to the exact time at which the experience took place, \emph{discrete event information} summarizes references to a concrete occurrence that influenced the experience, and lastly, \emph{expectations} summarize references to participants' expectations about the reported experience.

Sketching was found to have a significant effect on the number of contextual cues referring to a discrete event in the constructive mode of reconstruction ($\chi^{2}$=4.07, p$<$.05) where 45 out of 146 reports contained at least one cue referring to a discrete event in the sketching condition as opposed to 20 out 103 in the no-sketching condition (see table \ref{tab:Richness} line b), but not in the value-account condition ($\chi^{2}$=0.21, p=.650) where 27 out of 118 reports referred to discrete event in the sketching condition as opposed to 18 out of 89 in the no-sketching condition. No significant effect was found on the number of temporal cues present in experience reports both for the constructive (Pearson's $\chi^{2}$=2.13, p=.14), and for the value-account mode of reconstruction ($\chi^{2}$=2.48, p=.12). Last, no significant differences were found on the number of experience reports containing cues about participants' expectations before the experienced event.

In a similar vein, one might expect that recalling more contextual cues about the experienced event will result in an increase in design-relevant information. We distinguished four groups in the design relevance of an experience report. First, an experience report might provide \emph{no design relevant information}, e.g. the participant only reflects on her attitude towards the product without any further reasoning. Second, the participant may \emph{pinpoint a specific feature }of the product that an evaluative judgment is made for without any reasoning motivating her attitude. Third, the report might also \emph{contain reasoning about the reported attitude} such as the nature of an interaction problem or the exact actions that the user performed and which lead to the realization of the problem. Fourth, the participant might \emph{suggest a design solution }for solving the reported problem. Overall, no significant differences were found in the design relevance of the experience reports across the different conditions (see table \ref{tab:Richness}). Thus, while sketching seems to impact the richness of elicited experience reports, it does not necessarily lead to significantly more design-relevant information.

\begin{table}
\centering
\caption{Number of experience reports judged for the three dimensions of richness and for the four levels of design relevance, for the four conditions resulting from presence or absence of sketching and constructive or value-account mode of reconstruction} \label{tab:Richness}
\begin{tabular}{p{8cm}@{\hspace{1em}}ccccc} \toprule
 & & \multicolumn{2}{c}{Constr.} & \multicolumn{2}{c}{V-A} \\
Name & & S & NS & S & NS \\ \midrule
\emph{Contextual information} &  &  &  & \\ \midrule
\multirow{2}{8cm}{a. Temporal: Does the participant recall temporal information about the reported experience?} & Y: & 20 & 8 & 16 & 6 \\
 & N: & 126 & 95 & 102 & 83 \\
\multirow{2}{8cm}{b. Event: Does the participant recall one or more discrete events that lead to the realization of the reported experience?} & Y: & 45 & 20 & 27 & 18 \\
 & N: & 101 & 83 & 91 & 71 \\

\multirow{2}{8cm}{c. Expectation: Does the participant recall his/her expectations about the reported experience?} & Y: & 10 & 12 & 11 & 4 \\
 & N: & 136 & 91 & 107 & 85 \\ \midrule
\emph{Design Relevance}  &  &  &  & \\\midrule
 0. The report contains no design relevant information & & 12 & 5 & 5 & 3 \\
 1. The report pinpoints a product feature that the user is positive/negative about & & 67 & 41 & 64 & 42 \\
 2. The report contains reasoning about the users' attitude towards the feature & & 66 & 55 & 49 & 44 \\
 3. The report suggests a design solution & & 1 & 2 & 0 & 0 \\ \midrule
Total & & 146 & 103 & 118 & 89 \\ \bottomrule
\end{tabular}
\end{table}

\subsubsection{Does sketching result in more reliable recall of temporal structure?}
In the introduction of this paper we argued that while the veridicality of reconstructed experiences, i.e. the convergence of memory and actuality, is not that important, their reliability, i.e. the convergence of the same memory across multiple recalls, is. In other words, we argued that even if what we remember might be different from what we experienced, as long as these memories are consistent over multiple recalls, they provide valuable information.

In this section we assess the test-retest reliability of the recalled experiences based on their temporal information. Participants joined two distinct sessions with the second session being no less than seven and no more than 10 days after the first. In both sessions participants were asked to perform the same task, to report their experiences with the product. The conditions (i.e. presence or absence of sketching and product quality) remained the same across the two sessions, but the order of reporting was counterbalanced. For example, participants that used the iScale tool (sketching condition) to report on ease-of-use followed by the Experience Reporting tool (no-sketching condition) to report on innovativeness, in the second session they first used the Experience Reporting tool to report on innovativeness followed by the iScale tool to report on ease-of-use. These two sessions are expected to more or less result in the same experiences, thus experiences across two sessions may be coupled. For each reported experience participants estimated the time (i.e. days, weeks or months after the purchase of the product) at which this experience took place. In this test we use this temporal information, the convergence of the two reported times of two coupled experience reports elicited from the two sessions, as a metric of the reliability of the reconstruction process.

\begin{figure*}
\centerline{\includegraphics[width=6cm]{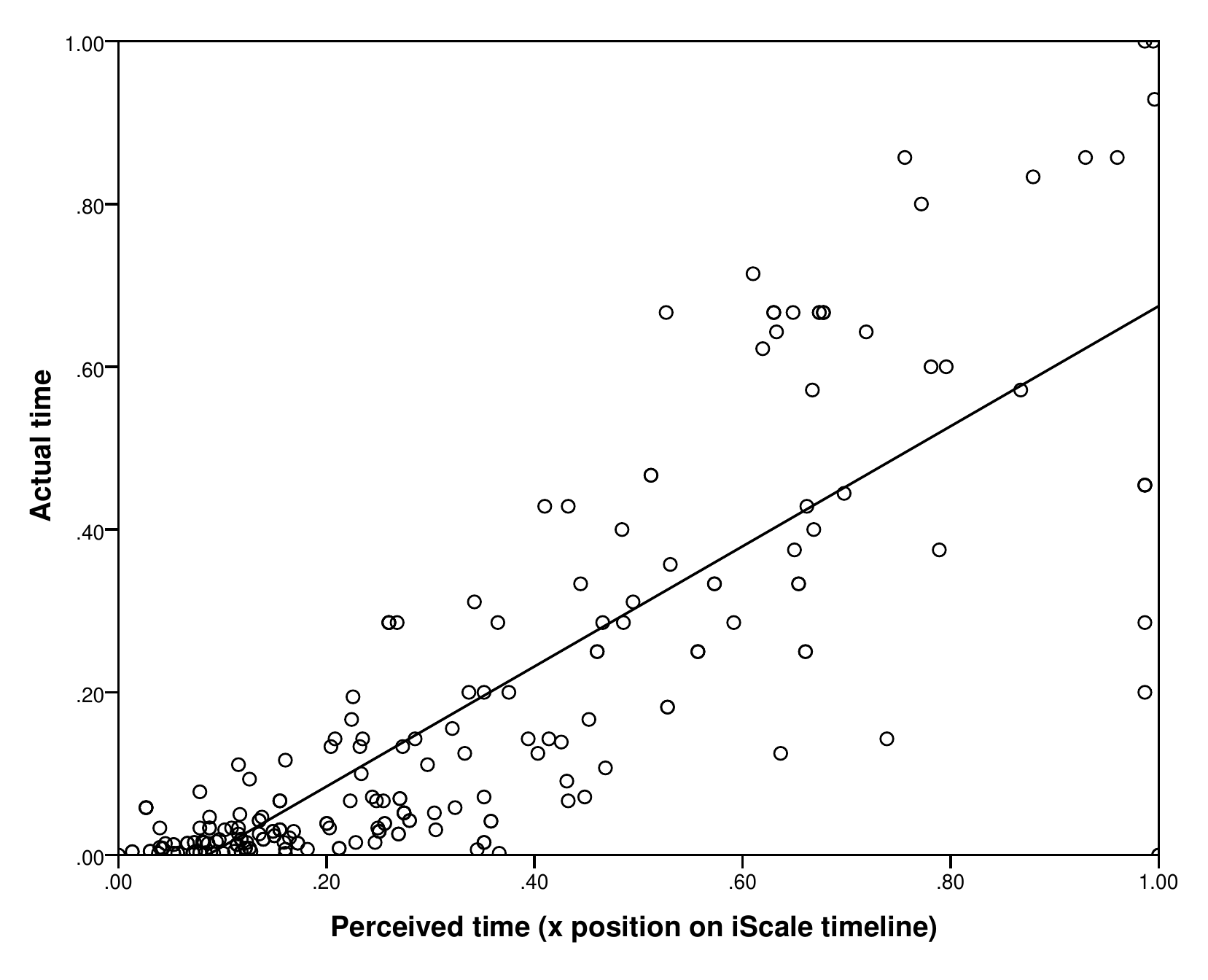}\includegraphics[width=6cm]{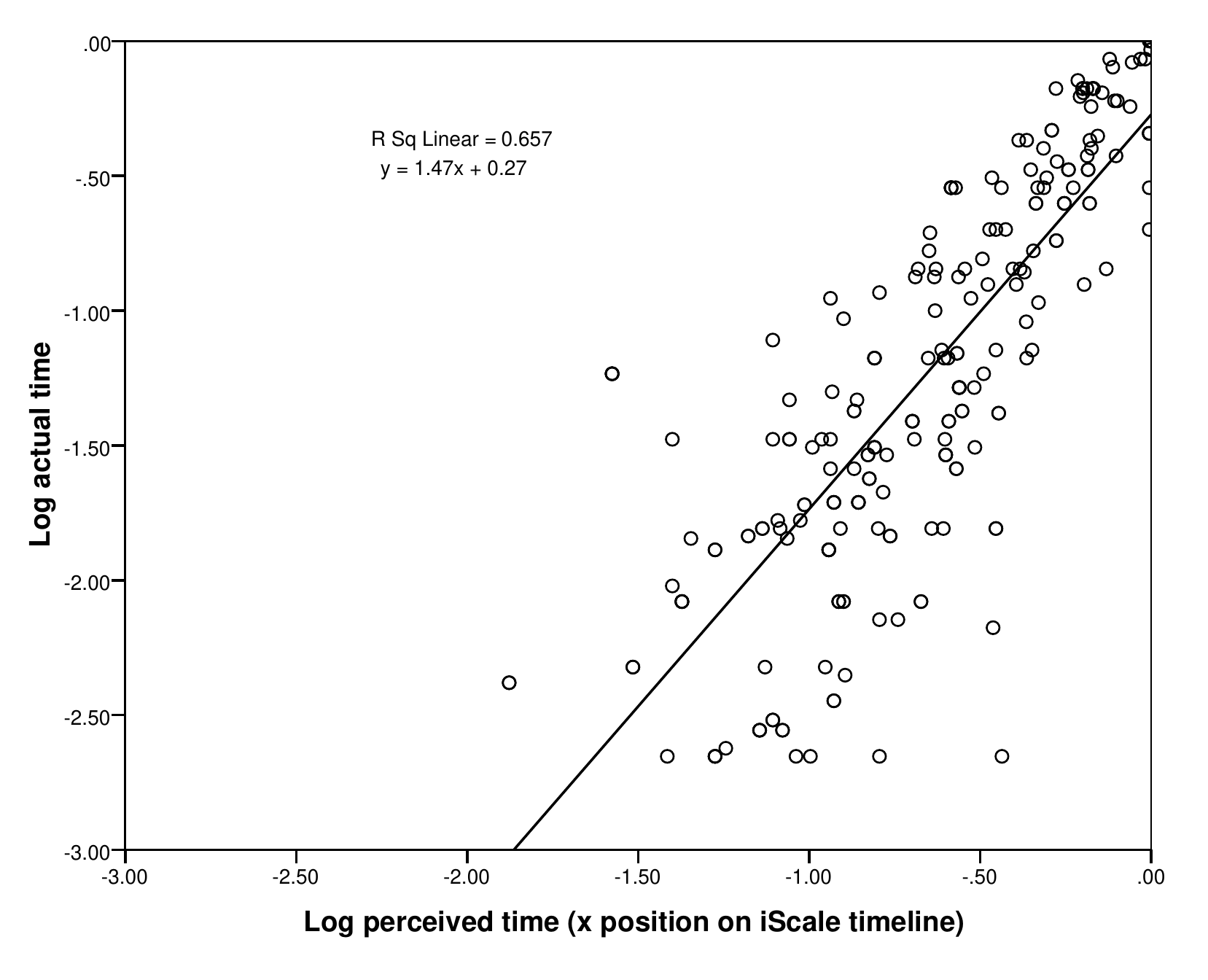}}
  \caption{Relation between actual time (reported time for a sketched node) and position of node along the iScale's x-axis: (a) linear relation, (b) power-law relation. Actual time (days) is adjusted to the full time of ownership of the product for each participant.}
  \label{fig:TimePowerLaw}
\end{figure*}

One question, however, relates to whether participants' accuracy in recalling temporal information remains constant across the full timeline, i.e. from the moment of purchase of the product to the present time. The participant's accuracy might be affected by the amount of contextual information surrounding the experience that is available at the moment of recalling. Theories of recall have suggested that recent experiences \cite{Koriat2000}, or experiences associated with important milestones (e.g. the purchase of the product) \cite{Barsalou1988} might be more easily accessible. If such biases exist, they will affect the reliability test as differences in the reliability of experiences reports might be due to pertaining to more or less salient periods and not due to the reconstruction process. In the presence of such biases, the temporal distance between the two coupled experience reports elicited in the two distinct sessions should be transformed to account for the accessibility biases.

We attempt to assess the existence of accessibility biases through examining the way in which participants used the timescale of the tool, i.e. iScale's x-axis. Participants sketched linear curves through adding nodes in the graph (see figure \ref{iScaleP1}a). Each node can be characterized by two properties: a) the actual time (participants explicitly annotated for each node the approximate amount of days, weeks, or months after purchase that this node represents, and b) the perceived time (the position of the node along the x-axis of iScale).

Figure \ref{fig:TimePowerLaw} depicts the relationship between the reported (actual) time versus the perceived time, i.e. the position of the node along iScale's x-axis. To enable an across-subject comparison, we normalized the reported (actual) time variable by the total time of ownership of the product for each participant, resulting to an index from 0 to 1. Given no accessibility bias, one would expect a linear relationship between these two pieces of information. One might note in figure \ref{fig:TimePowerLaw}a, however, that the variance in the dependent variable (actual time) is not uniformly distributed across the range of the independent variable (position along the x-axis of iScale). If one transforms the variables by the logarithmic function, the data become much more uniformly distributed. A linear regression on these transformed variables shows a significant prediction accounting for 66\% of the variance in the dependent variable. This suggests a power law relation between the recalled actual time of the experienced event and its position along the sketching tool's timeline with a power equal to 1/1.47=0.68 (i.e. perceived-time = actual-time$^{0.68}$). In other words, participants had a tendency to use a substantial fraction of the x-axis of iScale to map their initial experiences. In a similar vein, 95\% of all experience narratives related to the first six months of use and 75\% of all experience narratives related to the first month of use whereas the median time of ownership was 10 months. It thus becomes evident that experiences pertaining to initial use are more accessible in participants' memory. To account for this accessibility bias we compute the temporal distance between two events through the following formula:

\begin{equation}\label{distance}
\Delta = Abs ( log(t_{2}) - log(t_{1}) )
\end{equation}

where t is the reported time that has elapsed from the moment of purchase of  the product.

We used formula \ref{distance} to compute the temporal distance between the recalled time of occurrence of an experience, across the two sessions that each individual participated in. In the second session participants were asked to repeat the same tasks as in session 1. Participants were expected to elicit approximately the same experiences. The elicited experience reports from the two sessions were coupled based on their textual descriptions. The temporal distance was then calculated between each experience report elicited in session 1 and its coupled report elicited in session 2.

Following existing evidence, we assumed that that serial reconstruction, i.e. reconstructing in a chronological order, helps participants in eliciting the surrounding context of each experience \cite{Anderson1993}, and this positively affects participants' ability to recall more reliably contextual cues of the experience, such as when exactly did the experience take place \cite{Kahneman2004}. Similarly, we assumed that sketching (even in the value-account condition), when compared to the no-sketching condition, would result in more reliable recall of such temporal information as sketching provides a temporal context for all recalls, i.e. each experience report is related to others in a single timeline.

An analysis of variance with temporal distance $\Delta$ between experience reports from session 1 and session 2 as dependent variable and mode of reconstruction, presence or absence of sketching and quality of reporting as independent variables displayed significant effects for the presence or absence of sketching, F(1, 265) = 9.61, p$<$.01, where the sketching condition resulted in significantly lower temporal distance (higher consistency) than the no-sketching condition (sketching: $\mu_{\Delta}$=0.178, no-sketching: $\mu_{\Delta}$=0.272),  and the quality of reporting, F(1,265) = 7.43, p$<$.01, where narratives of ease-of-use were temporally more consistent than narratives of innovativeness (ease-of-use: $\mu_{\Delta}$=0.183, innovativeness: $\mu_{\Delta}$=0.266). All other main effects and interaction remained insignificant.

\begin{figure*}
\centerline{\includegraphics[width=9cm]{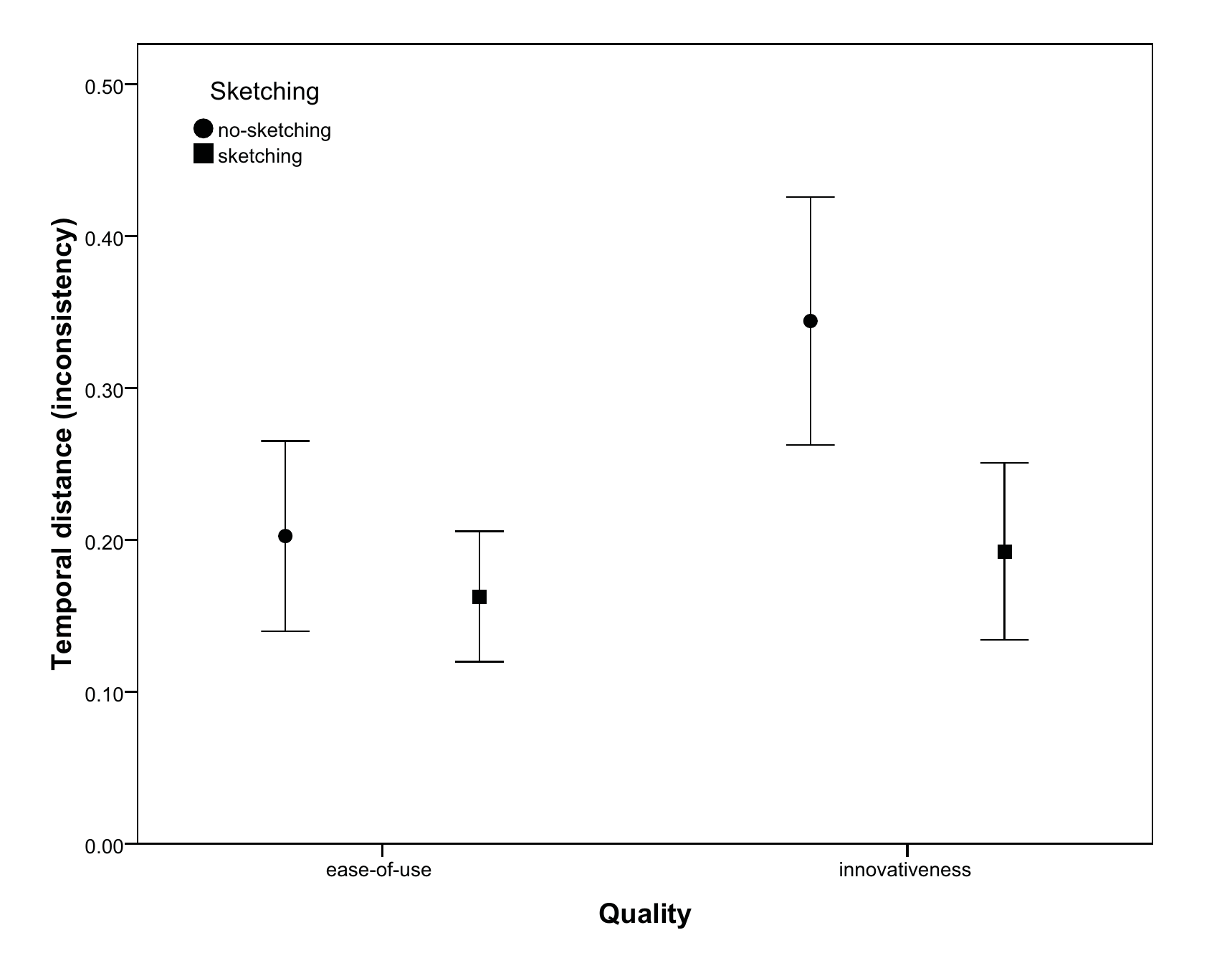}}
  \caption{Temporal inconsistency in the sketching and the no-sketching condition for the two respective qualities of reporting.}
  \label{fig:TemporalConsistency}
\end{figure*}

The results supported our initial expectations that sketching provides a temporal context and thus results in higher test-retest reliability of the reconstruction process, at least in terms of temporal information. Moreover, similarly to the analysis of the total number of experience narratives, experiences relating to the ease-of-use of the product were more reliably recalled in comparison to ones relating to the products' innovativeness. This may be attributed to the same reasoning that experiences of ease of use are tied to concrete action, whereas experiences of innovativeness cannot be allocated to specific events. Thus, participants might be better able to reconstruct such temporal information in experiences relating to concrete events where more contextual information is available than in more abstract reports of innovativeness. Last, contrary to our initial expectation, no significant difference was found between the constructive and the value-account condition. A marginal interaction effect between the presence or absence of sketching and the quality of reporting was found, F(1,265) = 3.32, p=.069 . The presence of sketching had a stronger impact when participants were reporting on innovativeness than when reporting on ease-of-use (figure \ref{fig:TemporalConsistency}a). For experience reports that related to the products' ease-of-use, sketching had an impact in the constructive condition, but not in the value-account condition. This was not observed in experience reports of innovativeness for which sketching had a strong impact in both the constructive and the value-account conditions.

\subsubsection{Does Value-Account result in higher sketched pattern consistency?}
In the previous section we tested the reliability of recalled temporal information of experience reports across the two sessions of the study. In this section we will test the consistency of the sketched patterns, the overall evaluative judgment on participants' experiences over time. We expect that the value-account condition will result in a higher consistency in the sketched pattern. It is sensible to assume that given that this value-charged information is assumed to be cued directly from a hypothetical memory structure \cite{Betsch2001} in the value-account condition and thus would be more consistent across repeated recalls. Contrary, in the constructive condition we assume that this information is reconstructed from contextual details that are recalled from episodic memory and thus will result in lower consistency across repeated recalls, as repeated chronological reconstruction might cue a different set of experiences and thus follow a different path in reconstructing the overall pattern of how one's experience with a product developed over time.

\begin{figure*}
\centerline{\includegraphics[width=5.7cm, trim=0.625in 5in 2.75in 0.8in, clip=true]{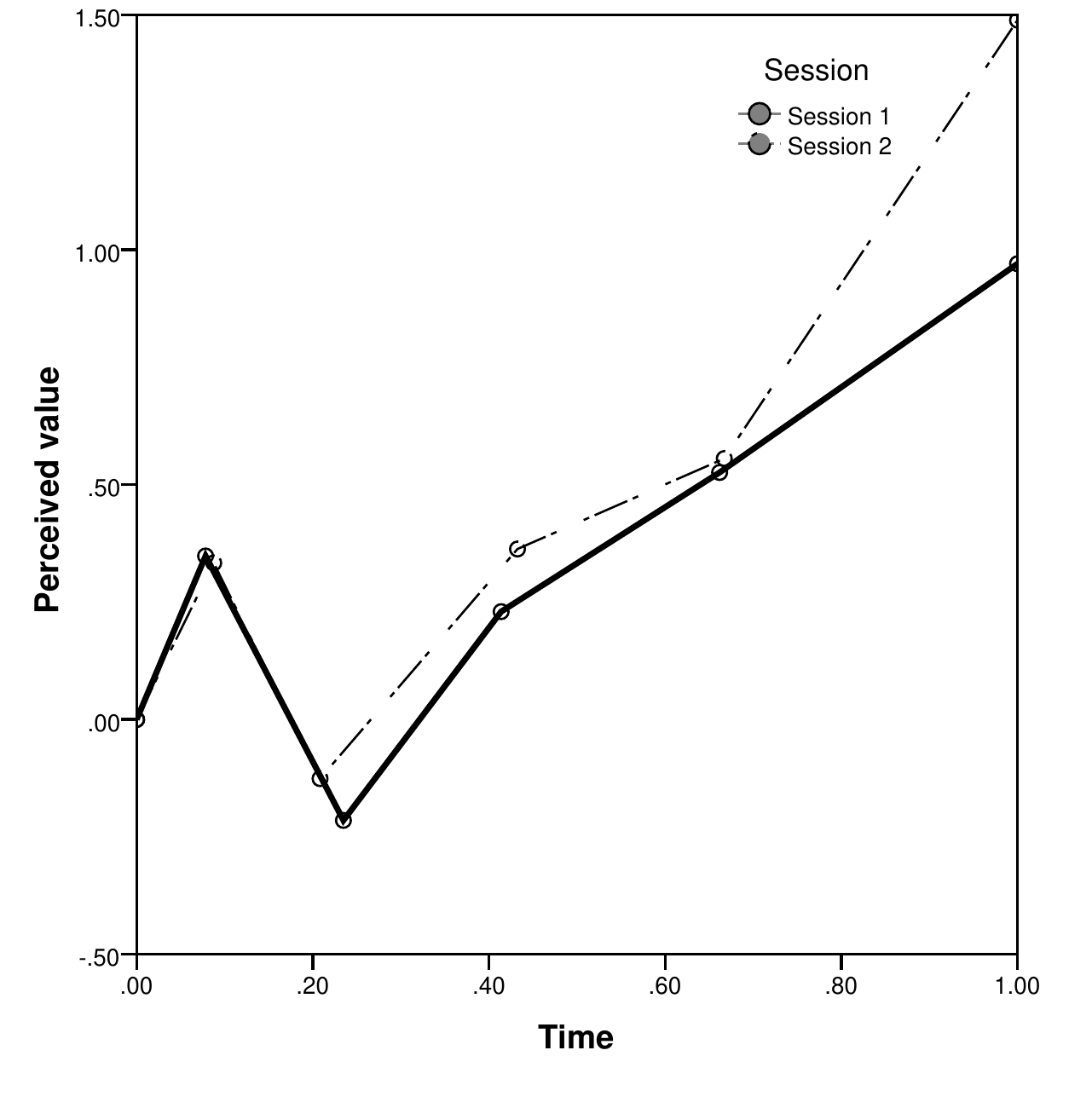} \includegraphics[width=5.7cm, trim=0.625in 5in 2.75in 0.8in, clip=true]{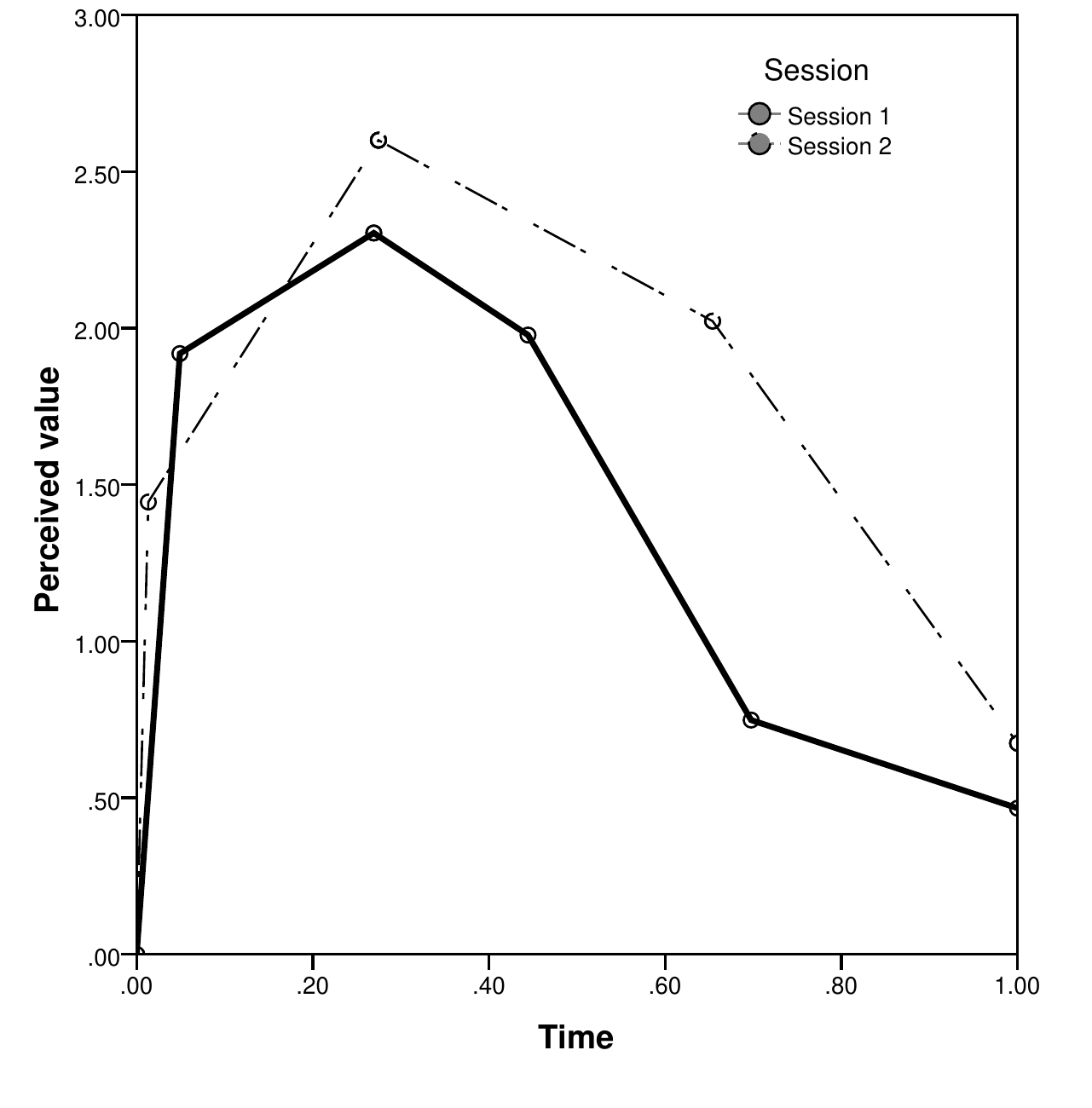}}
\centerline{\includegraphics[width=5.7cm, trim=0.625in 4.9in 2.75in 0.9in, clip=true]{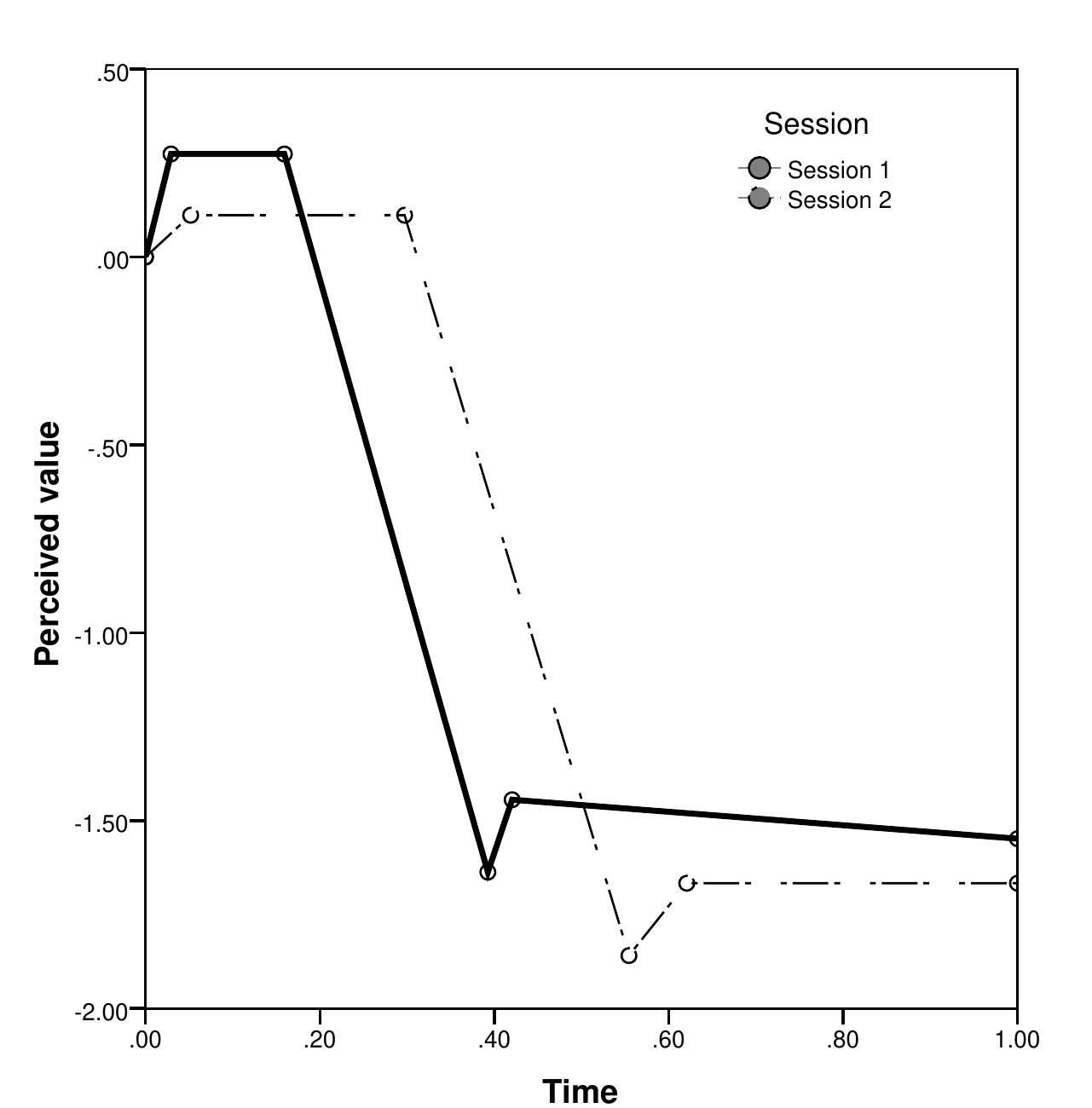} \includegraphics[width=5.7cm, trim=0.625in 4.9in 2.75in 0.9in, clip=true]{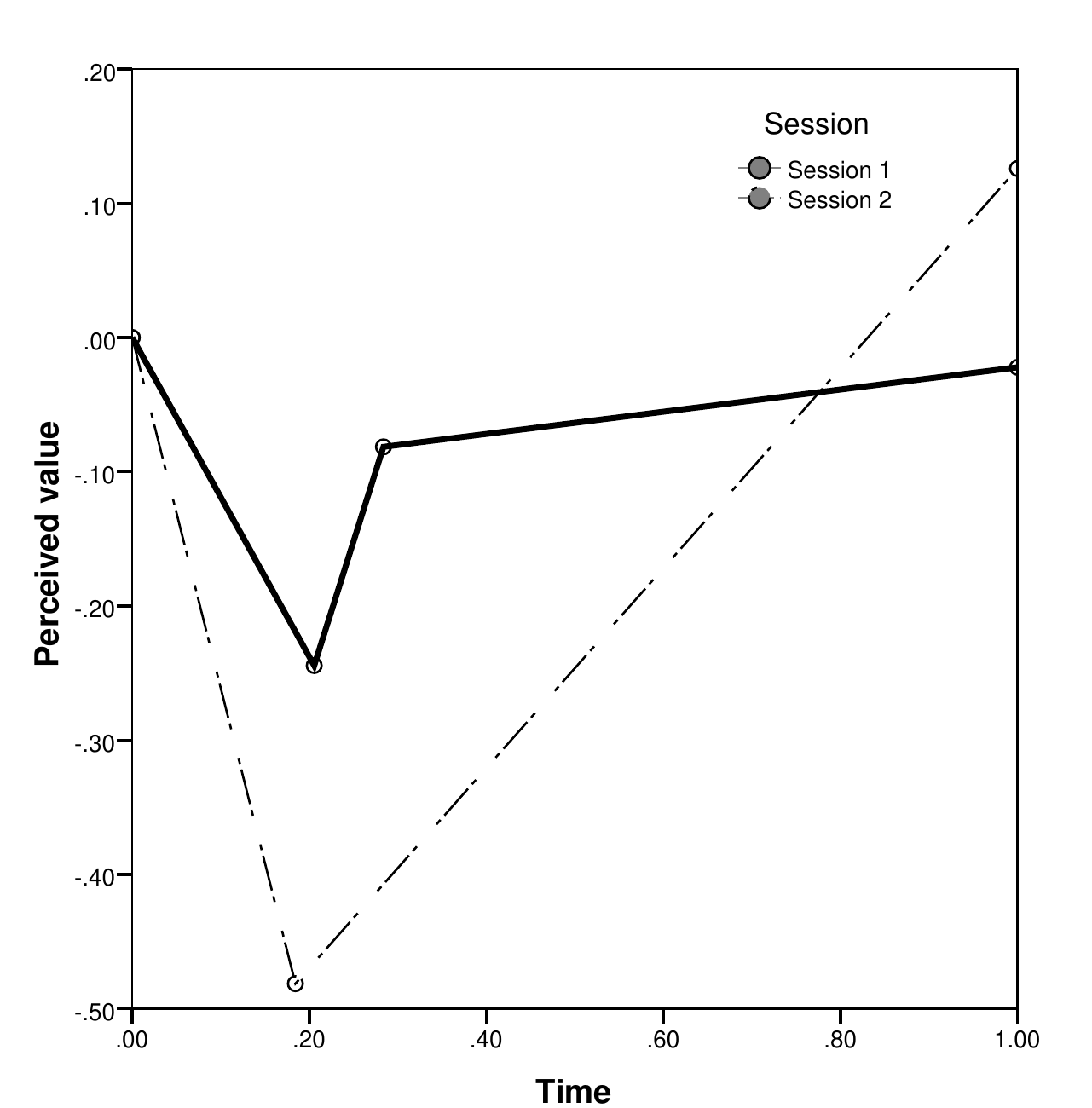}}
  \caption{Example sketches elicited in the constructive (top) and the value-account (bottom) conditions during the two sessions.}
  \label{fig:SketchesISCALE}
\end{figure*}

Figure \ref{fig:SketchesISCALE} displays example graphs sketched by two participants in two respective sessions in the constructive condition and two participants in the value-account condition (see appendix \ref{sec:elecappendix} for the complete set of elicited sketches). The area A between the two sketches is a simple measure for the inconsistency in participants' sketched pattern over repeated recalls. It was calculated through sampling the graphs in 100 steps.

An analysis of variance with this area measure as dependent variable and mode of reconstruction and product quality being reported as independent variables displayed significant effects for the mode of reconstruction, F(1, 45) = 6.05, p$<$.05, while all other main effects and interactions remained insignificant. Contrary to what we expected, participants were more inconsistent across the two sessions when employing the value-account tool (constructive: $\mu_{A}$= 30.2, value-account: $\mu_{A}$=50.8). This might relate to the finding of \citeN{Reyna1994} that participants may falsely recognize the overall gist of a memory while correctly recalling its exact details.

\subsubsection{How does the perceived quality of mobile phones change over time?}
While presenting a detailed analysis of the obtained data is not within the scope of this paper, this section provides a short overview of how users' reported experiences and perceptions developed over time.

First, participants' elicited experiences were content analyzed \cite{Hsieh2005} using an existing classification scheme proposed in \cite{Karapanos2009UXOT}. Due to the limited scope of the elicited data in this study, as participants were restricted to reporting only two distinct qualities of the product, we identified four different types of reports. Reports relating to learnability and stimulation and thus signifying the existence of an orientation phase in the adoption of the product, as well as reports relating to long-term usability and usefulness signifying the existence of the incorporation phase \cite{Karapanos2009UXOT}.

\begin{figure*}
\centerline{\includegraphics[width=6cm]{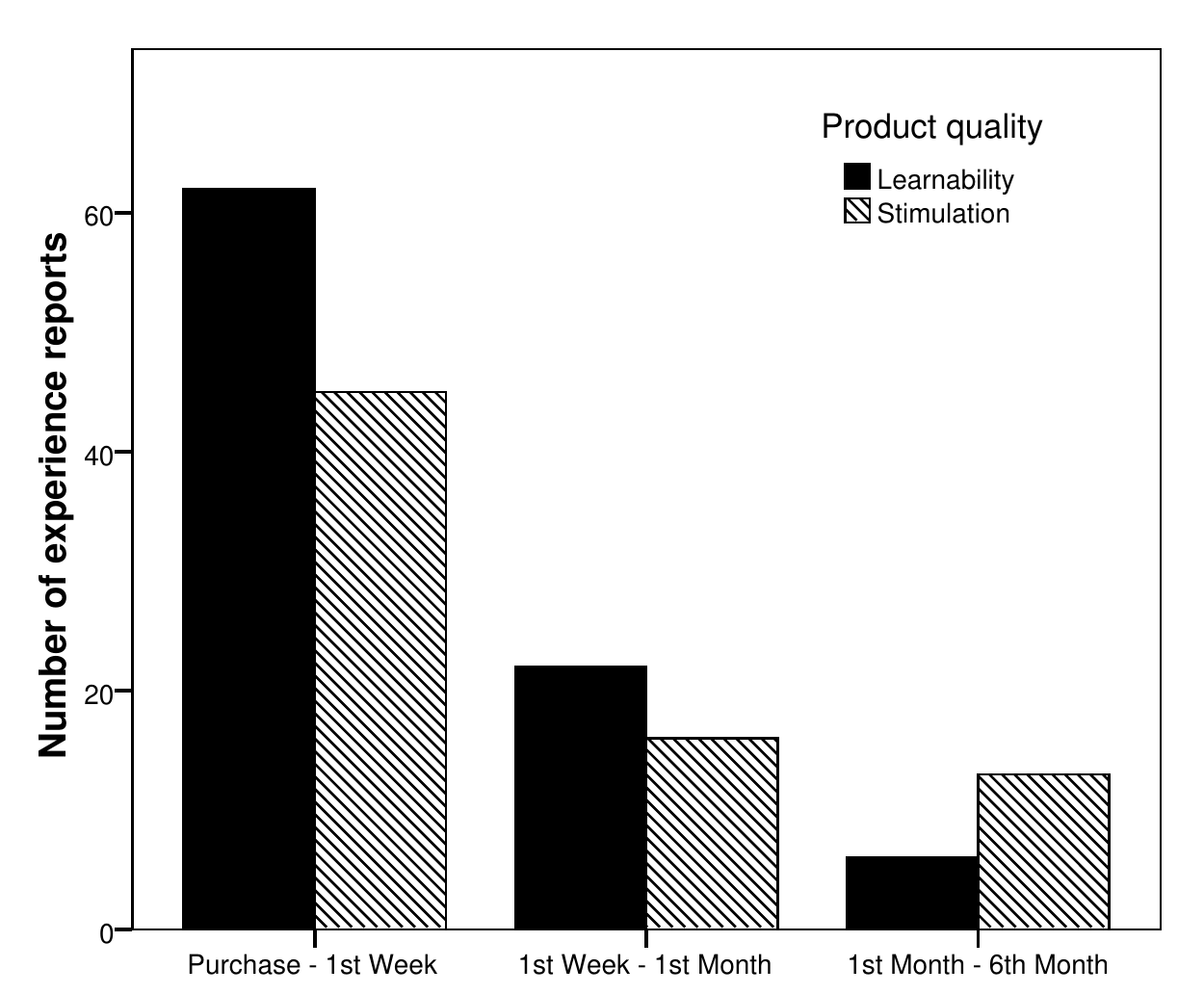}\includegraphics[width=6cm]{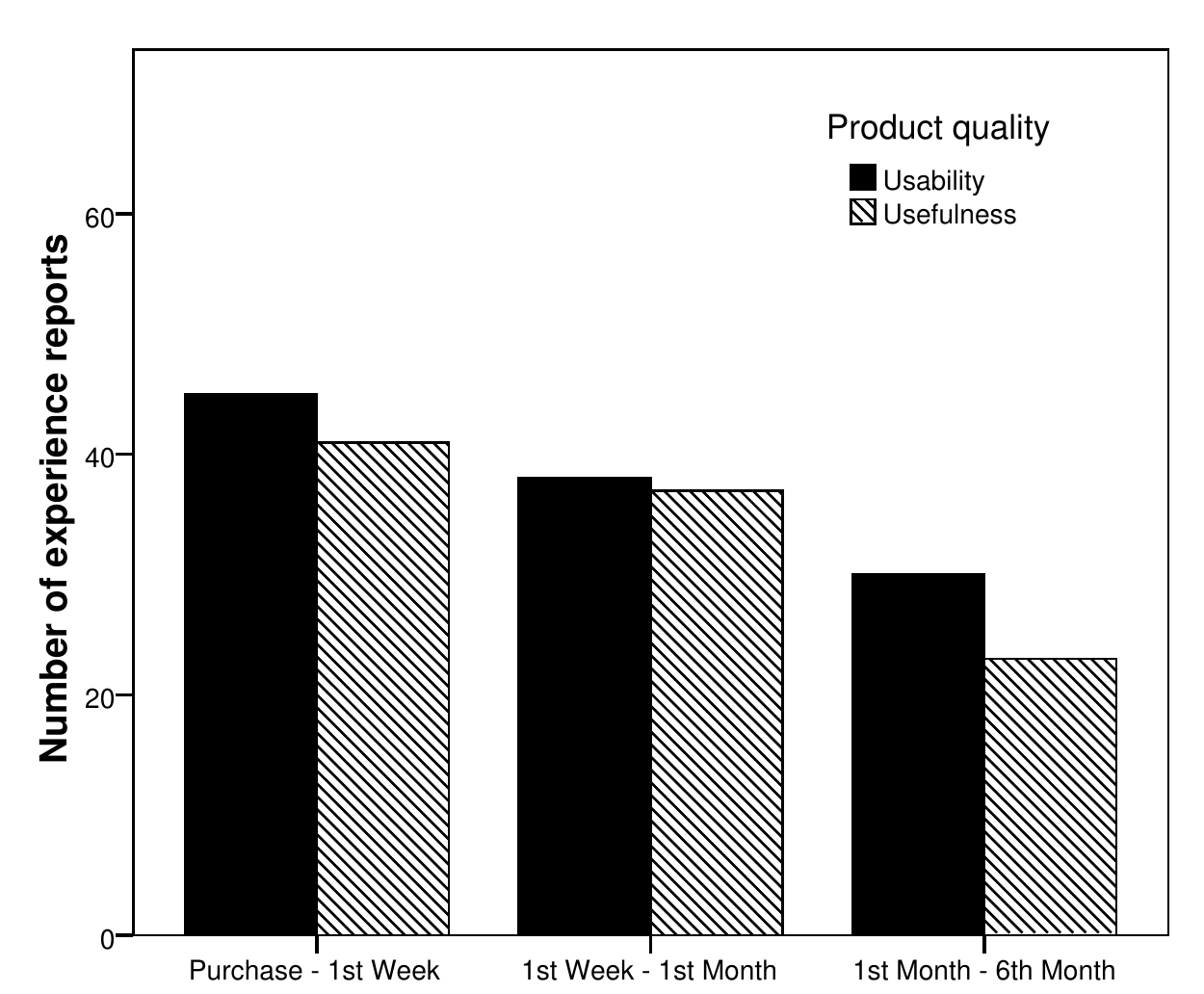}}
  \caption{Number of narratives in a certain period relating to learnability, stimulation, long-term usability and usefulness as a function of time of ownership.}
  \label{fig:Graphs}
\end{figure*}

Figure \ref{fig:Graphs} illustrates the number of experience reports for the four categories of experiences and over three time-periods: during the first week, during the first month, and after 6 months. In agreement with prior studies \cite{Karapanos2009UXOT,Wilamowitz06,wilamowitz2007}, the dominance of learnability and stimulation experiences decrease over time, while those associated with long-term usability and usefulness increase. Most learnability and stimulation experiences are associated with the first week of use. Experiences relating to long-term usability and usefulness also decrease over time, but their relative dominance increases. It is surprising that 95\% of users remembered experiences relate to the first 6 months of using the product.

Participants' sketched patterns were sampled according to the actual time, i.e. participants' reported days in the timeline, to construct average graphs of how two product qualities, i.e. ease of use and innovativeness, changed over time. In this sampling, each participant's sketched pattern is transformed to an actual timeline by mapping participants' reported days to a continuous actual time scale and then all patterns are sampled to create an average pattern. Sampling followed a power law.

The resulting averaged pattern suggests that users' perception of the innovativeness of mobile phones increase over the course of the first month and then remain approximately stable for the first six months. On the contrary, users seem to experience usability problems mostly in the first week of use; after this period usability displays a sharp increase over the course of the first month while this increase continues with a lower magnitude till the end of the studied six-month period.

\begin{figure}
\centerline{\includegraphics[width=8cm]{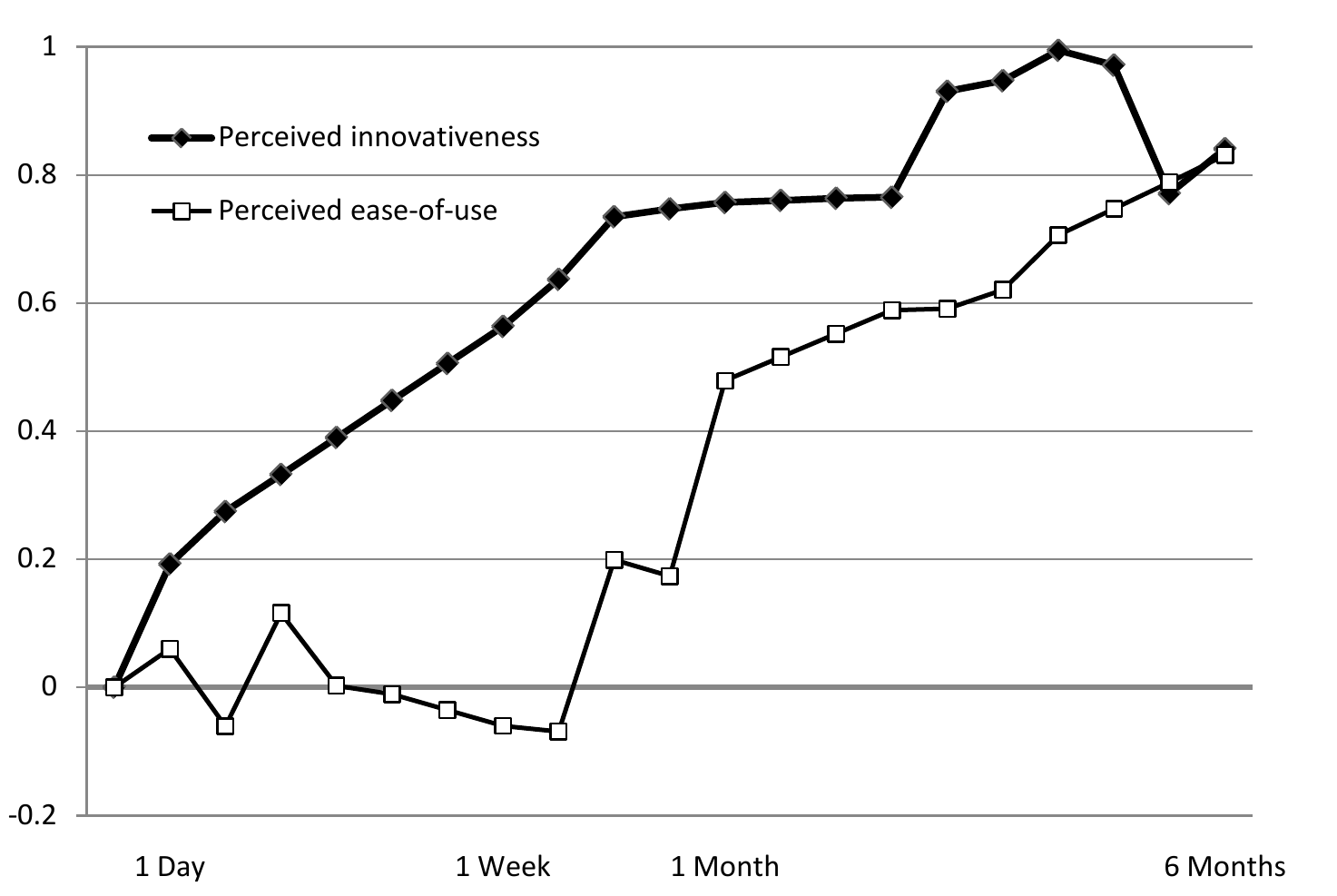}}
  \caption{Averaged across-subjects pattern sampled over actual time, i.e. participants reported days in the timeline.}
  \label{fig:SketchPattern}
\end{figure}

\section{Discussion \& Conclusion}
This paper has presented iScale, a sketching tool intended for longitudinal online surveys of user experiences. It took the general approach of retrospective elicitation of users' experiences as an alternative to longitudinal studies. More specifically, the tool was designed with the aim of increasing participants' effectiveness and reliability in recalling their experiences with a product. Two different versions of iScale were created based on two distinct theoretical approaches to the reconstruction of one's emotional experiences. The constructive iScale tool imposes a chronological order in the reconstruction of one's experiences. It assumes that chronological reconstruction results in recalling more contextual cues surrounding the experienced events and that the felt emotion is constructed on the basis of the recalled contextual cues. The value-account iScale tool aims at distinguishing the elicitation of the two kinds of information: value-charged (e.g. emotions) and contextual cues. It assumes that value-charged information can be recalled without recalling concrete contextual cues of an experienced event due to the existence of a specific memory structure that can store the frequency and intensity of one's responses to stimuli.

Study 2 tested the effectiveness of the two iScale tools against a control condition, that of reporting experiences without any form of sketching. We observed significant differences between iScale and the control condition. First, both sketching tools resulted in a higher number of experience reports when compared to the control condition, while the constructive sketching tool elicited significantly more experience reports than the value-account sketching tool. These results support the idea that sketching assists in the reconstruction of the context in which experiences took place, thus forming stronger temporal and semantic links across the distinct experiences \cite{Kahneman2004}. In addition, imposing a chronological order in the elicitation of experiences seems to have a positive impact in the effectiveness of recalling \cite{Anderson1993}.

Second, sketching was found to result in recalling significantly more contextual information regarding an experienced event when the participants followed a chronological order in the reconstruction of their experiences, i.e. in the constructive condition, but not when they followed a top-down reconstruction approach, i.e. in the value-account condition. These results provide further evidence for the claim that chronological reconstruction seems have a positive impact in the effectiveness of recalling \cite{Anderson1993}. This improved recall of contextual cues, however, did not impact participants' ability to report design relevant information that was approximately similar in all four conditions. Thus, while iScale does not influence the design relevance of elicited information, it may provide richer details about the context in which a given event was experienced.

Third, sketching seemed to have a significant impact on participants' test-retest reliability in recalling concrete contextual cues of the experienced events such as temporal information. Contrary to the initial expectations, the constructive sketching tool resulted in significantly higher consistency across two repeated recalls in the formation of overall evaluation, i.e. the sketched pattern, than the Value-Account sketching tool.

These results suggest that the constructive iScale tool outperforms the Value-Account iScale tool and offers a significant improvement in the amount, the richness and the reliability of recalled information when compared to conventional recall of experiences that does not involve any techniques for improving participants effectiveness and reliability in recalling their experiences. iScale is thus a promising tool for the elicitation of longitudinal data from memory. It provides a cost-effective solution, enabling the elicitation of large amounts of information from the field while increasing the reliability of the elicited information. One has to be aware, however, of the potential discrepancy between the actual experiences as elicited through longitudinal field studies and retrospective data elicited through iScale. These retrospections may span long periods of times and thus one would expect systematic biases to occur \cite{kahneman1999Peak-And-End,Bartlett1932}. In this paper we argued that veridicality may not be as important as the reliability of these data. This also implies that it is upon the researcher to define what she or he is interested to explore. In some cases, understanding what users remember from the temporal development of their experiences may be more important than the actual experiences, while in others the actual experiences may be what the researchers are interested in understanding.

Next, iScale will evidently result in large amounts of qualitative information that will require labor-intensive analysis given traditional human-performed qualitative data analysis procedures like Content Analysis \cite{Krippendorff2004,Hsieh2005} and Grounded Theory \cite{Strauss1998}. Novel techniques found in the field of information retrieval \cite{Landauer1997LSA,blei2003LDA} may prove especially fruitful in automating or semi-automating the qualitative analysis process. Finally, the interpersonal analysis of the sketched graphs is definitely a subject for further research and was addressed here only superficially.

\bibliographystyle{acmtrans}
\bibliography{C:/Aggelos/aggelos}

\section{Electronic Appendices}
\label{sec:elecappendix}

Participants' sketches across the two sessions, in the Constructive and in the Value-Account condition. Time (perceived) reflects the x-axis of the iScale tool. The area between the two sketches, the one elicited in the first session and the second in the second session, may be used as an index of (lack of) test-retest reliability of participants recall process.

\begin{figure*}
\centerline{\includegraphics[width=5.7cm, trim=0.625in 5in 2.75in 0.8in, clip=true]{images/perceptionsOverTime/con1.pdf} \includegraphics[width=5.7cm, trim=0.625in 5in 2.75in 0.8in, clip=true]{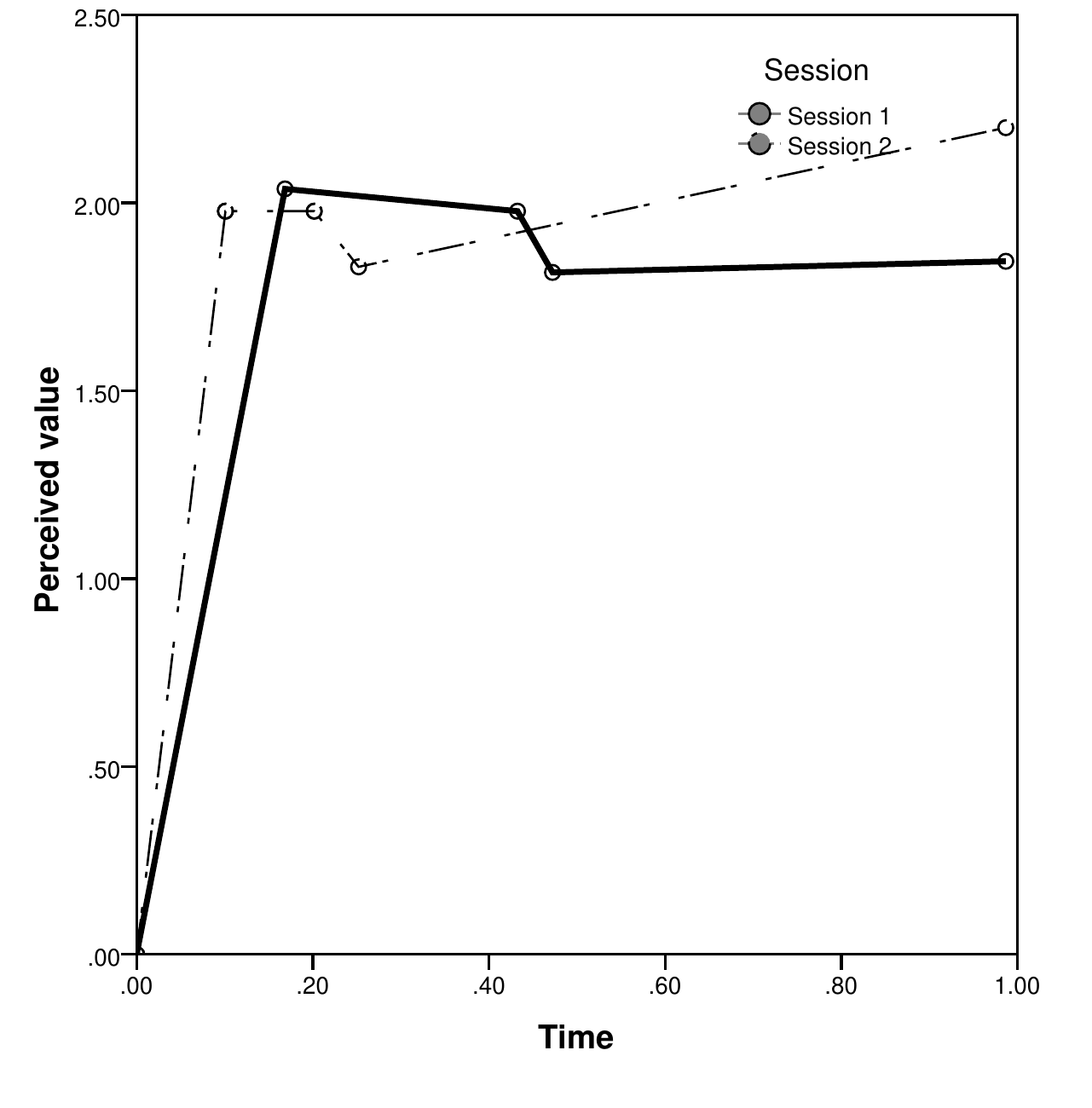}}
\centerline{\includegraphics[width=5.7cm, trim=0.625in 4.9in 2.75in 0.8in, clip=true]{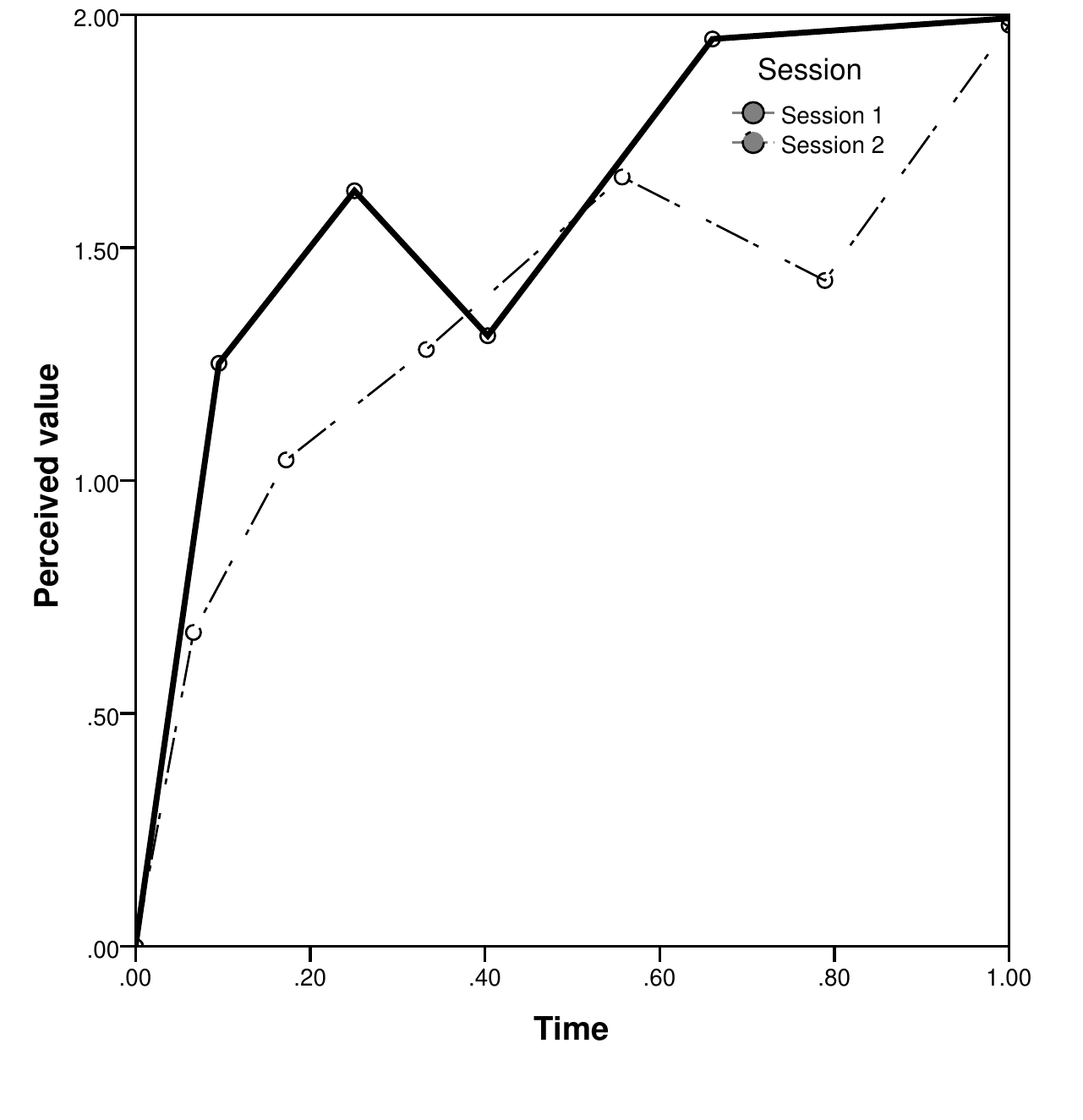} \includegraphics[width=5.7cm, trim=0.625in 4.9in 2.75in 0.8in, clip=true]{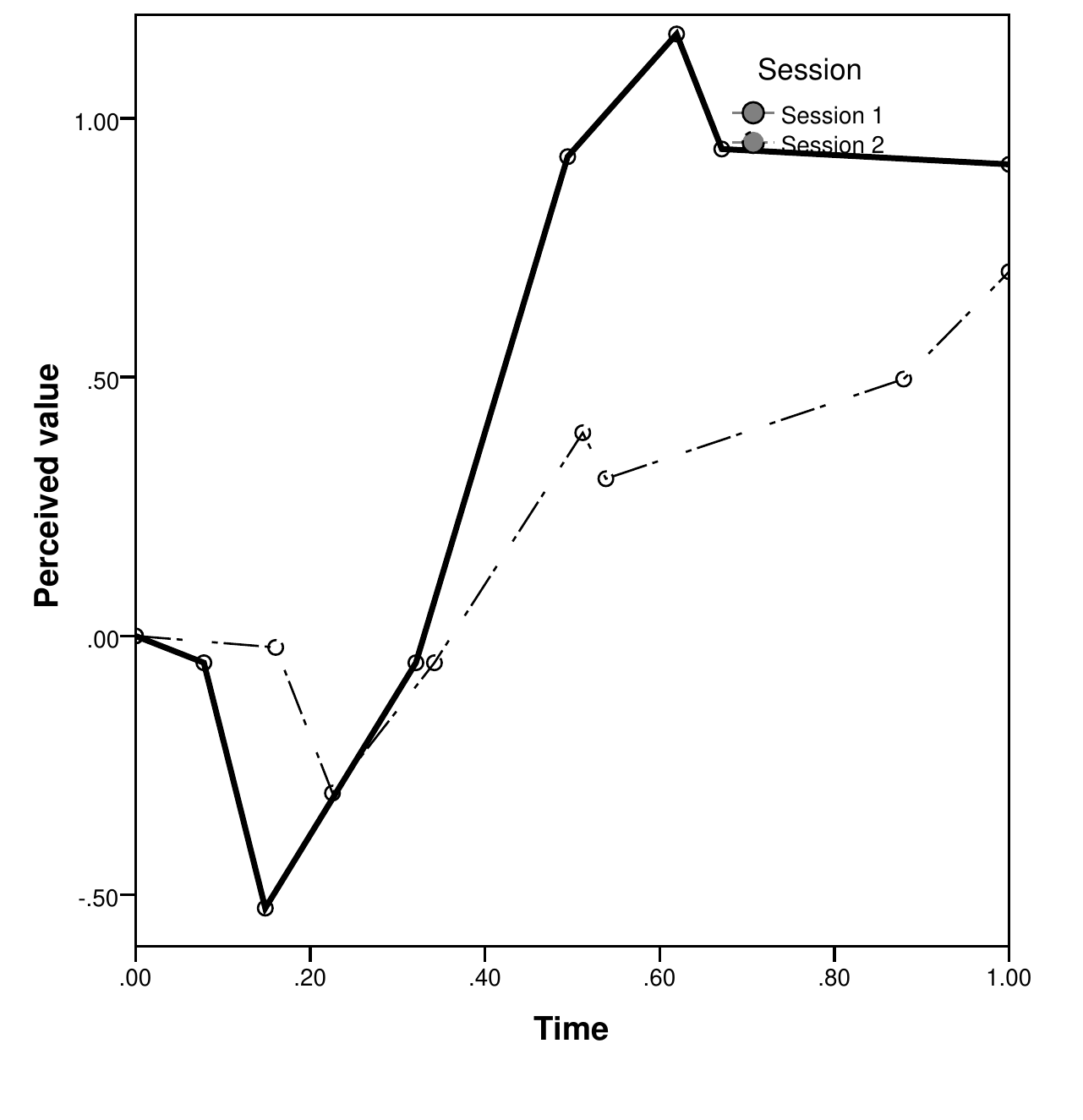}}
\centerline{\includegraphics[width=5.7cm, trim=0.625in 4.9in 2.75in 0.8in, clip=true]{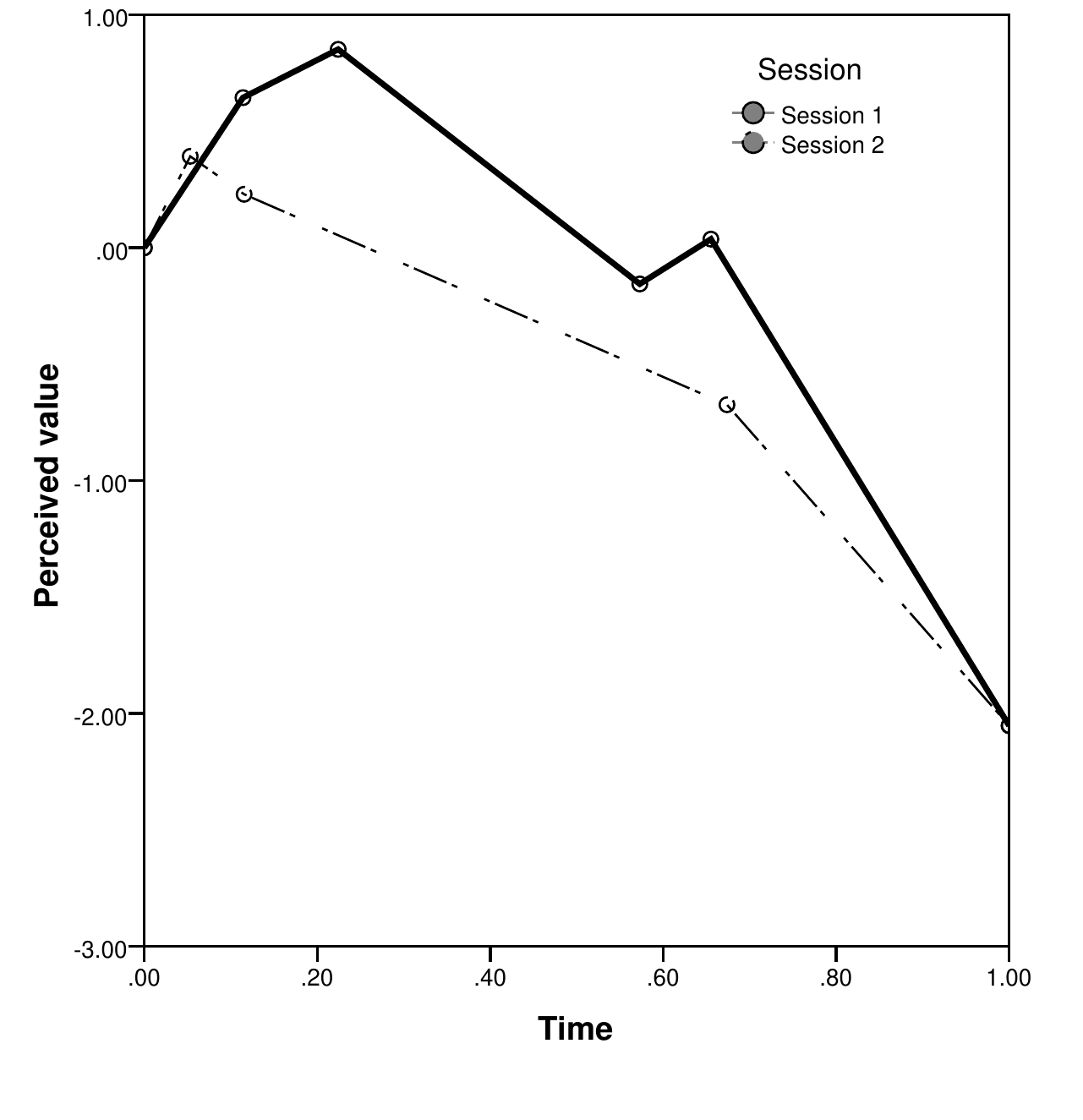} \includegraphics[width=5.7cm, trim=0.625in 4.9in 2.75in 0.8in, clip=true]{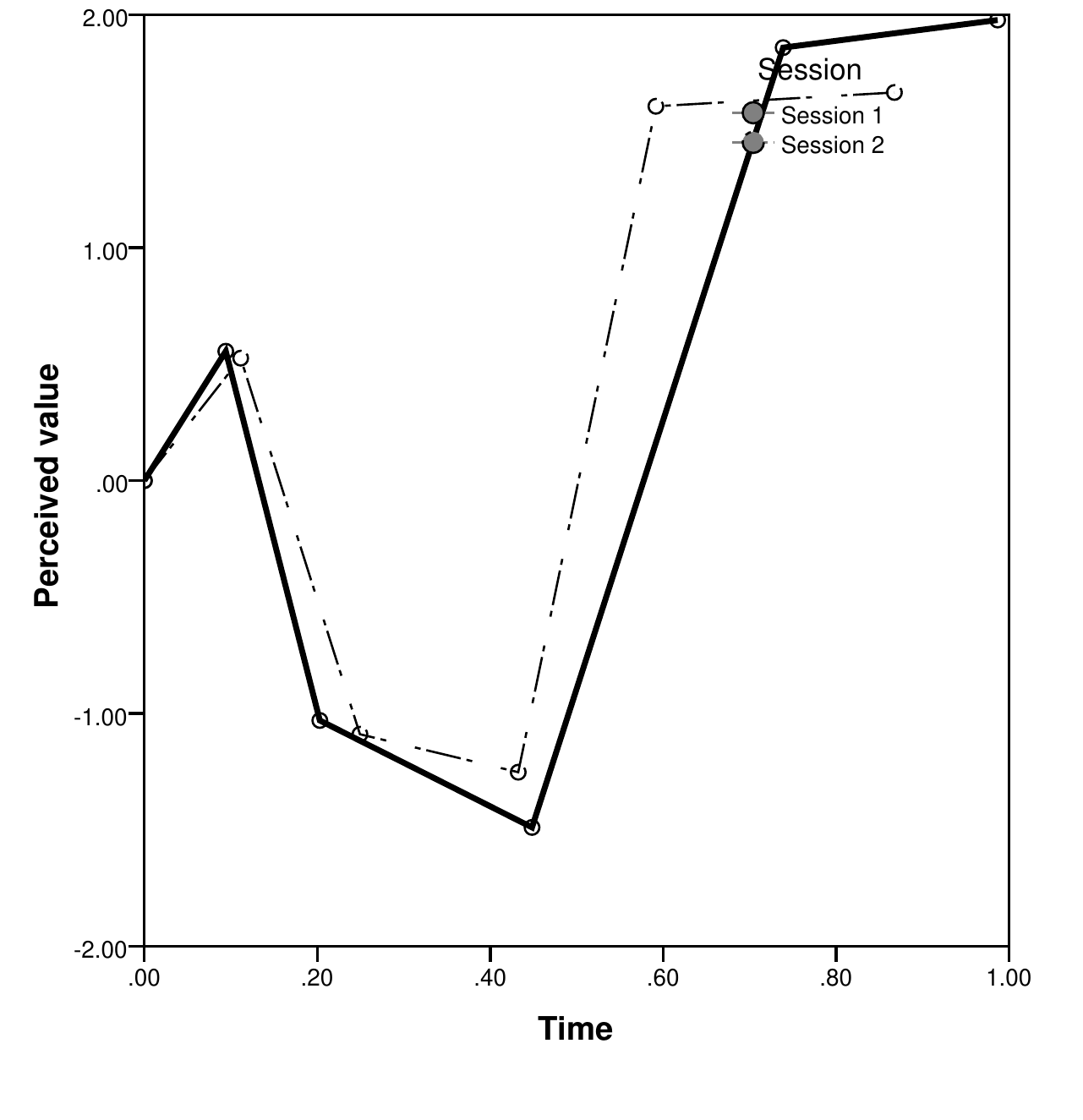}}
  \caption{Participants' sketches resulting from the Constructive iScale tool. Part A.}
  \label{fig:SketchesA}
\end{figure*}

\begin{figure*}
\centerline{\includegraphics[width=5.7cm, trim=0.625in 5in 2.75in 0.8in, clip=true]{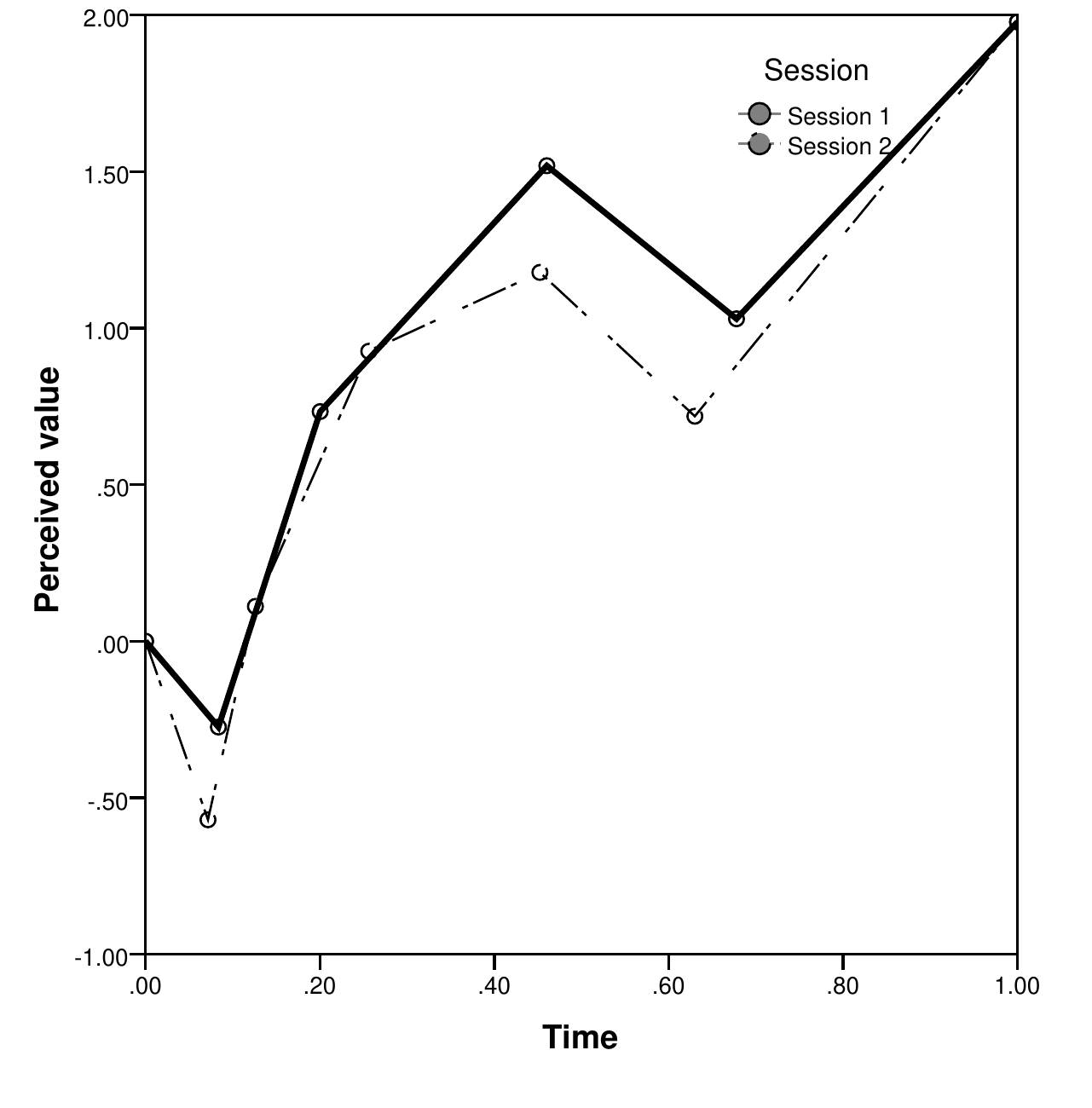} \includegraphics[width=5.7cm, trim=0.625in 5in 2.75in 0.8in, clip=true]{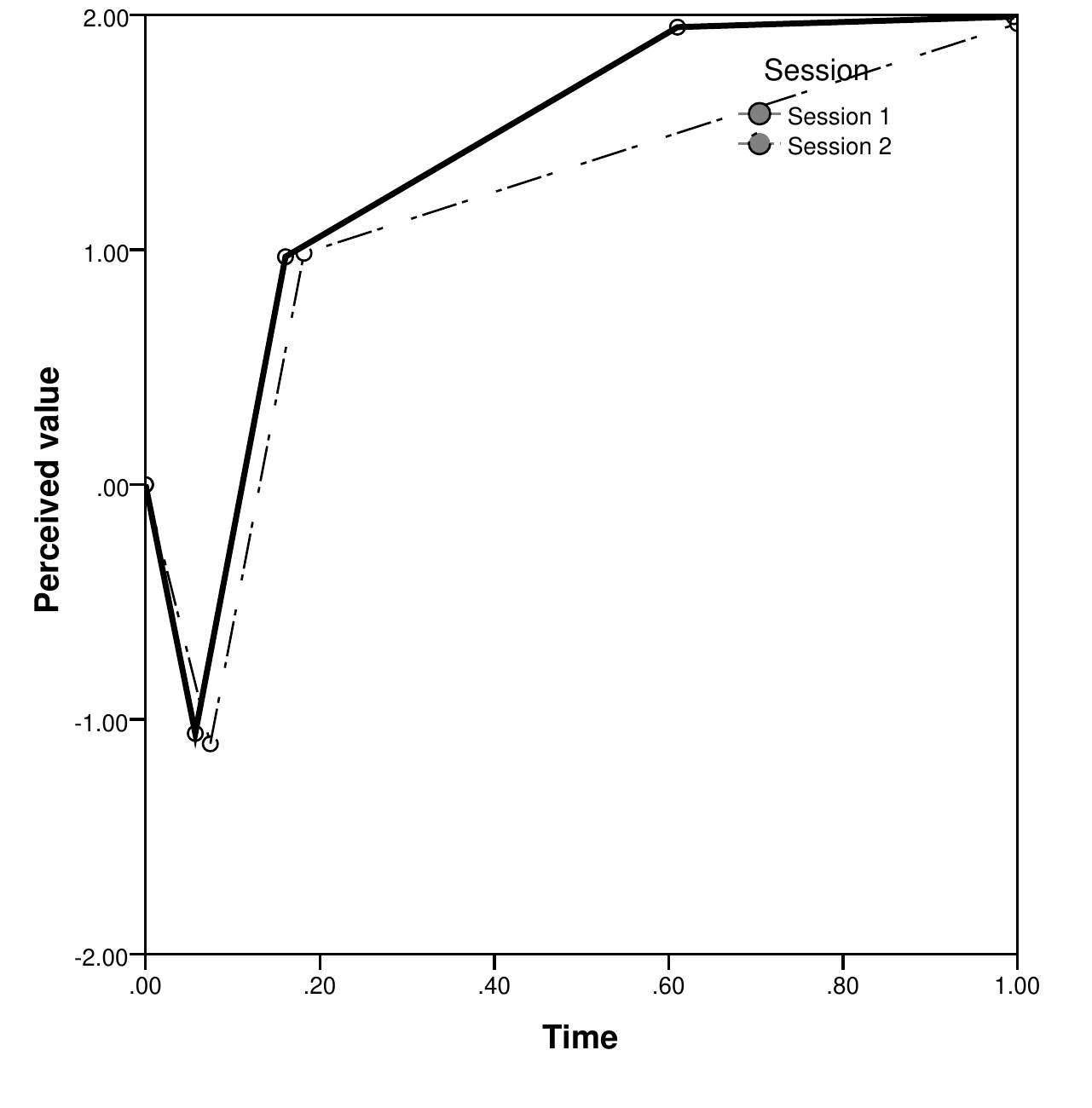}}
\centerline{\includegraphics[width=5.7cm, trim=0.625in 4.9in 2.75in 0.8in, clip=true]{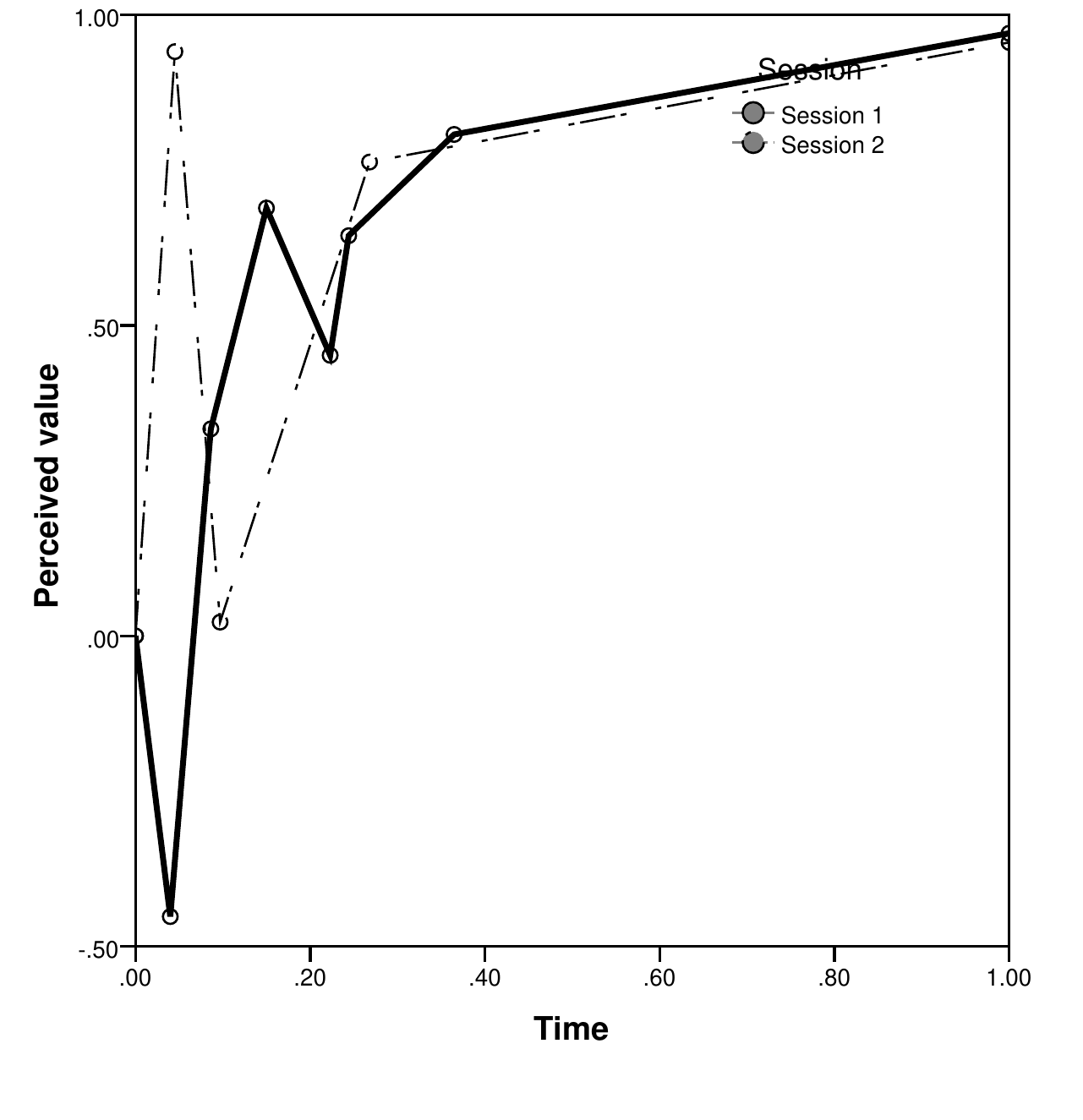} \includegraphics[width=5.7cm, trim=0.625in 4.9in 2.75in 0.8in, clip=true]{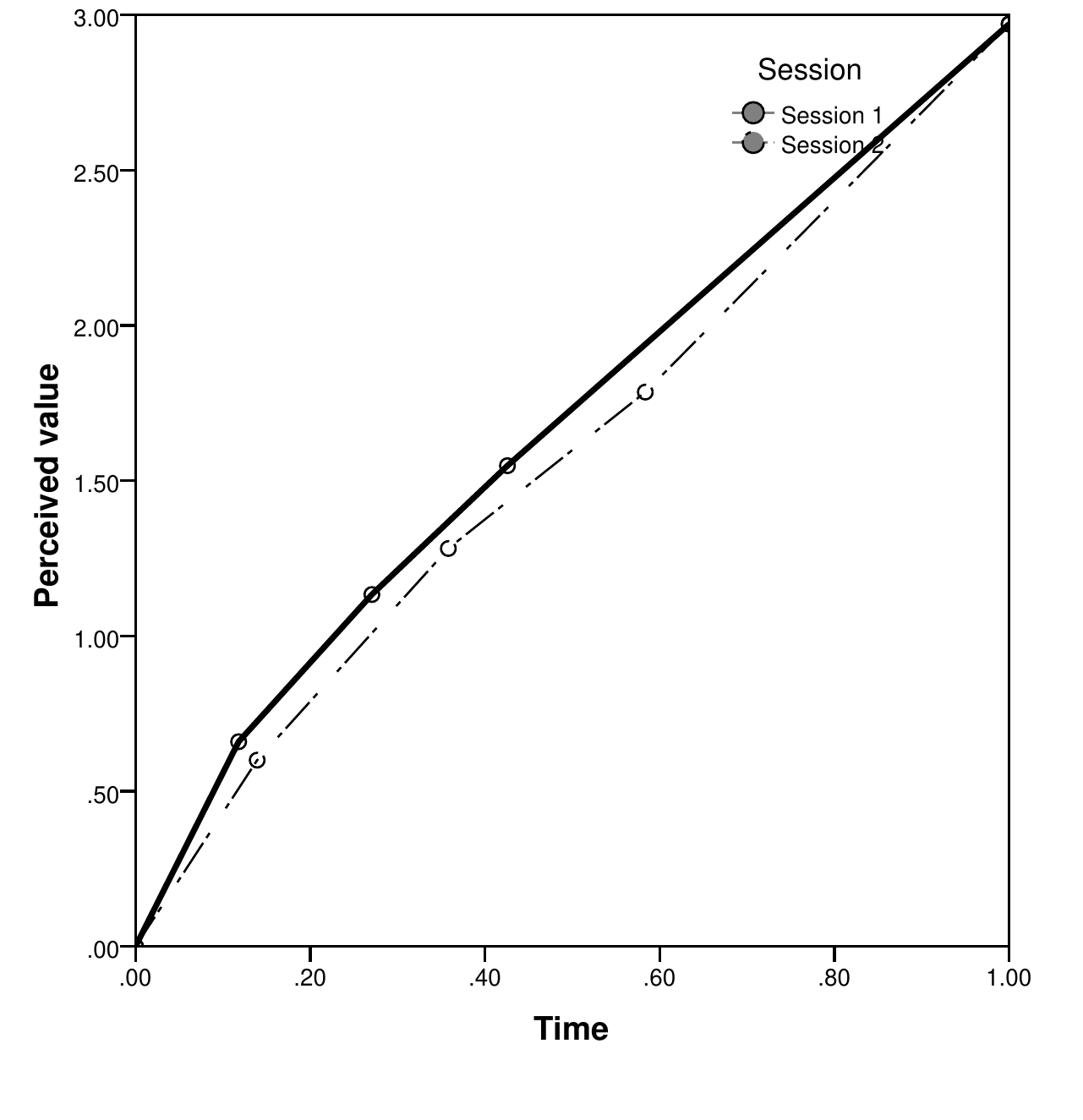}}
\centerline{\includegraphics[width=5.7cm, trim=0.625in 4.9in 2.75in 0.8in, clip=true]{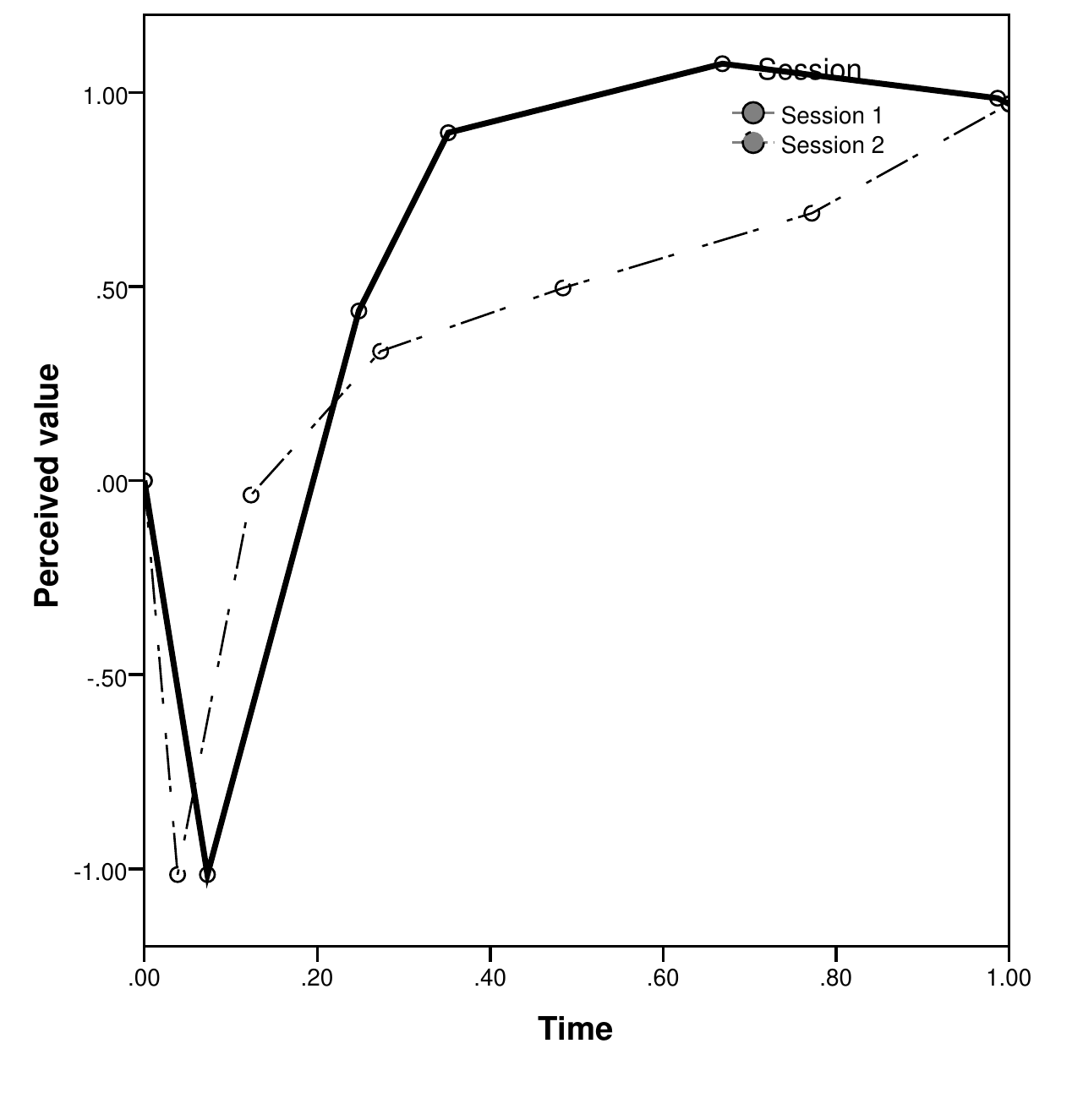} \includegraphics[width=5.7cm, trim=0.625in 4.9in 2.75in 0.8in, clip=true]{images/perceptionsOverTime/con12.pdf}}
  \caption{Participants' sketches resulting from the Constructive iScale tool. Part B.}
  \label{fig:SketchesB}
\end{figure*}

\begin{figure*}
\centerline{\includegraphics[width=5.7cm, trim=0.625in 5in 2.75in 0.8in, clip=true]{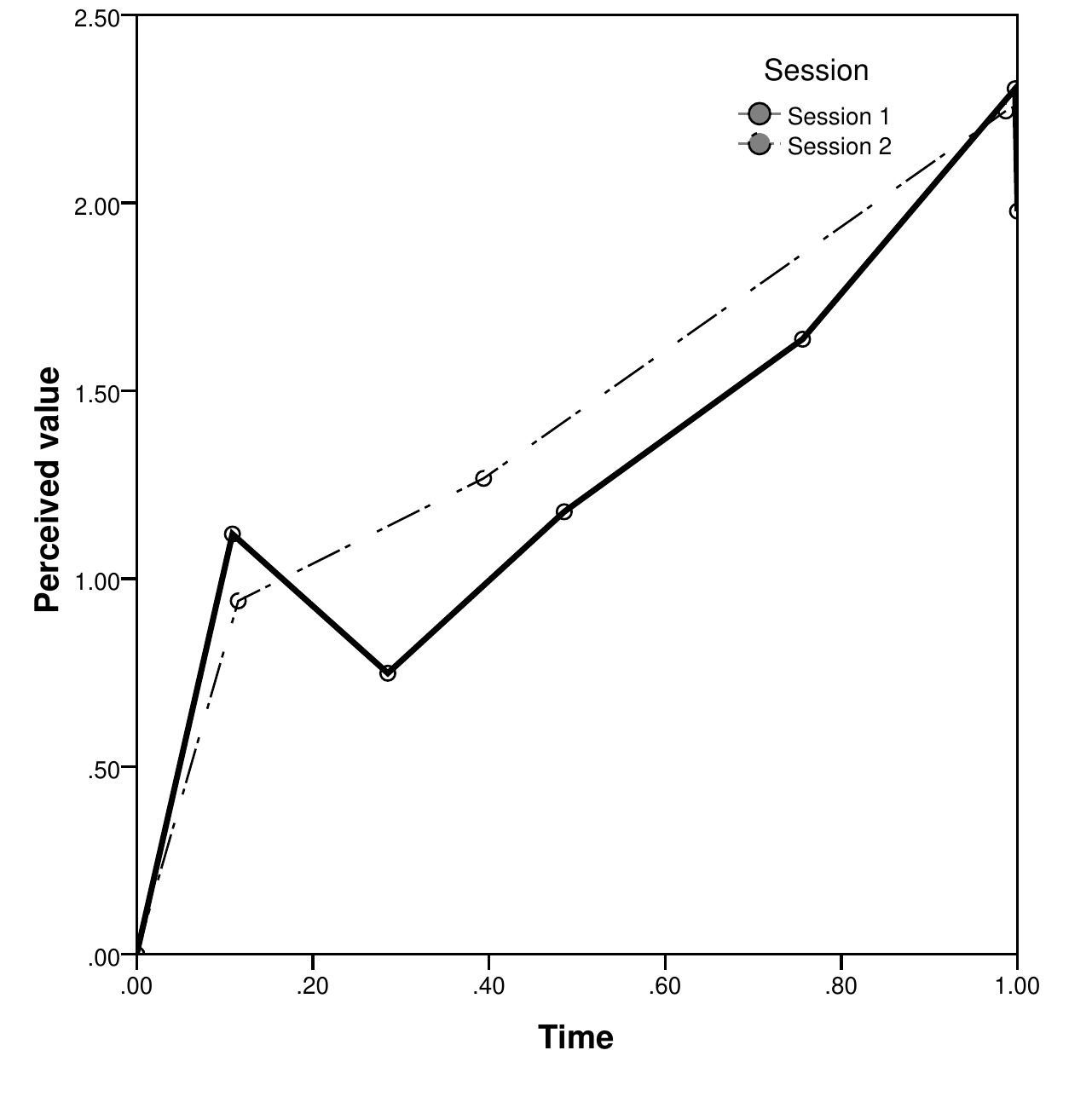} \includegraphics[width=5.7cm, trim=0.625in 5in 2.75in 0.8in, clip=true]{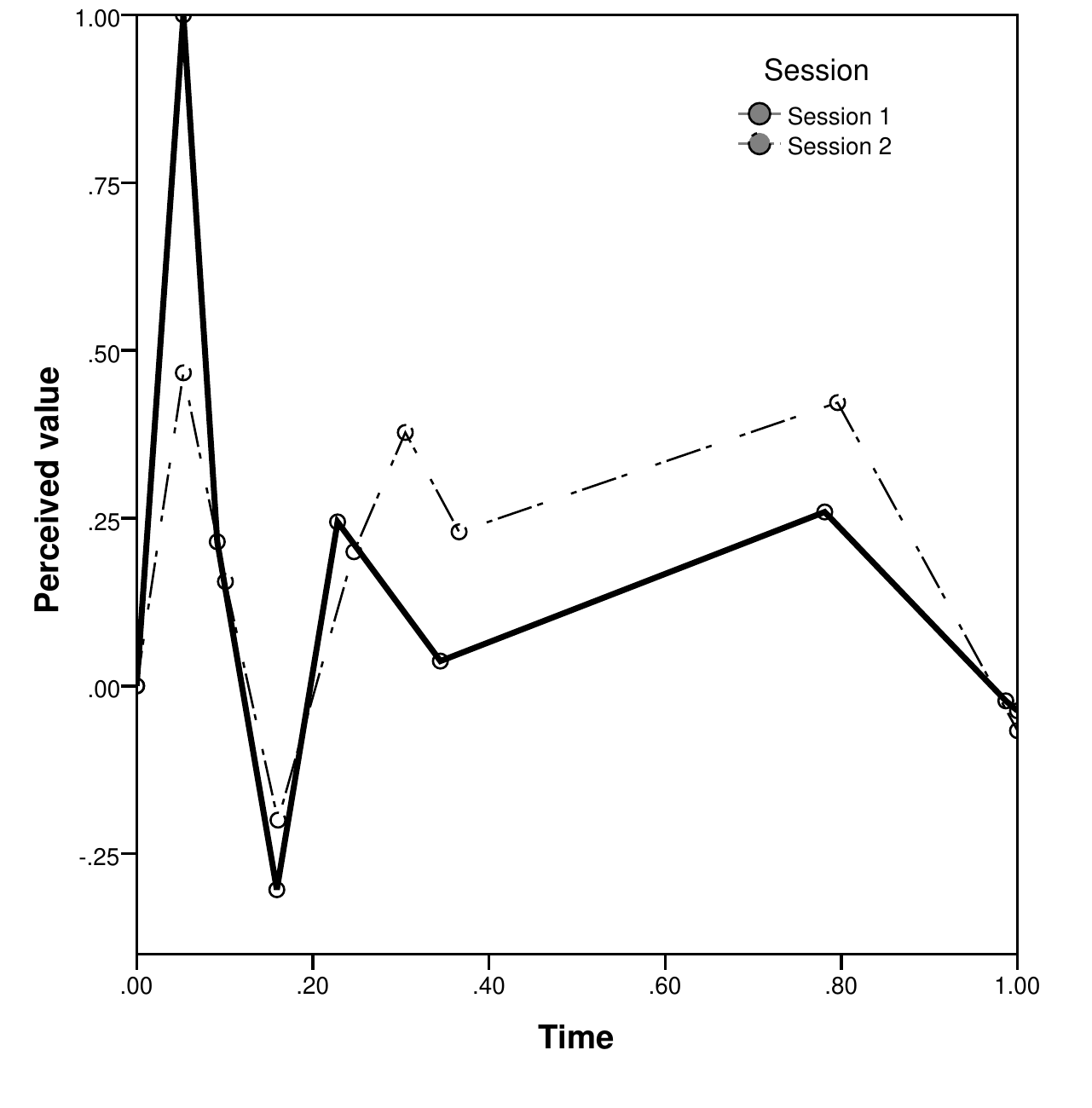}}
\centerline{\includegraphics[width=5.7cm, trim=0.625in 4.9in 2.75in 0.8in, clip=true]{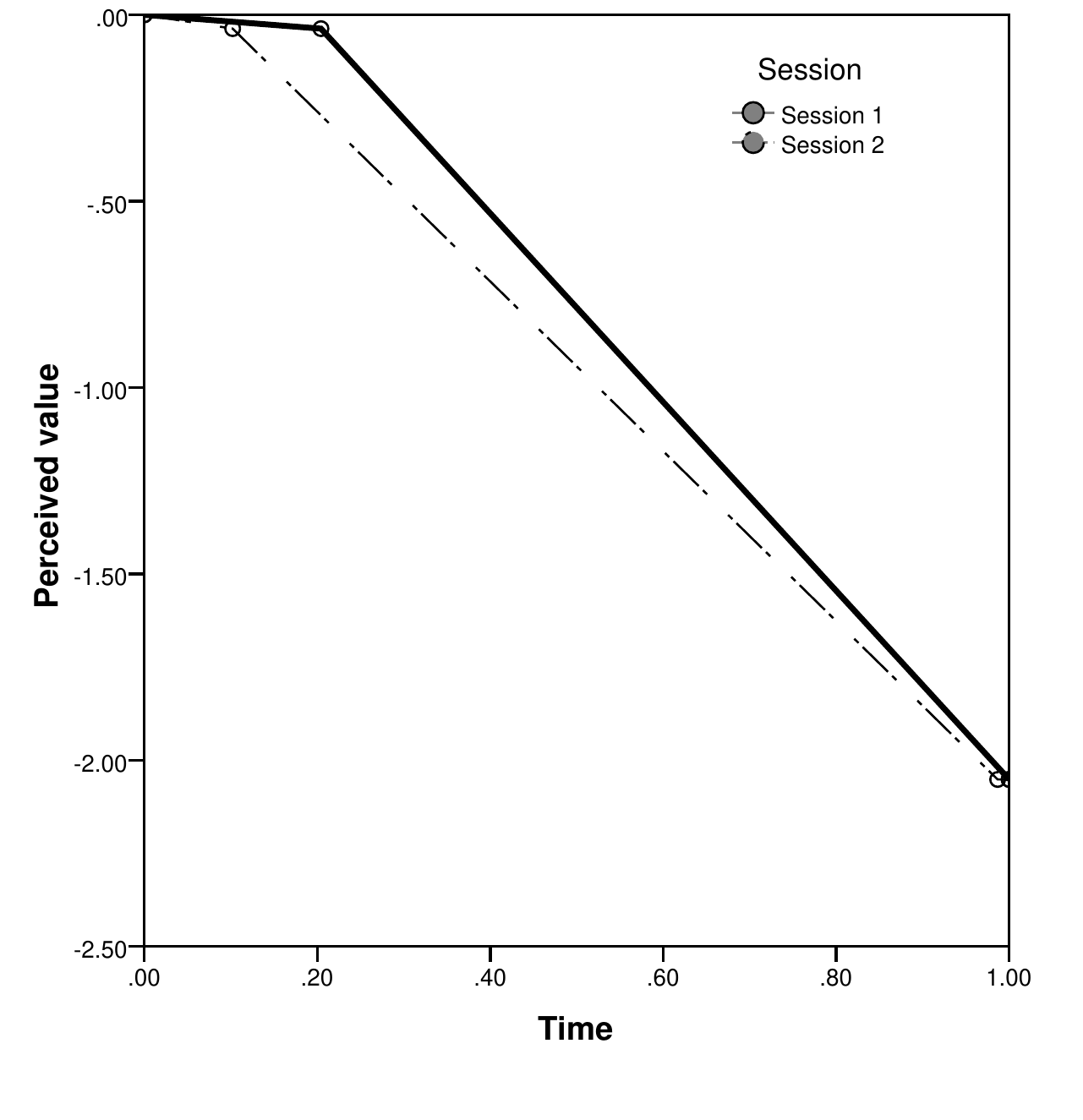} \includegraphics[width=5.7cm, trim=0.625in 4.9in 2.75in 0.8in, clip=true]{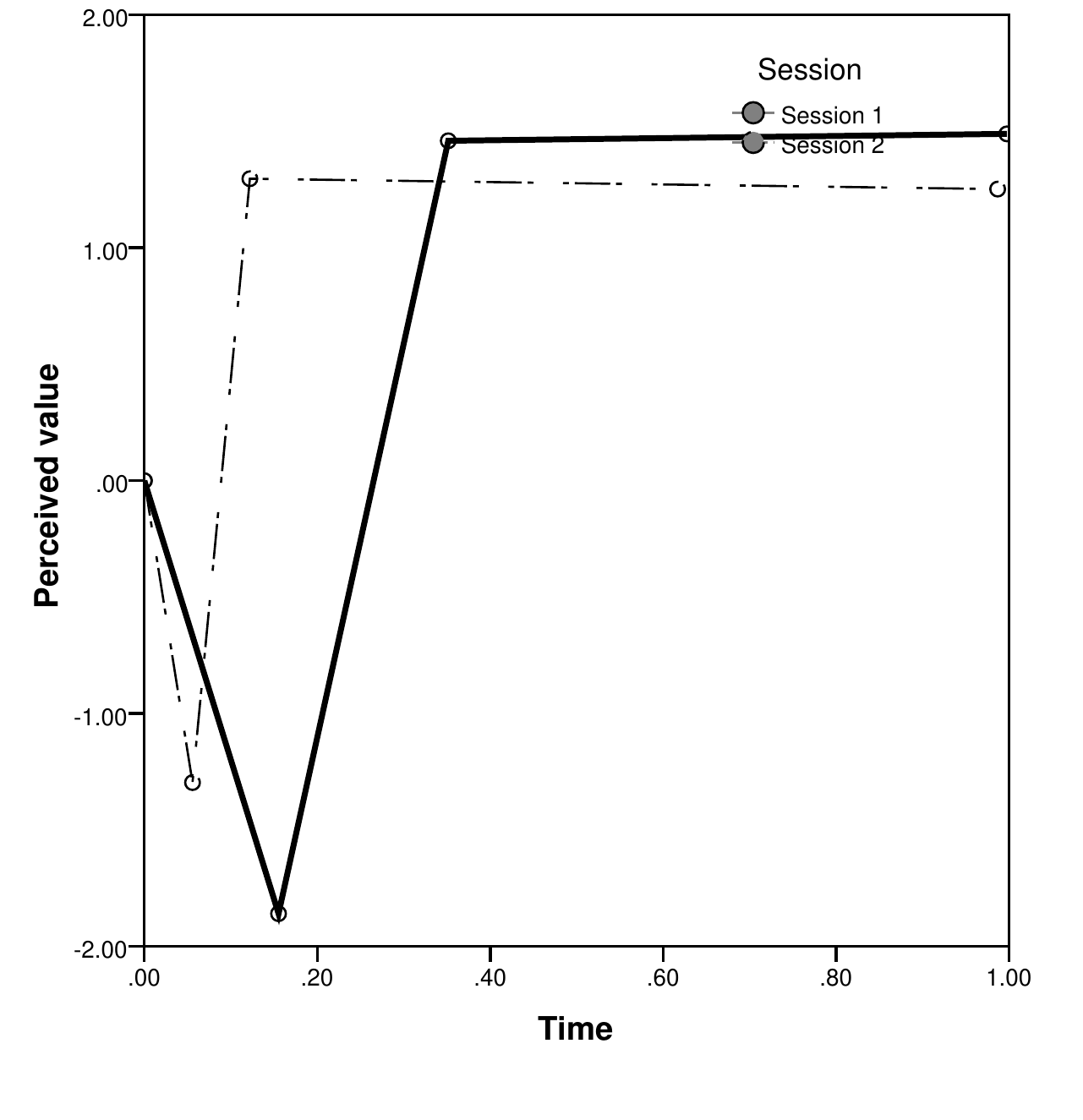}}
\centerline{\includegraphics[width=5.7cm, trim=0.625in 4.9in 2.75in 0.8in, clip=true]{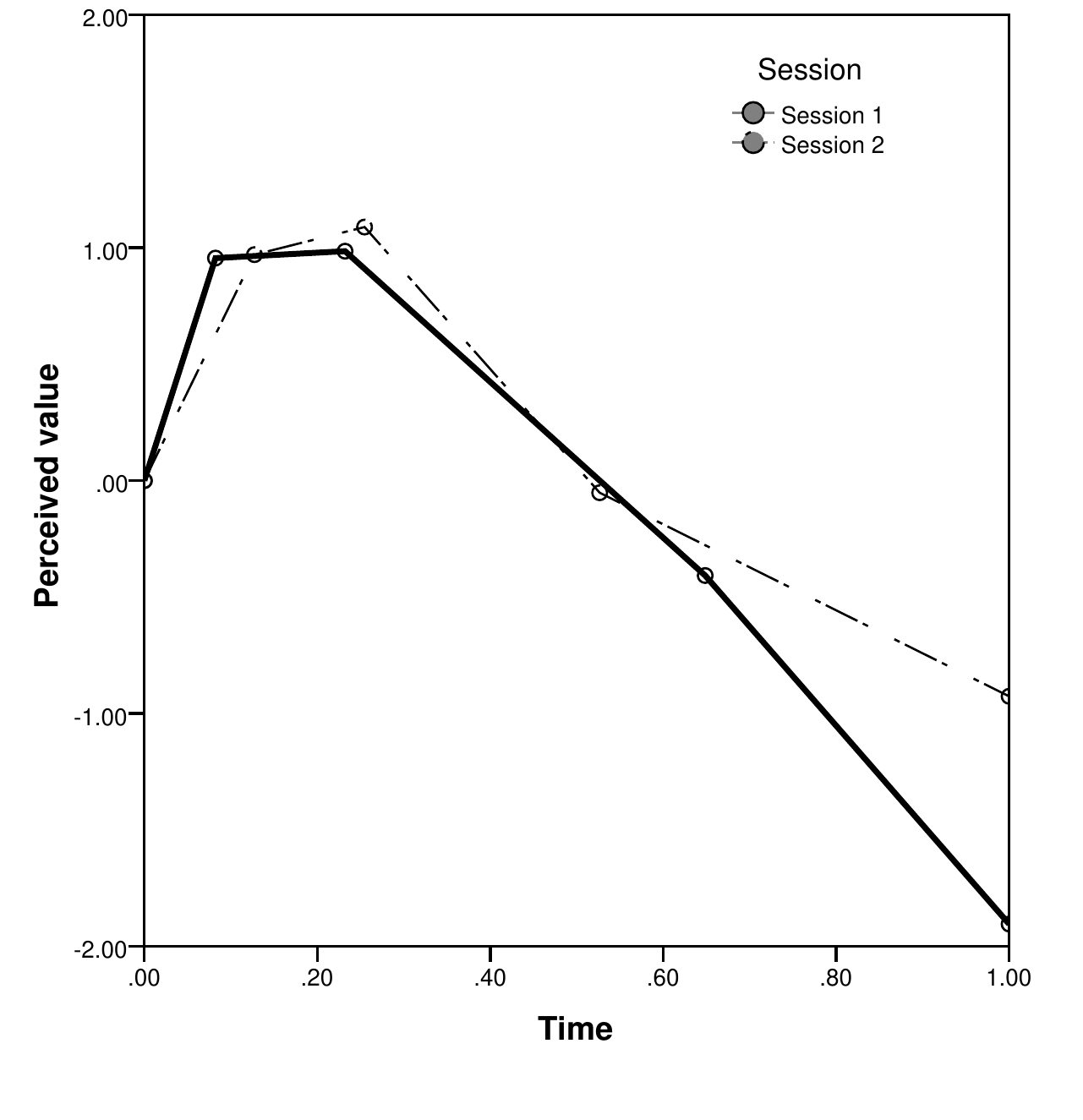} \includegraphics[width=5.7cm, trim=0.625in 4.9in 2.75in 0.8in, clip=true]{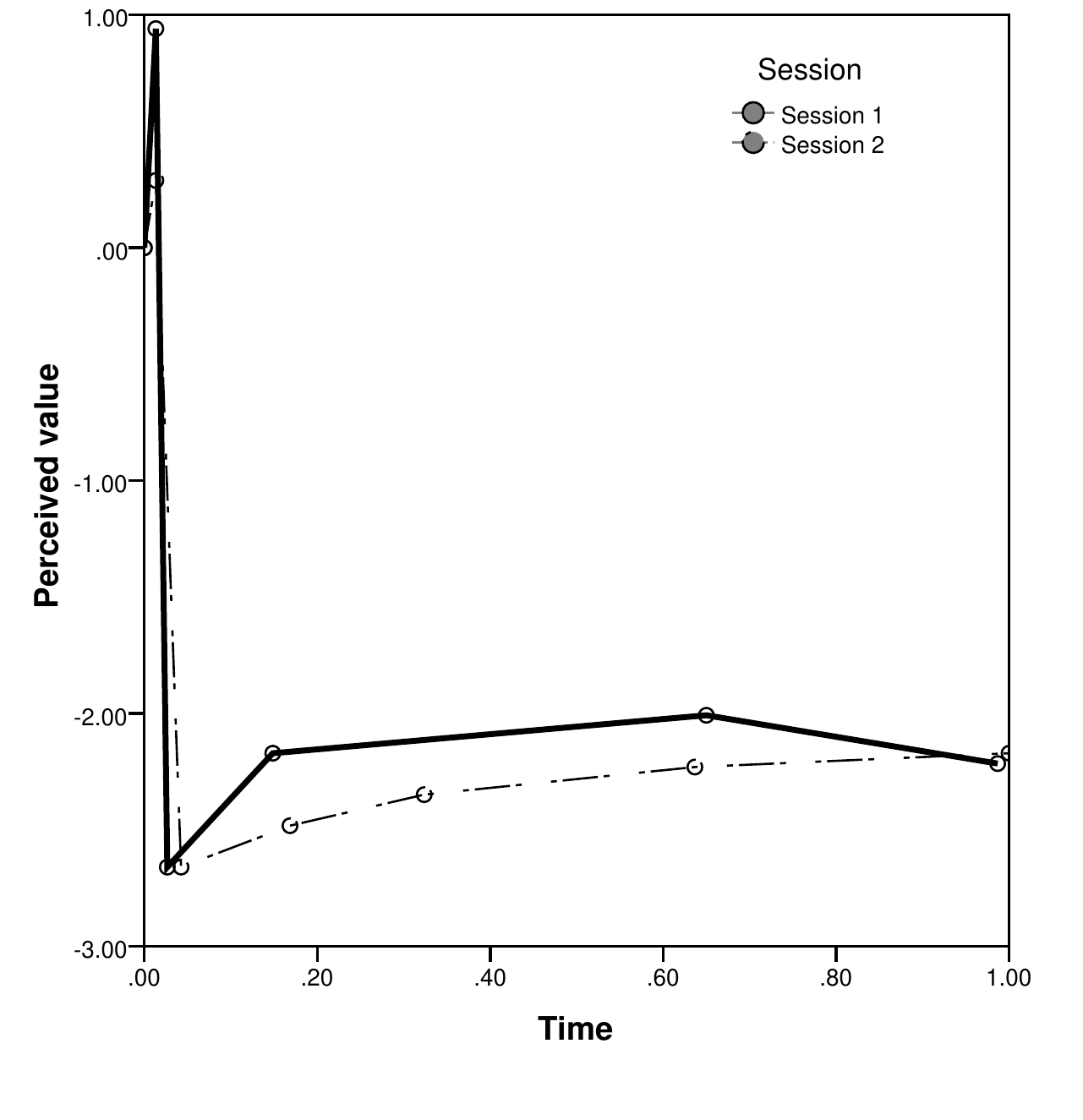}}
  \caption{Participants' sketches resulting from the Constructive iScale tool. Part C.}
  \label{fig:SketchesC}
\end{figure*}

\begin{figure*}
\centerline{\includegraphics[width=5.7cm, trim=0.625in 5in 2.75in 0.8in, clip=true]{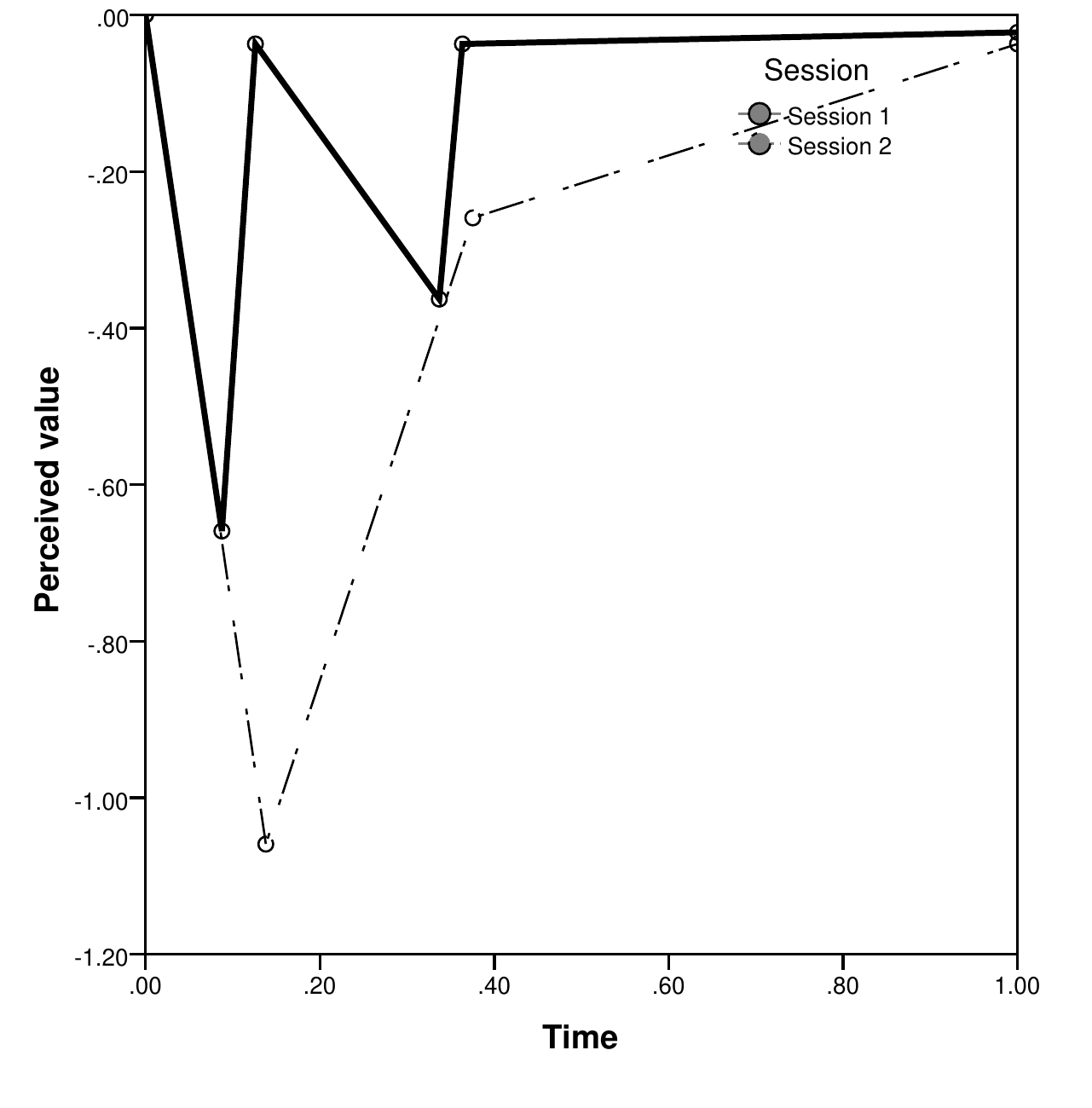} \includegraphics[width=5.7cm, trim=0.625in 5in 2.75in 0.8in, clip=true]{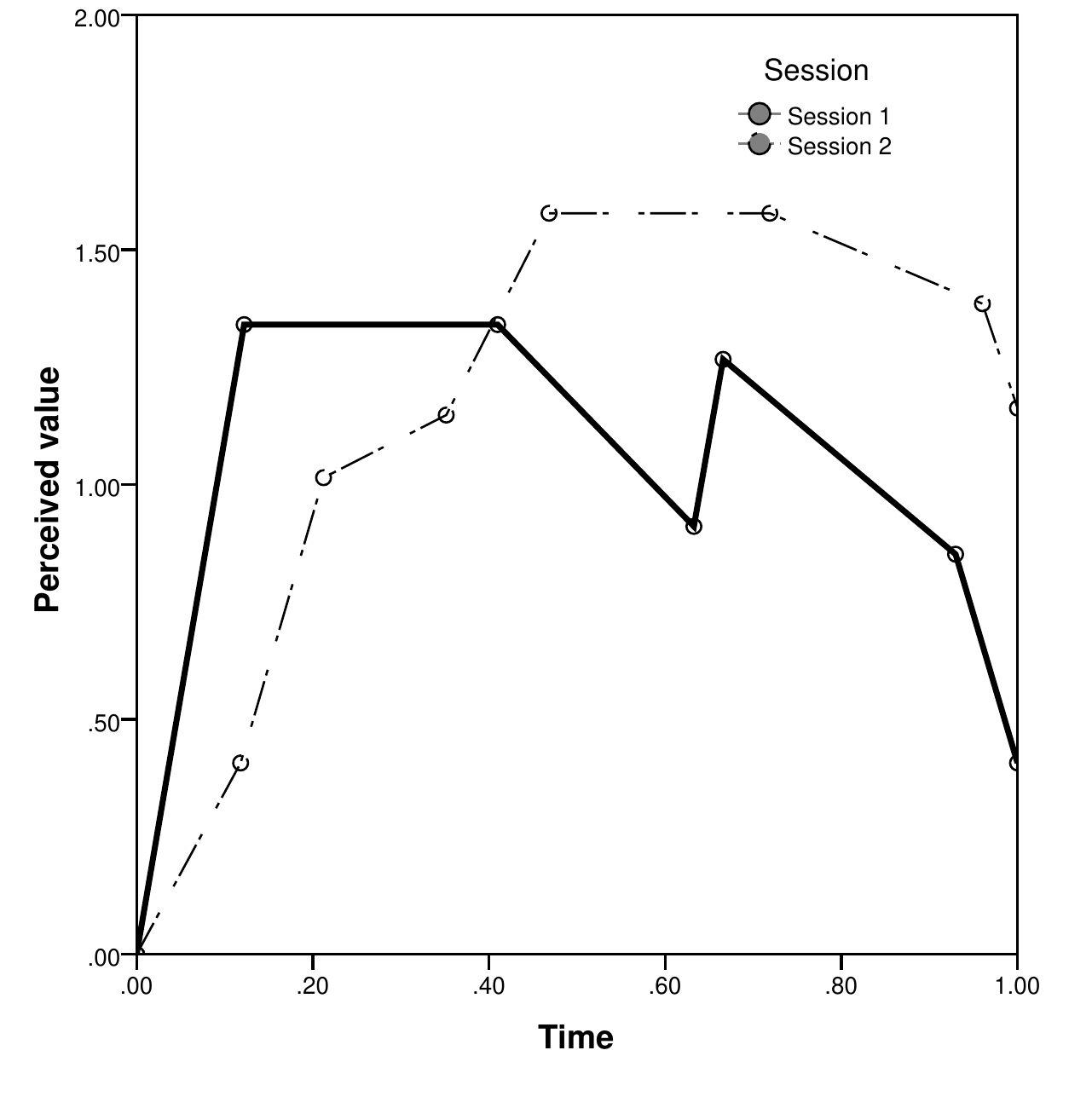}}
\centerline{\includegraphics[width=5.7cm, trim=0.625in 4.9in 2.75in 0.8in, clip=true]{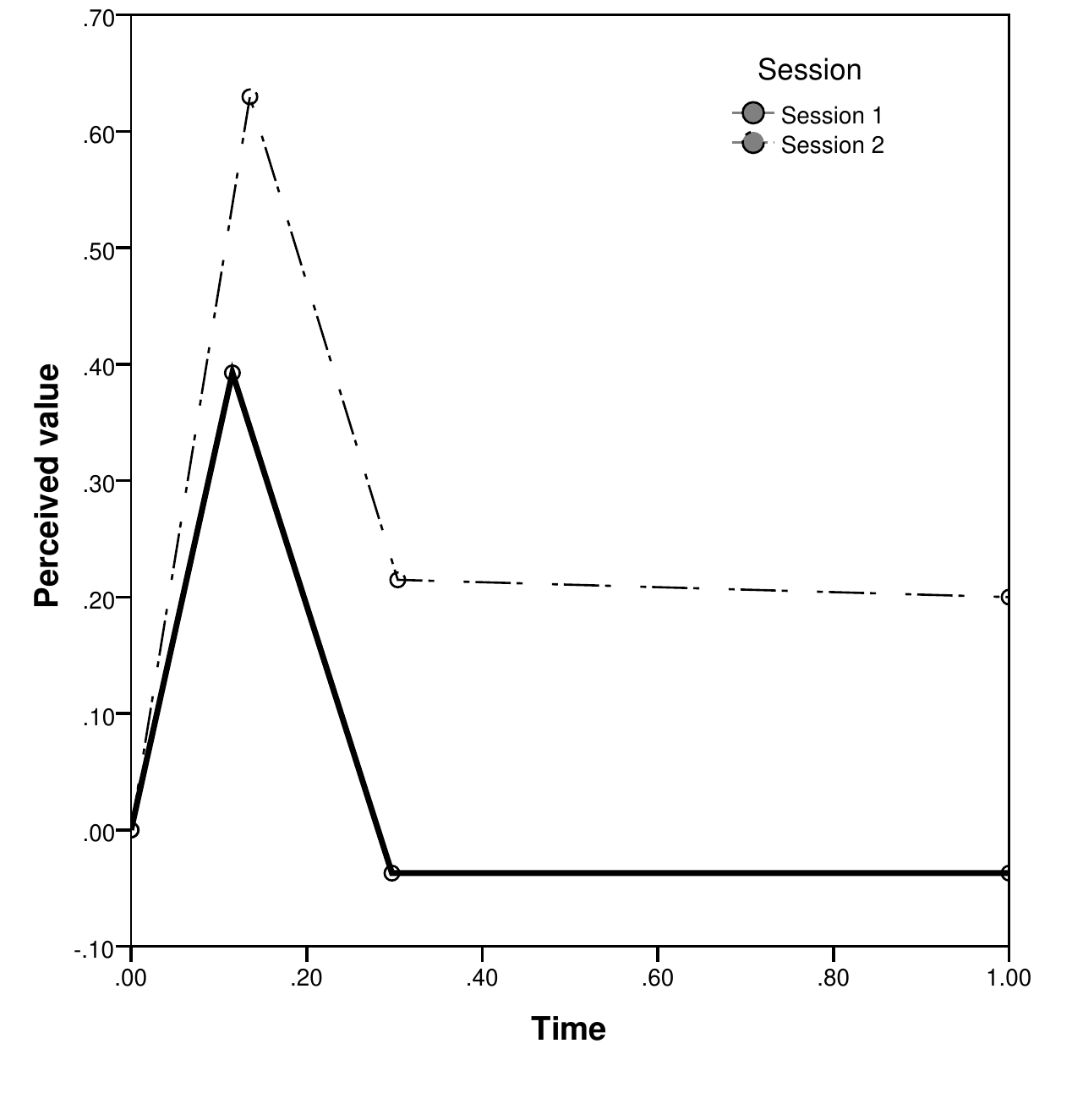} \includegraphics[width=5.7cm, trim=0.625in 4.9in 2.75in 0.8in, clip=true]{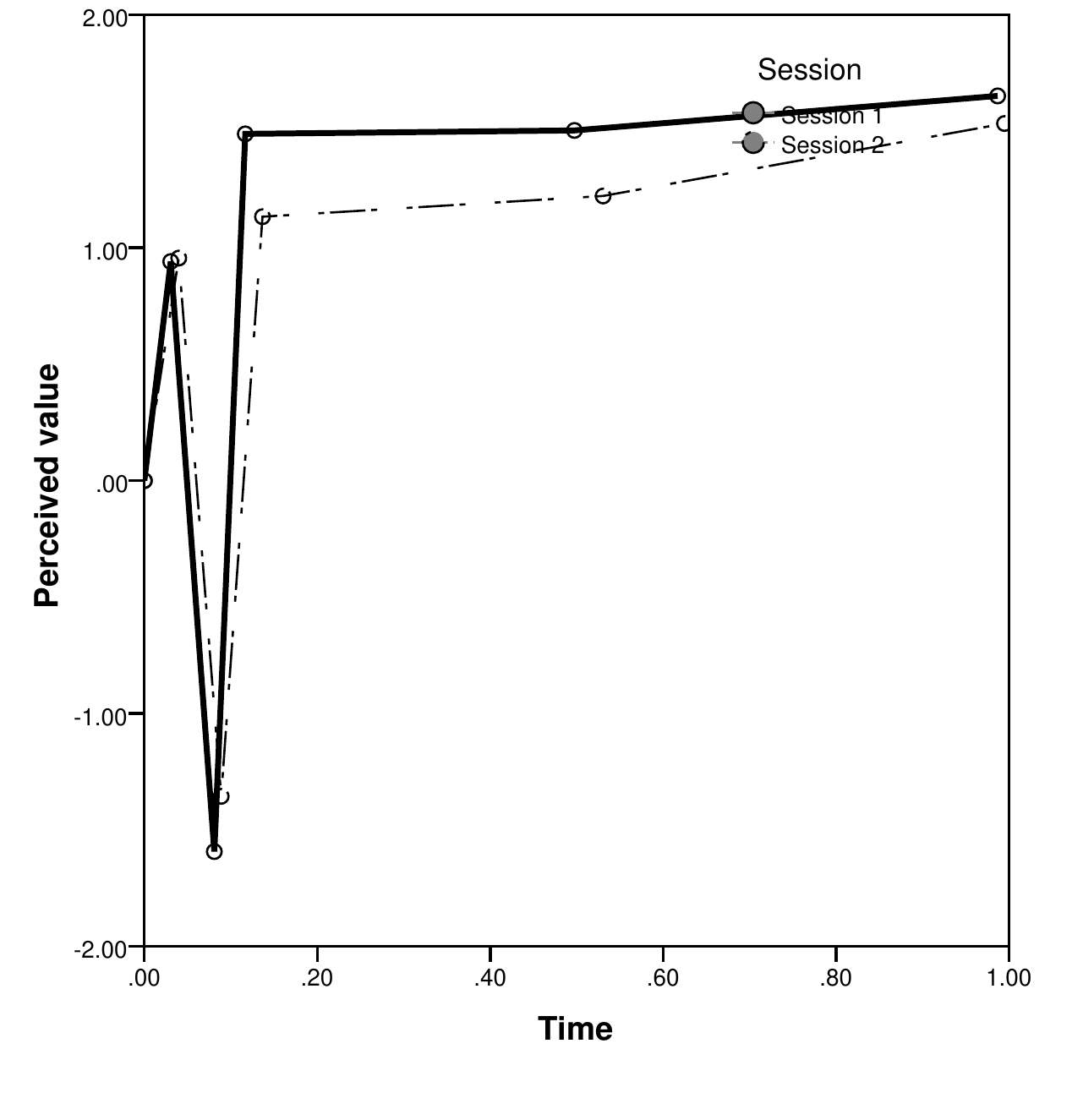}}
\centerline{\includegraphics[width=5.7cm, trim=0.625in 4.9in 2.75in 0.8in, clip=true]{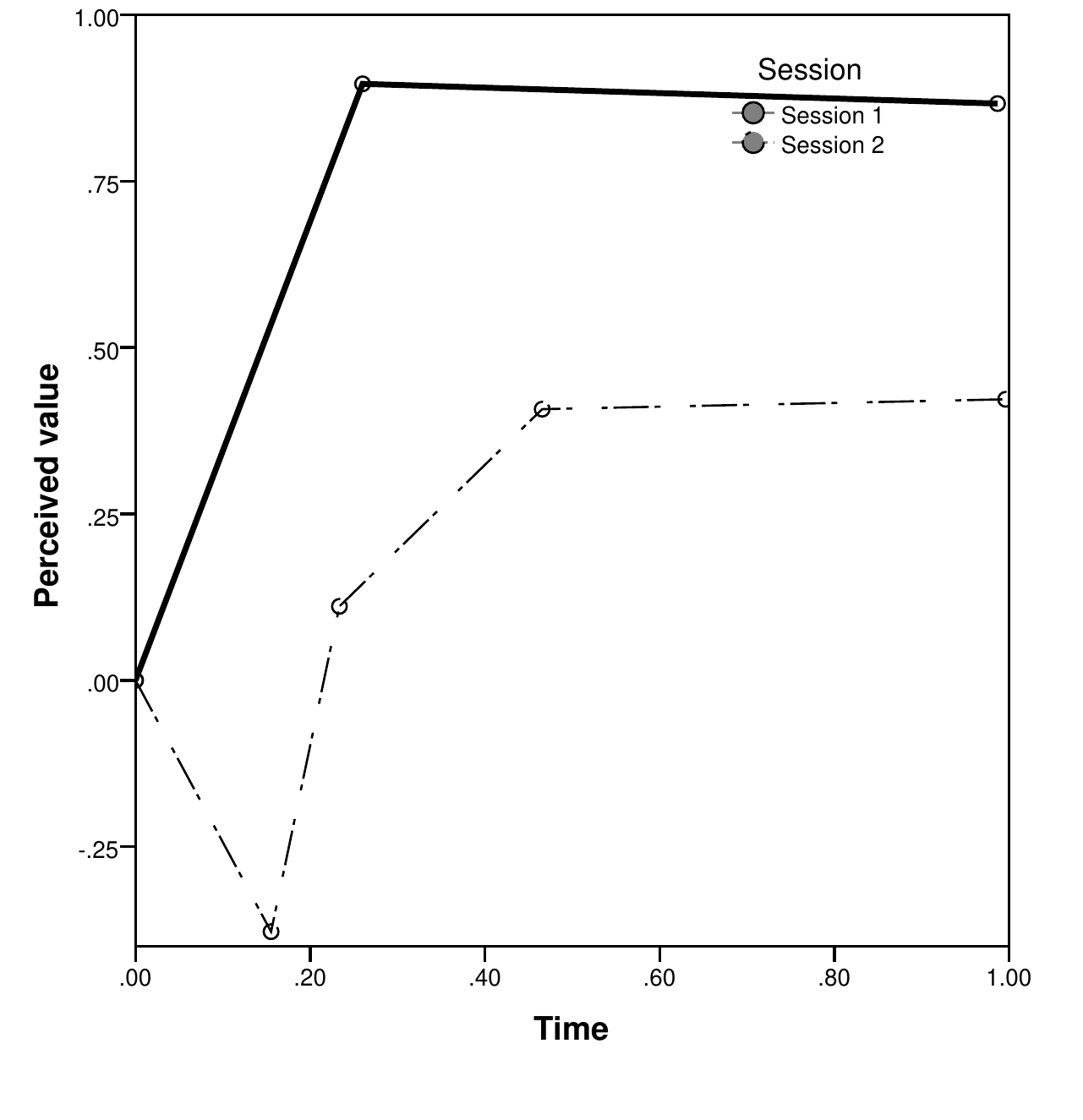} \includegraphics[width=5.7cm, trim=0.625in 4.9in 2.75in 0.8in, clip=true]{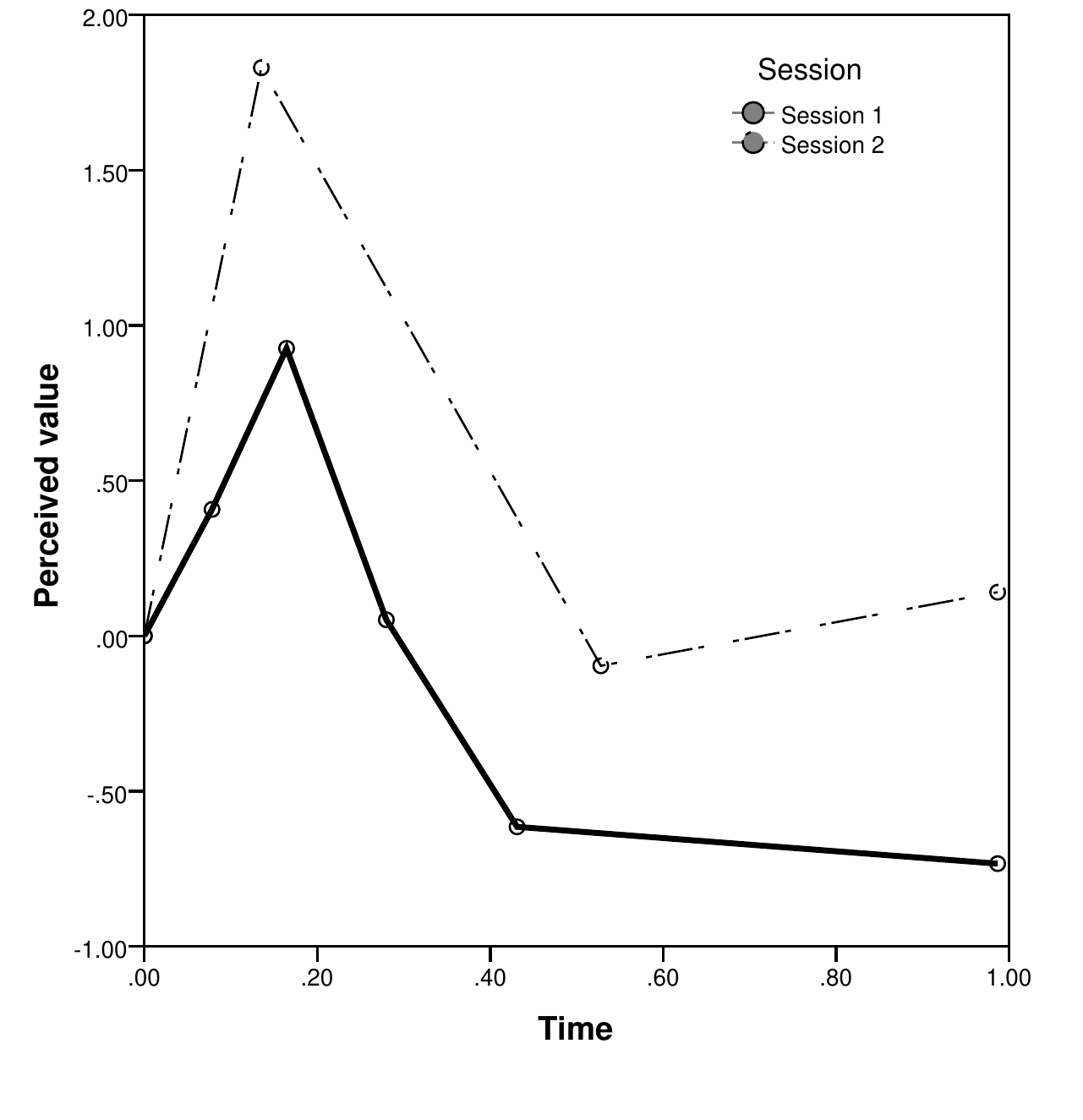}}
  \caption{Participants' sketches resulting from the Constructive iScale tool. Part D.}
  \label{fig:SketchesD}
\end{figure*}

\begin{figure*}
\centerline{\includegraphics[width=5.7cm, trim=0.625in 5in 2.75in 0.8in, clip=true]{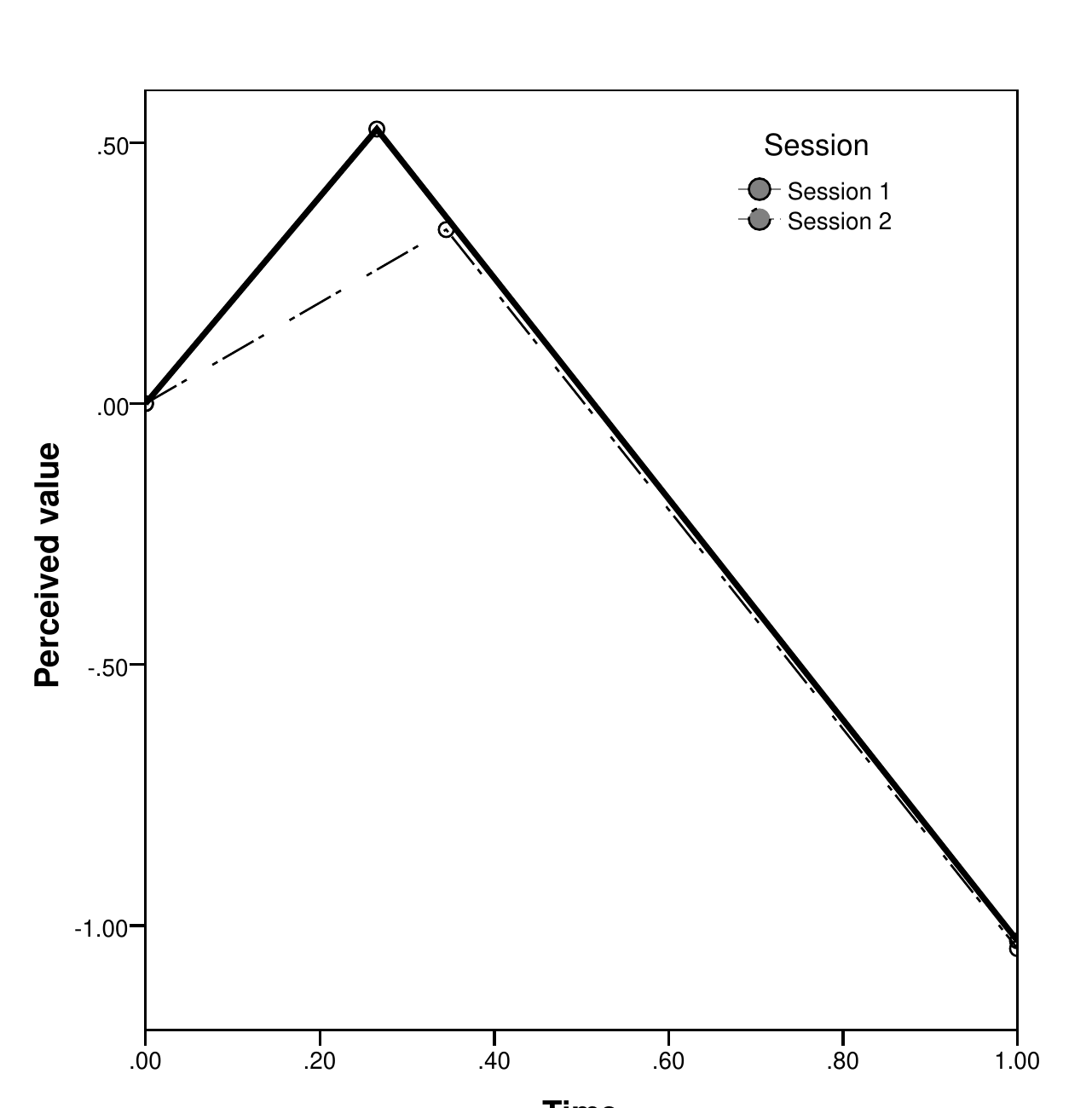} \includegraphics[width=5.7cm, trim=0.625in 5in 2.75in 0.8in, clip=true]{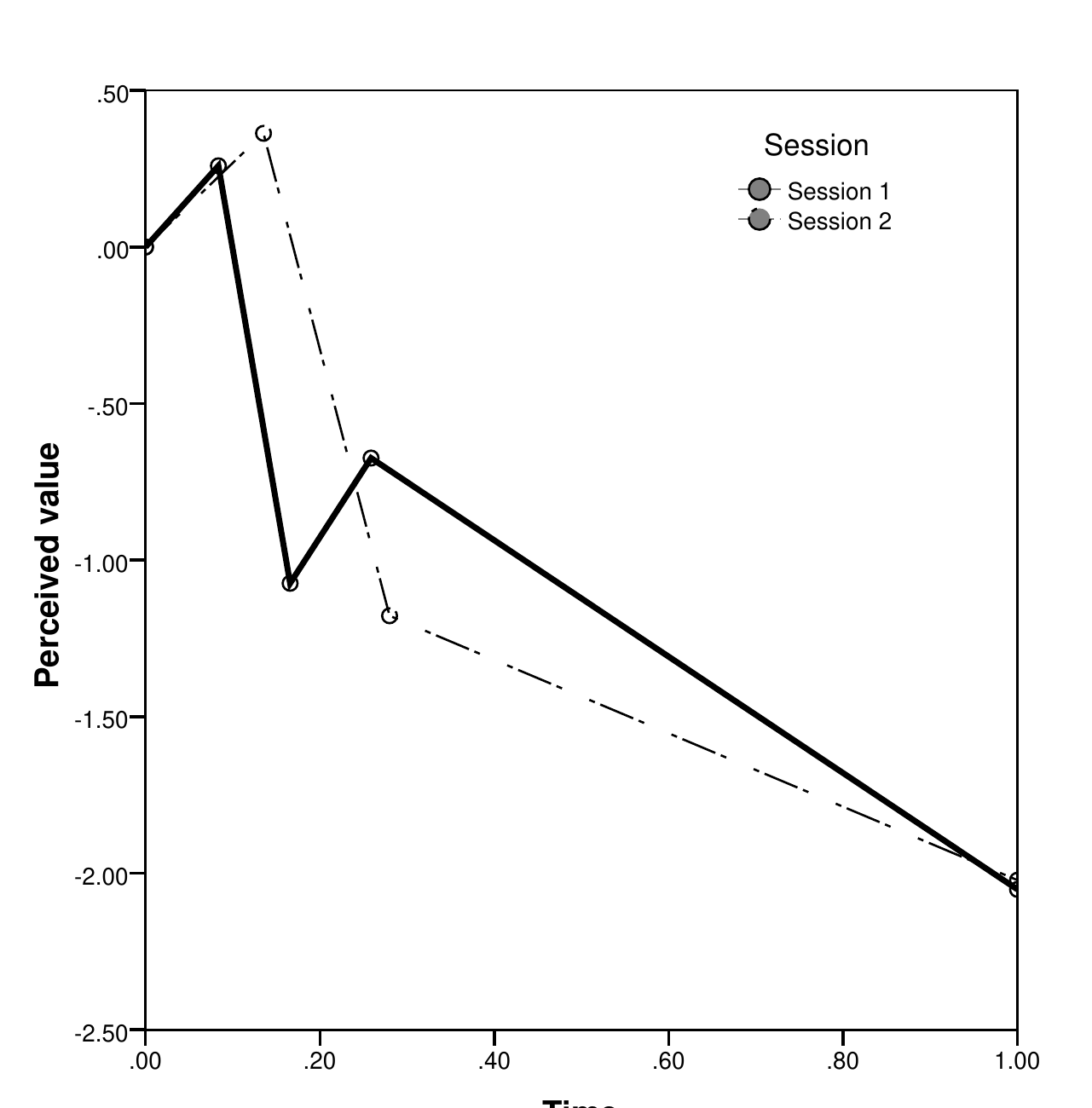}}
\centerline{\includegraphics[width=5.7cm, trim=0.625in 4.9in 2.75in 0.8in, clip=true]{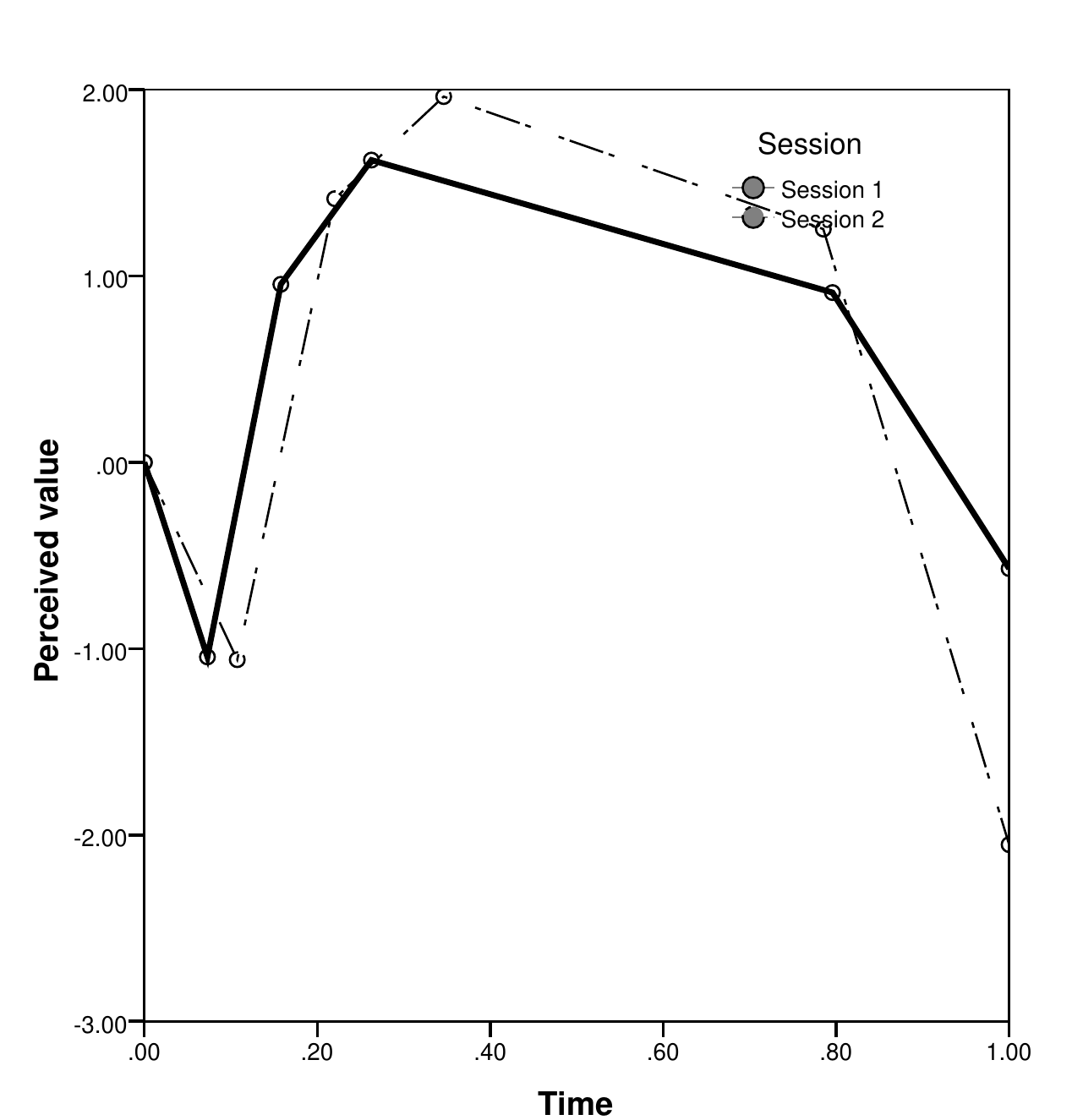} \includegraphics[width=5.7cm, trim=0.625in 4.9in 2.75in 0.8in, clip=true]{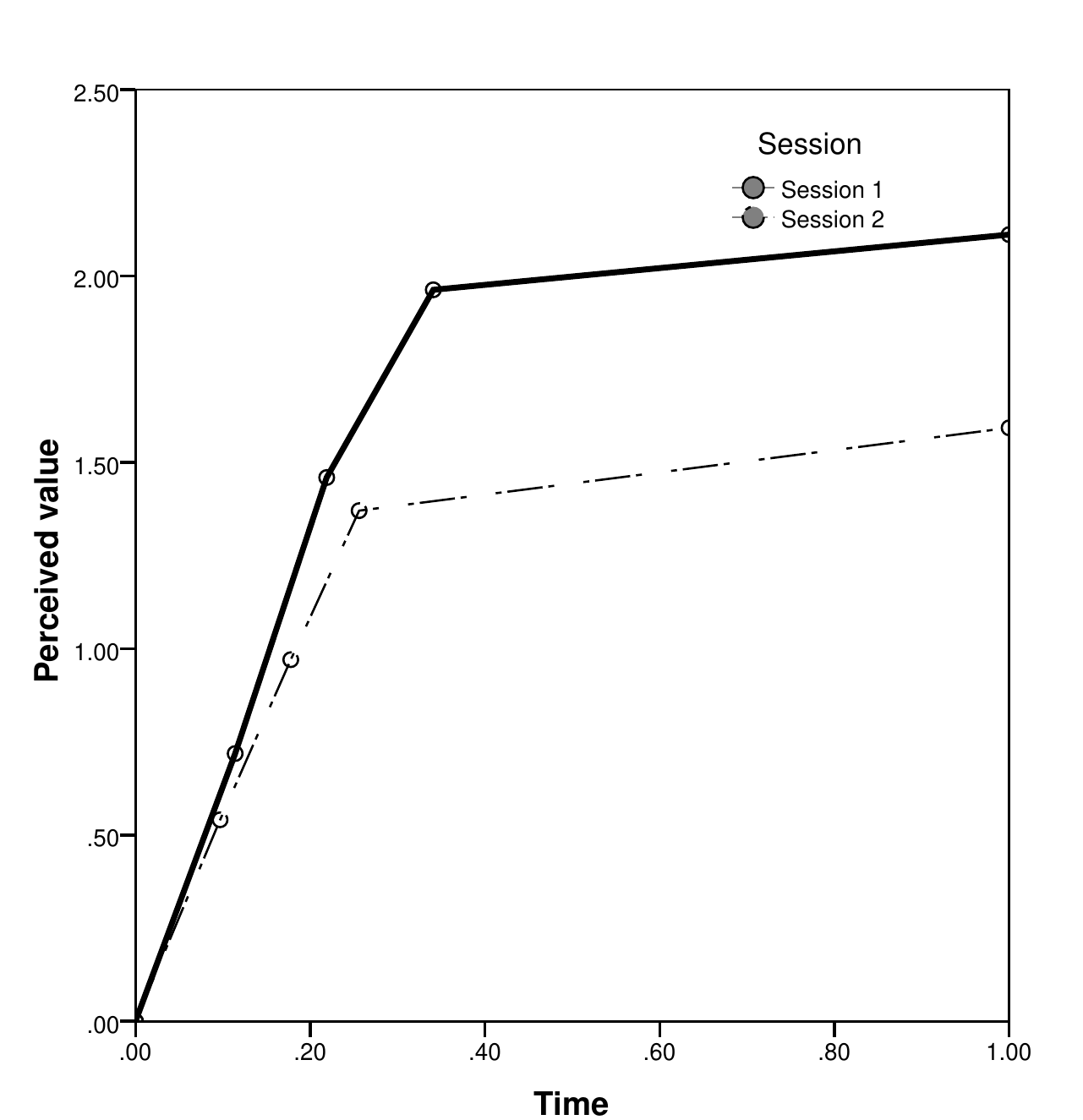}}
\centerline{\includegraphics[width=5.7cm, trim=0.625in 4.9in 2.75in 0.8in, clip=true]{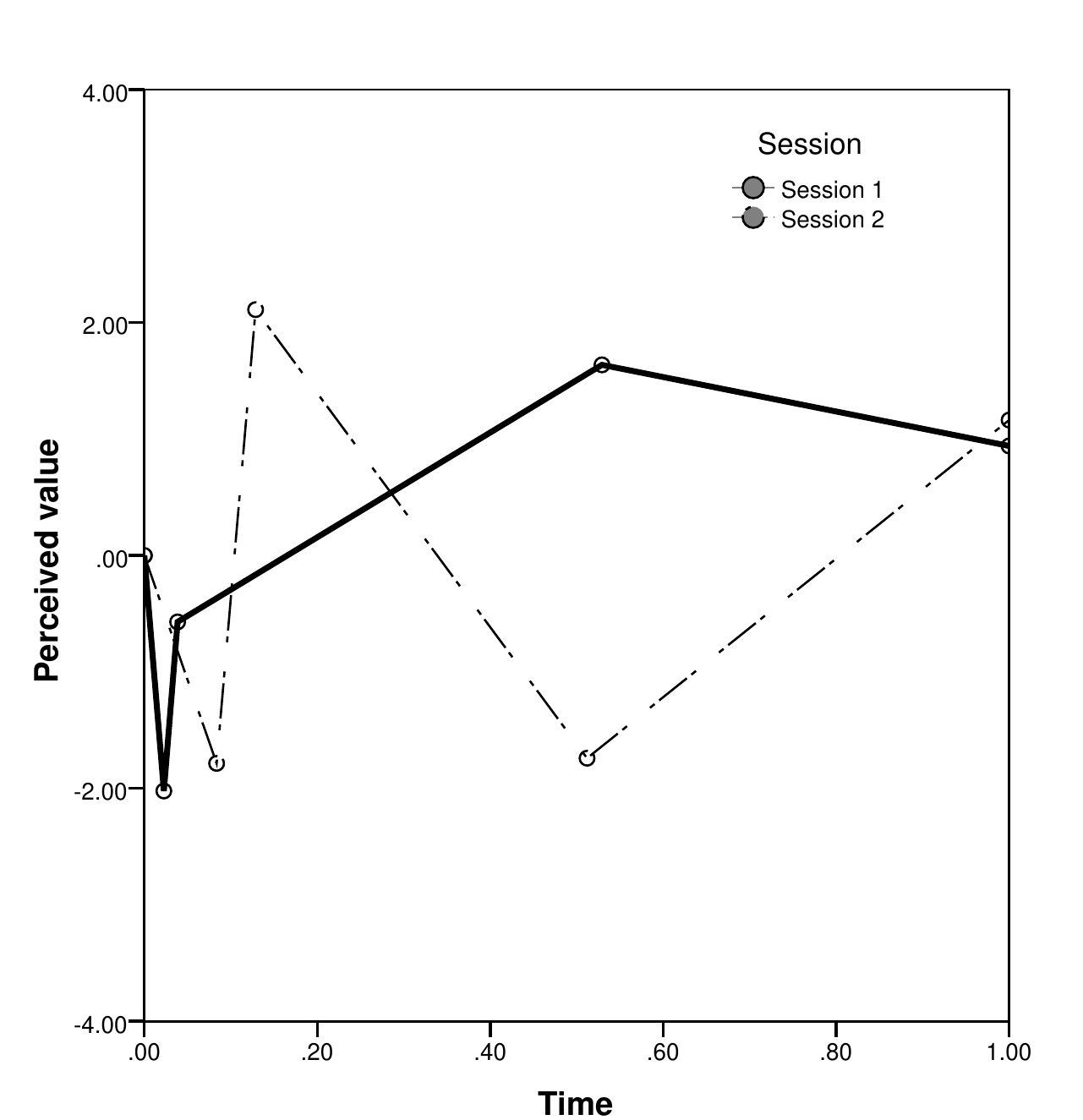} \includegraphics[width=5.7cm, trim=0.625in 4.9in 2.75in 0.8in, clip=true]{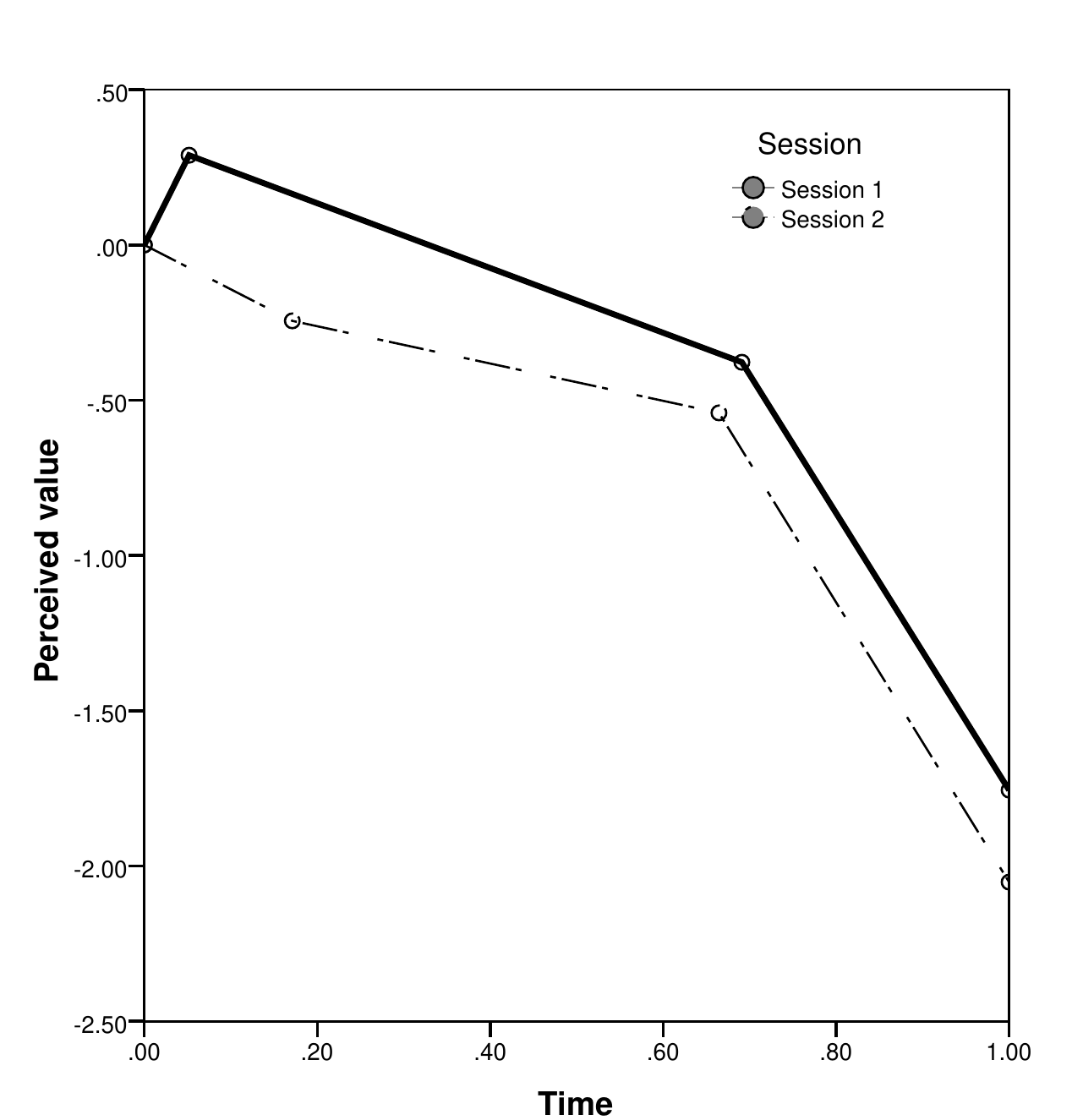}}
  \caption{Participants' sketches resulting from the Value-Account iScale tool. Part A.}
  \label{fig:SketchesAva}
\end{figure*}

\begin{figure*}
\centerline{\includegraphics[width=5.7cm, trim=0.625in 5in 2.75in 0.8in, clip=true]{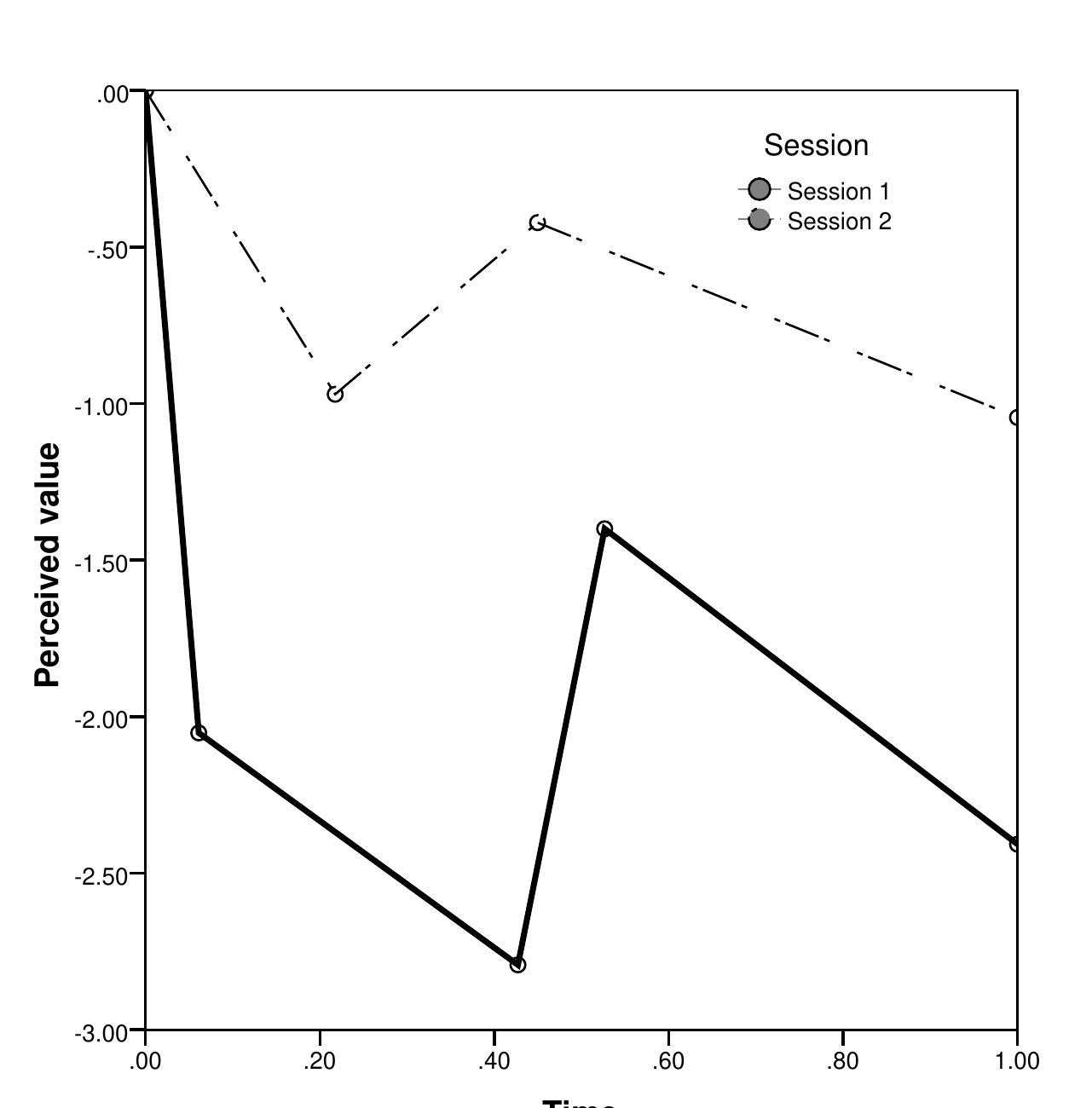} \includegraphics[width=5.7cm, trim=0.625in 5in 2.75in 0.8in, clip=true]{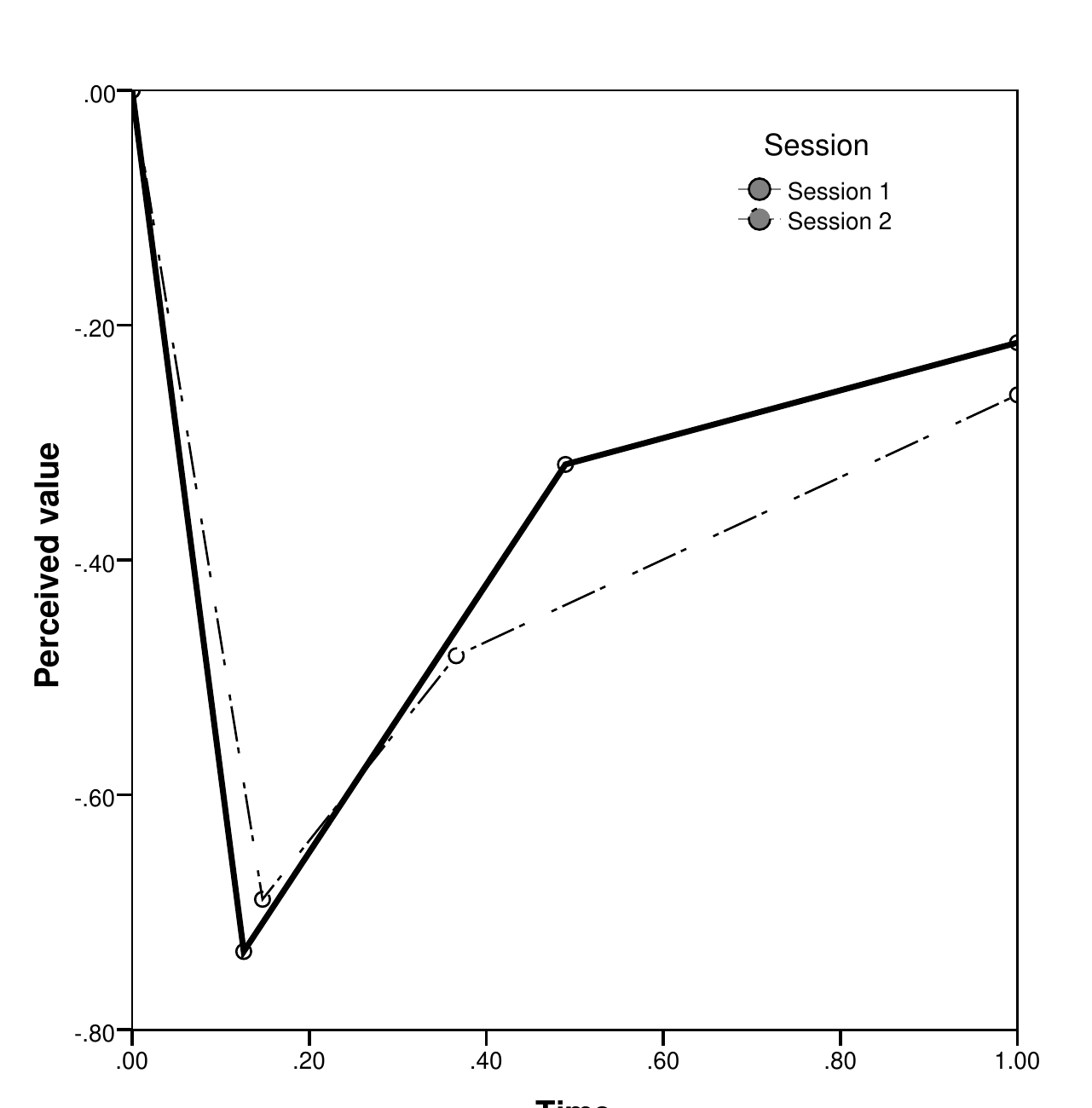}}
\centerline{\includegraphics[width=5.7cm, trim=0.625in 4.9in 2.75in 0.8in, clip=true]{images/perceptionsOverTime/va10.pdf} \includegraphics[width=5.7cm, trim=0.625in 4.9in 2.75in 0.8in, clip=true]{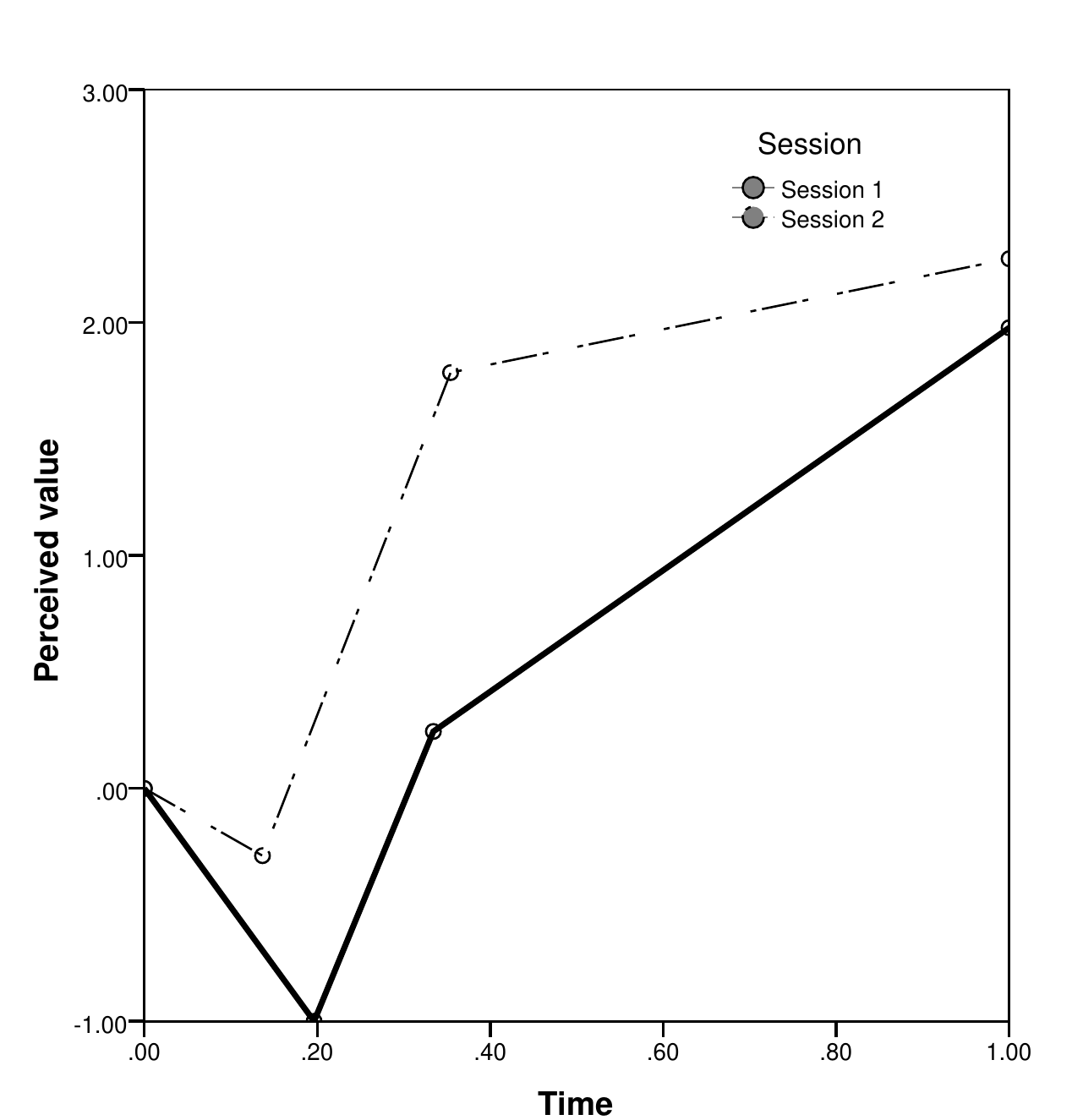}}
\centerline{\includegraphics[width=5.7cm, trim=0.625in 4.9in 2.75in 0.8in, clip=true]{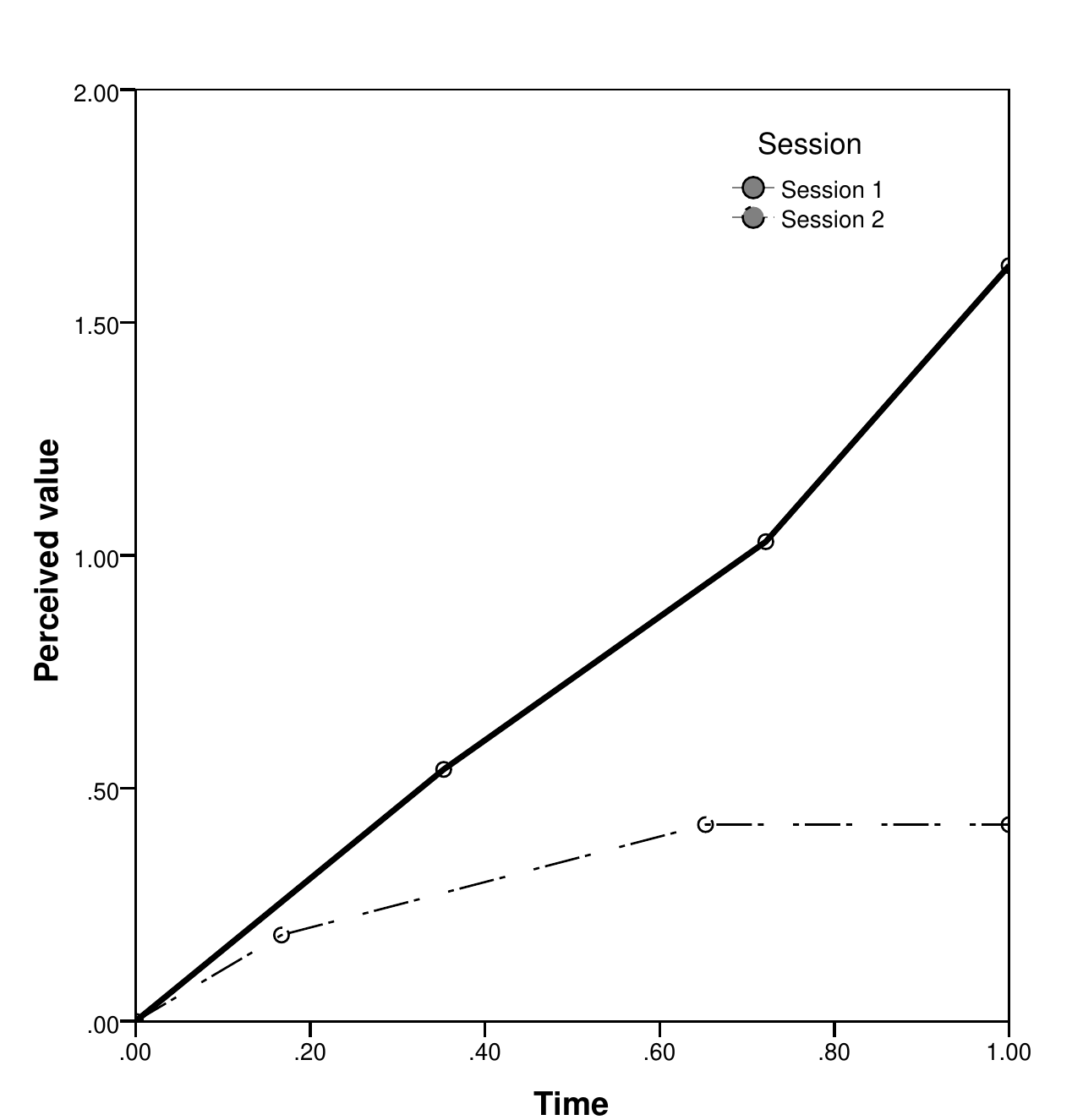} \includegraphics[width=5.7cm, trim=0.625in 4.9in 2.75in 0.8in, clip=true]{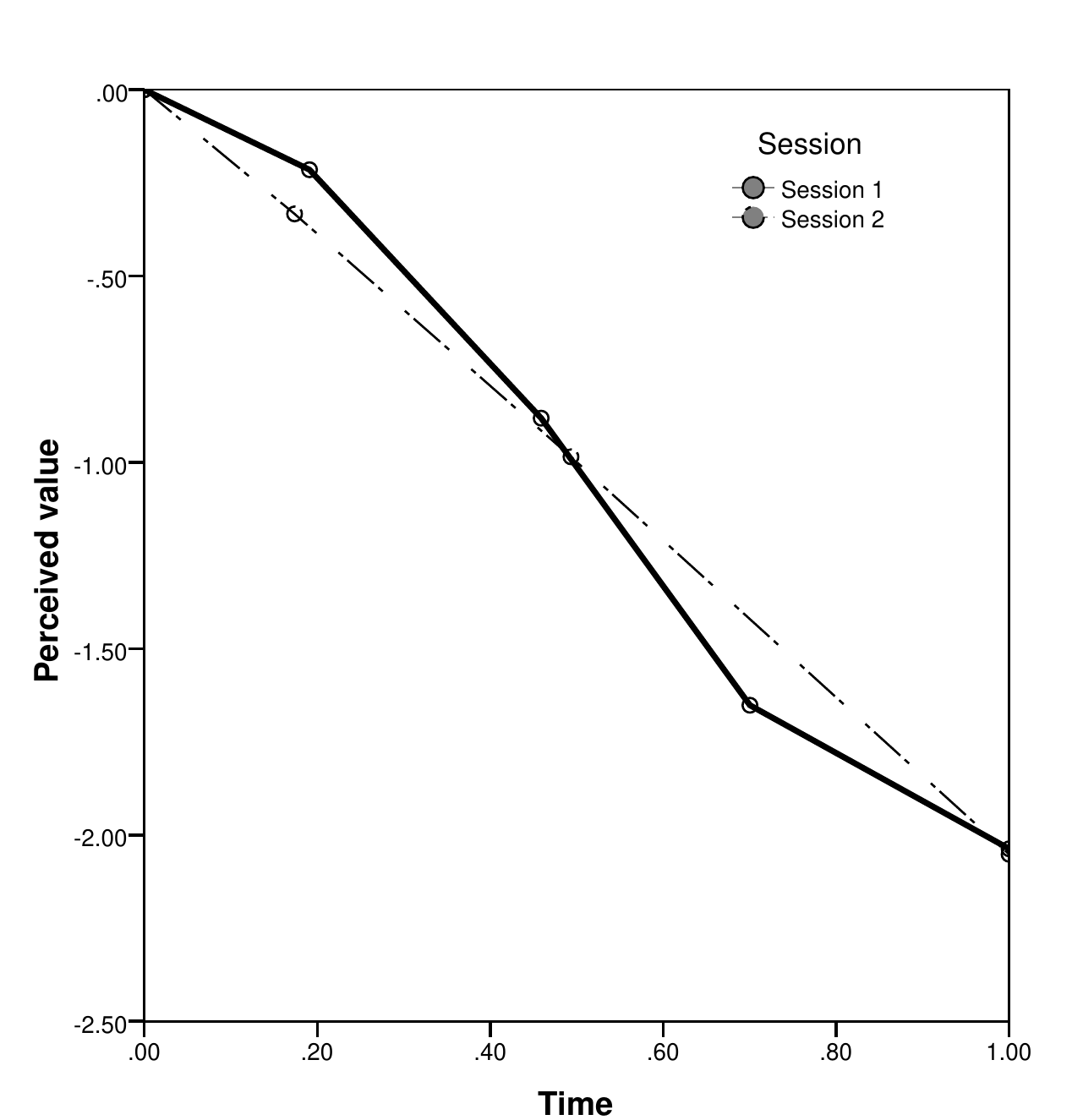}}
  \caption{Participants' sketches resulting from the Value-Account iScale tool. Part B.}
  \label{fig:SketchesBva}
\end{figure*}

\begin{figure*}
\centerline{\includegraphics[width=5.7cm, trim=0.625in 5in 2.75in 0.8in, clip=true]{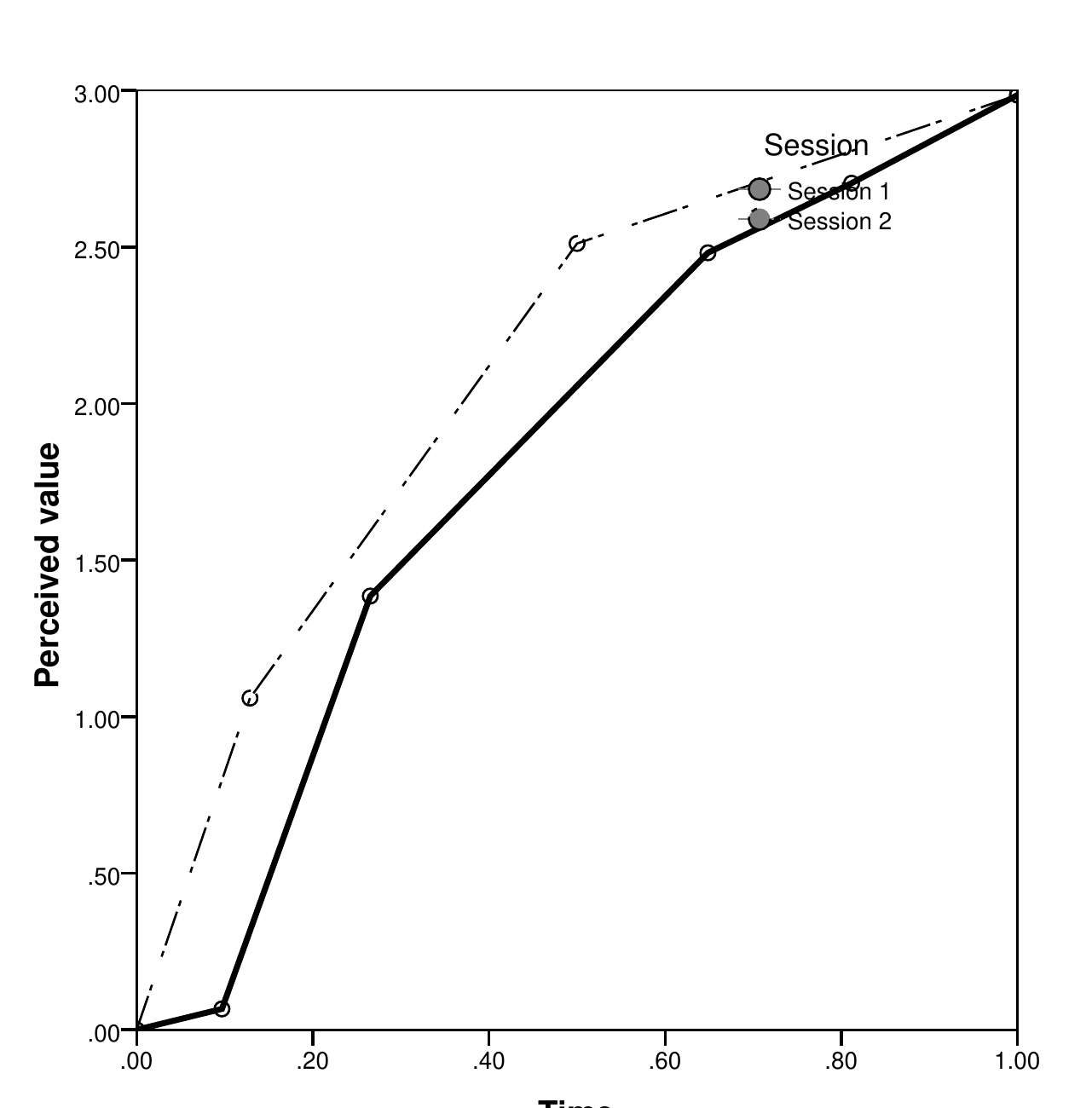} \includegraphics[width=5.7cm, trim=0.625in 5in 2.75in 0.8in, clip=true]{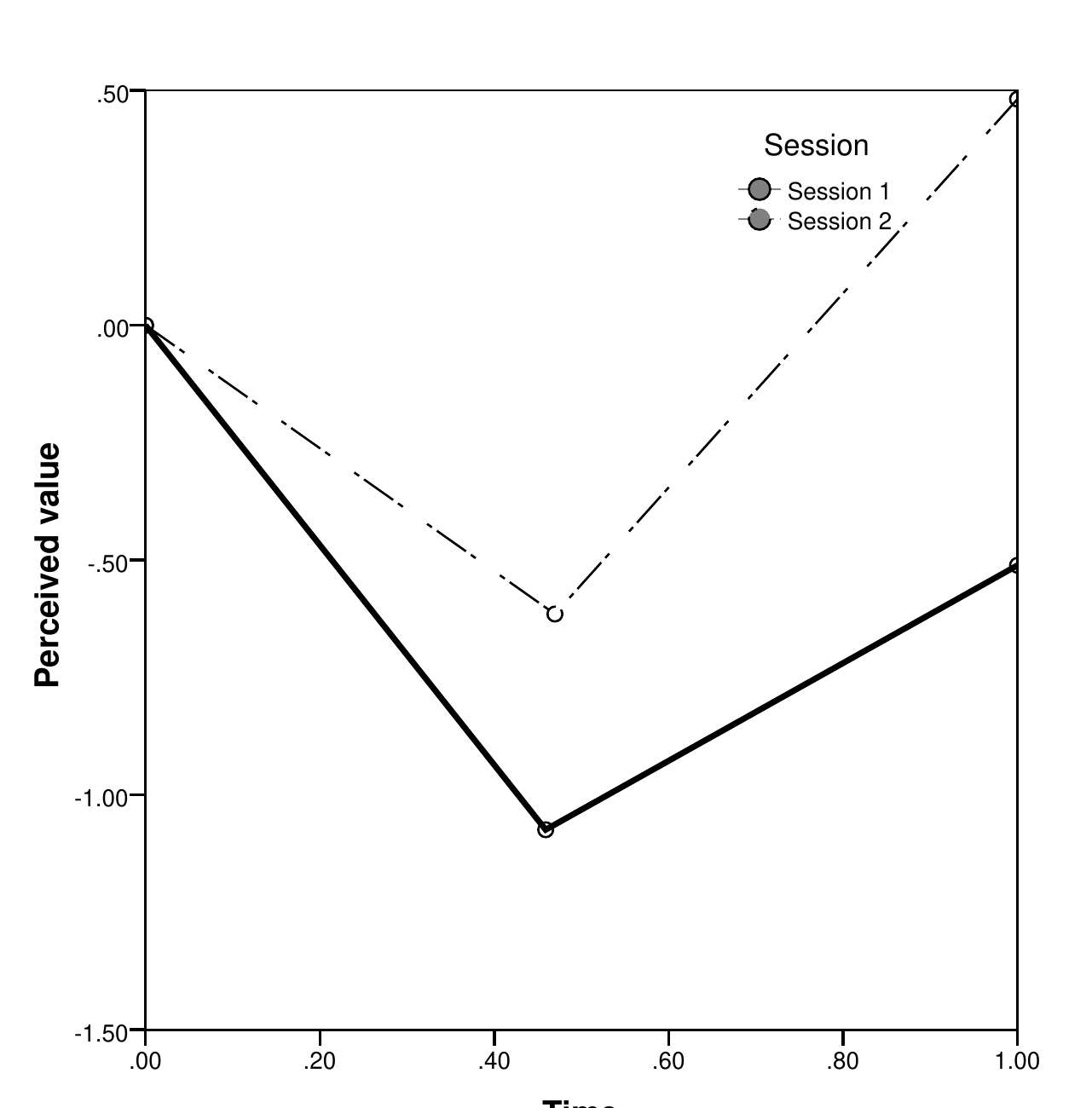}}
\centerline{\includegraphics[width=5.7cm, trim=0.625in 4.9in 2.75in 0.8in, clip=true]{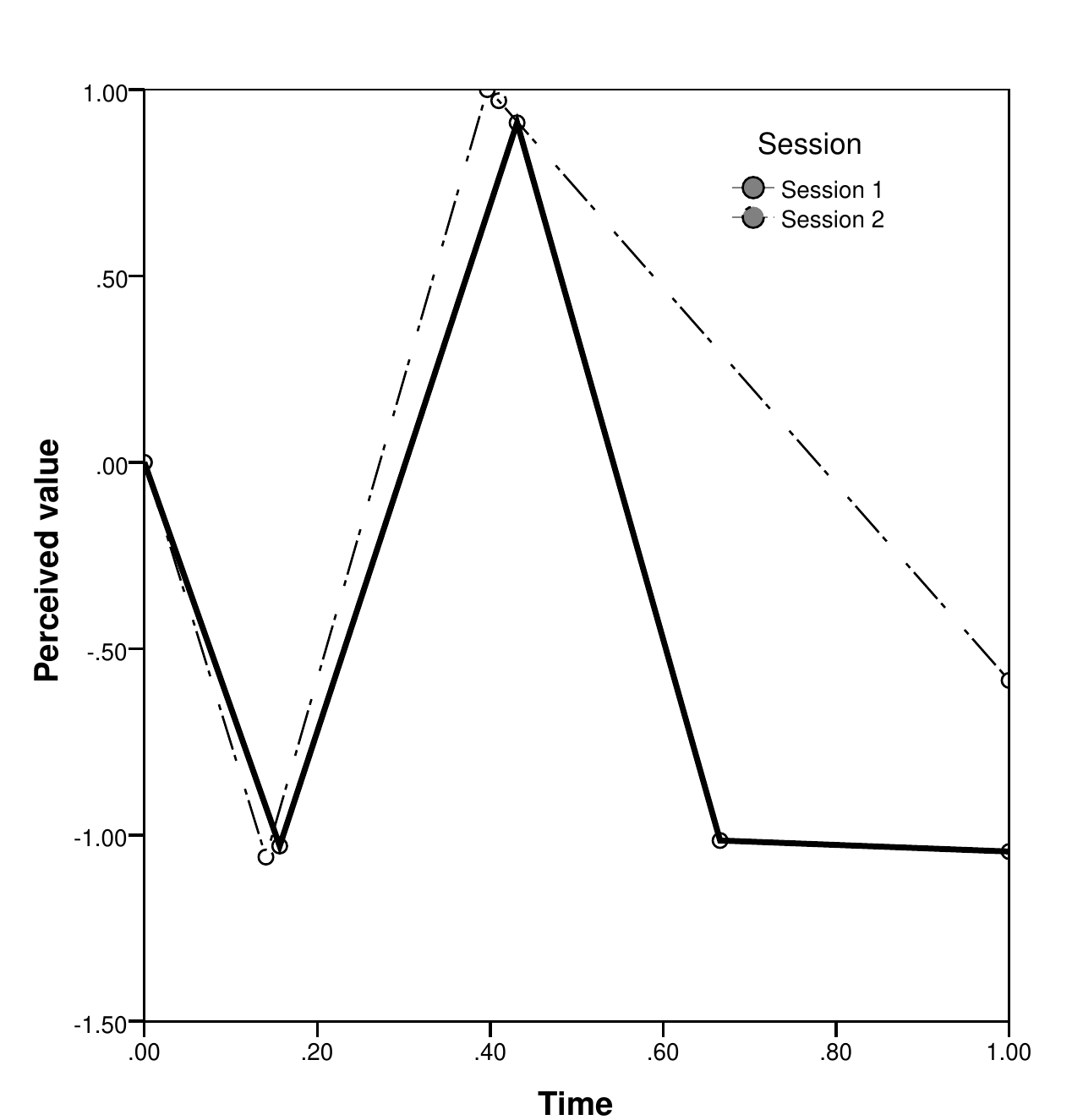} \includegraphics[width=5.7cm, trim=0.625in 4.9in 2.75in 0.8in, clip=true]{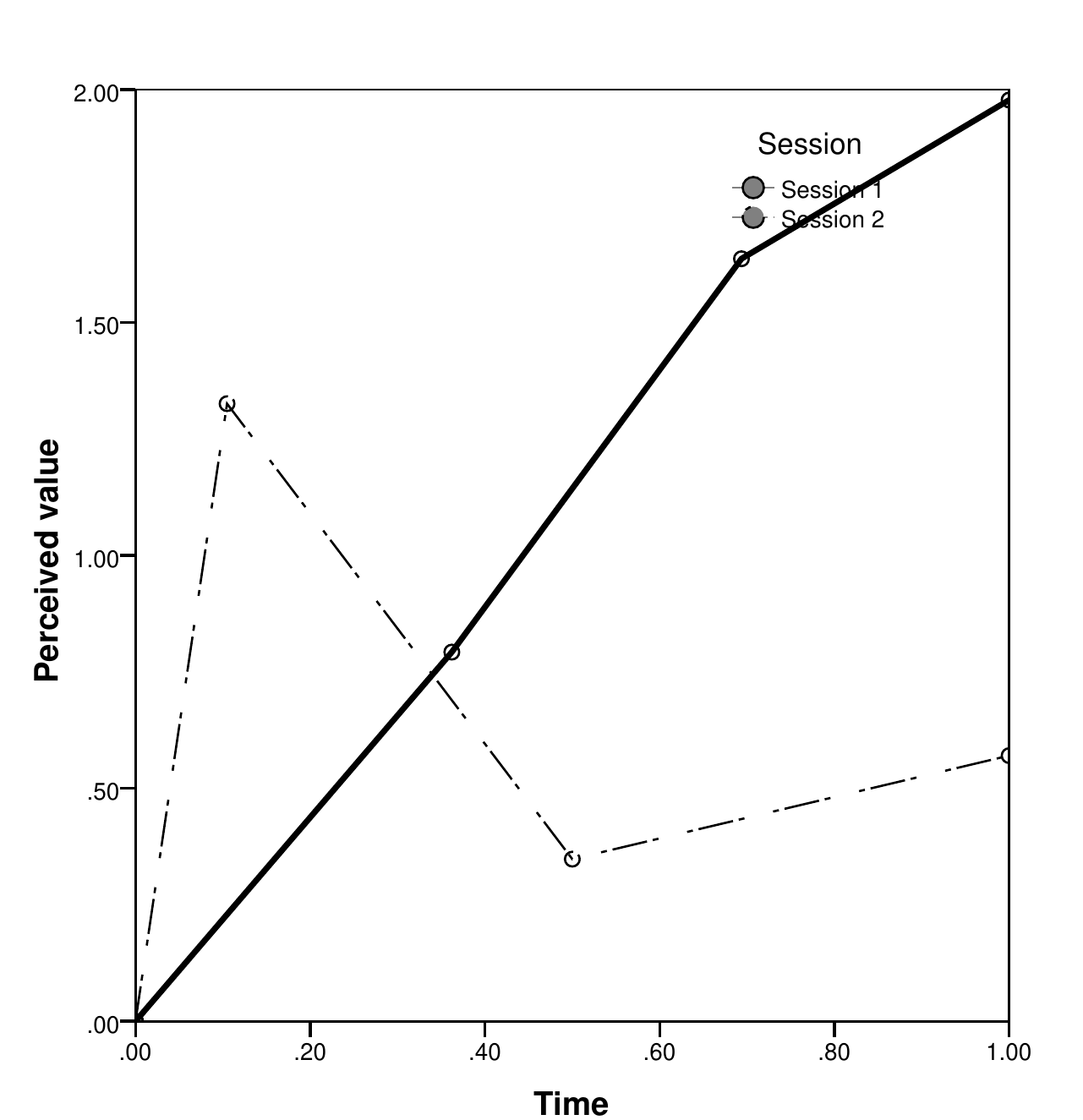}}
\centerline{\includegraphics[width=5.7cm, trim=0.625in 4.9in 2.75in 0.8in, clip=true]{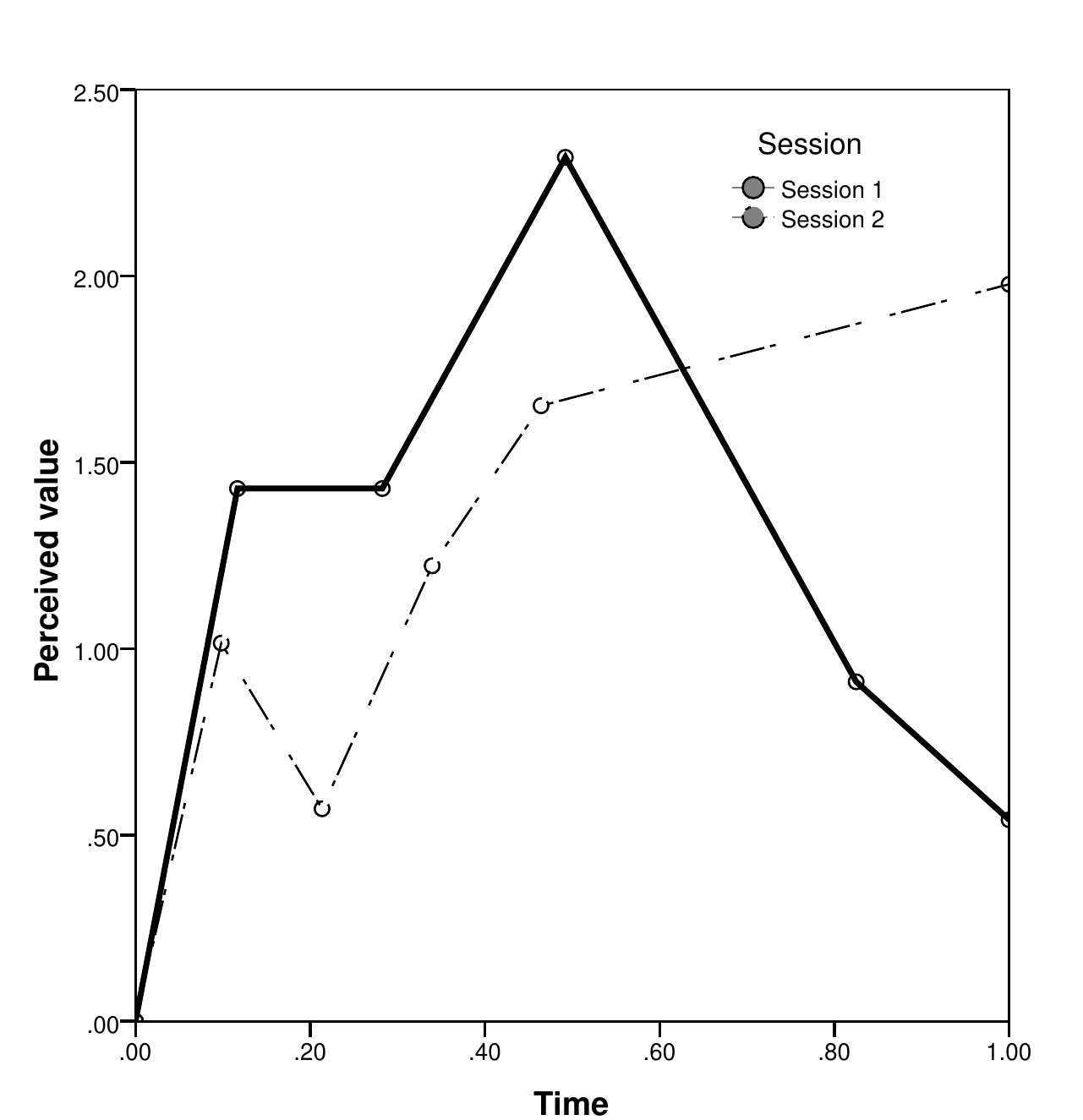} \includegraphics[width=5.7cm, trim=0.625in 4.9in 2.75in 0.8in, clip=true]{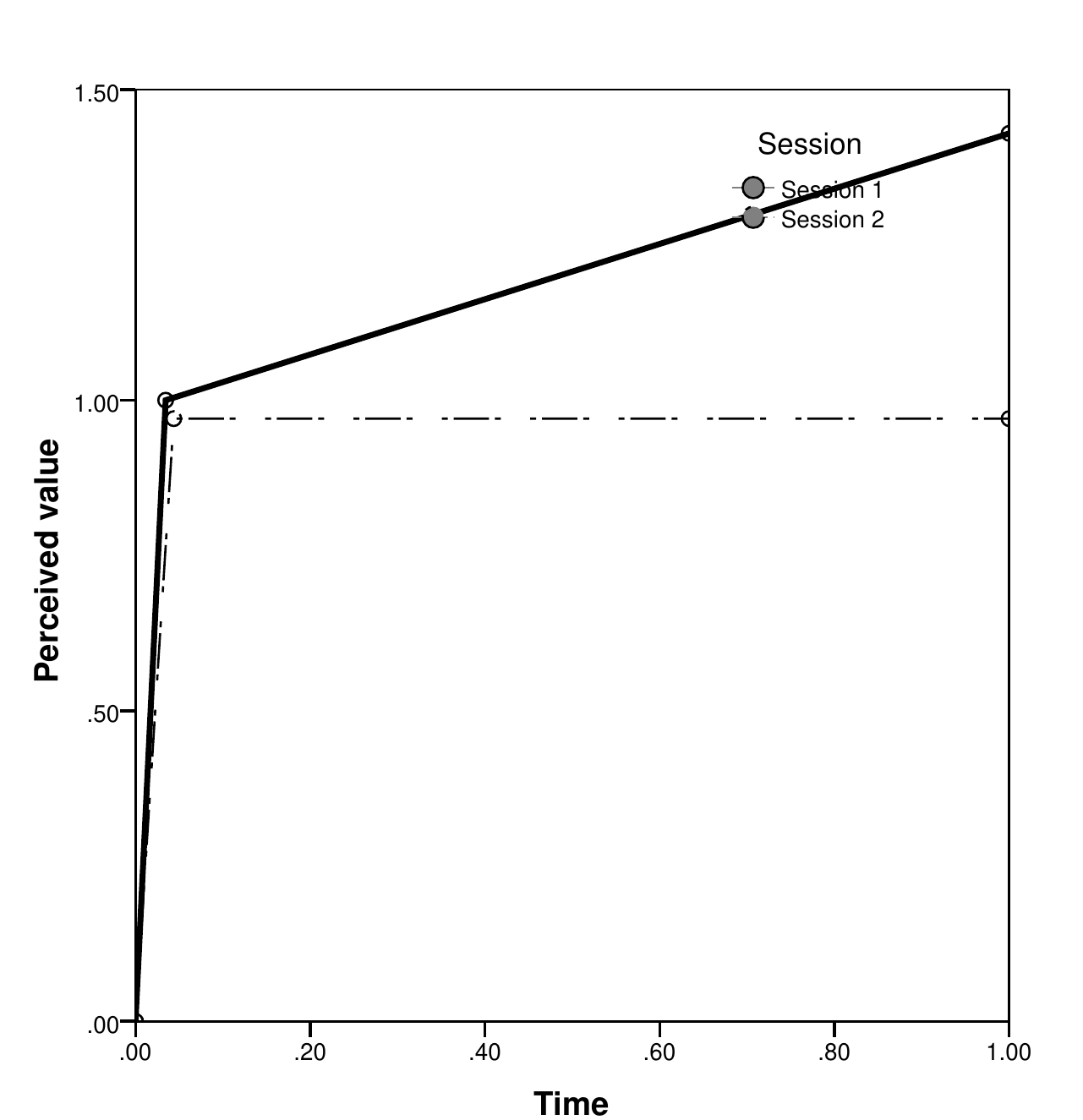}}
  \caption{Participants' sketches resulting from the Value-Account iScale tool. Part C.}
  \label{fig:SketchesCva}
\end{figure*}

\begin{figure*}
\centerline{\includegraphics[width=5.7cm, trim=0.625in 5in 2.75in 0.8in, clip=true]{images/perceptionsOverTime/va20.pdf} \includegraphics[width=5.7cm, trim=0.625in 5in 2.75in 0.8in, clip=true]{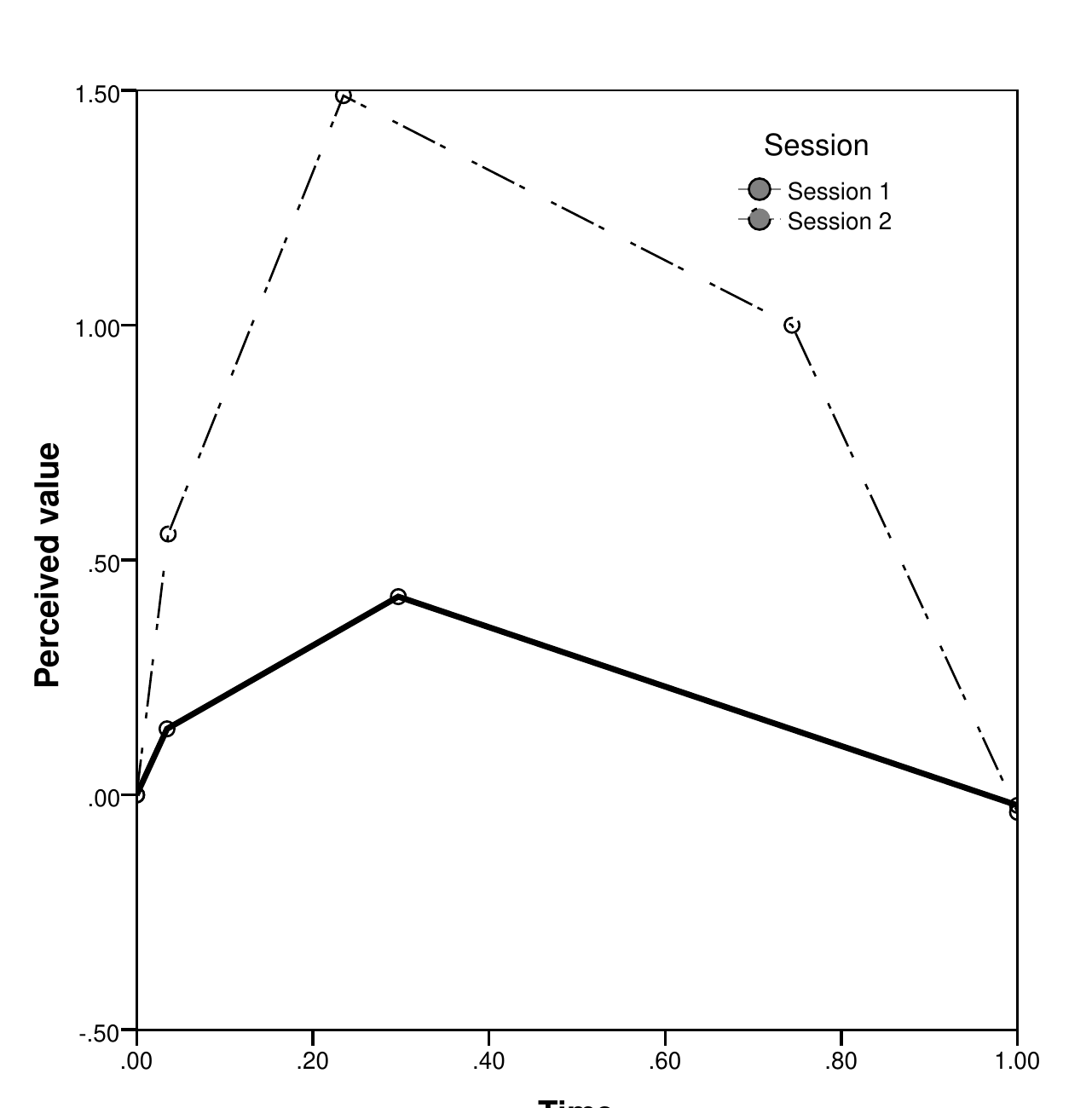}}
\centerline{\includegraphics[width=5.7cm, trim=0.625in 4.9in 2.75in 0.8in, clip=true]{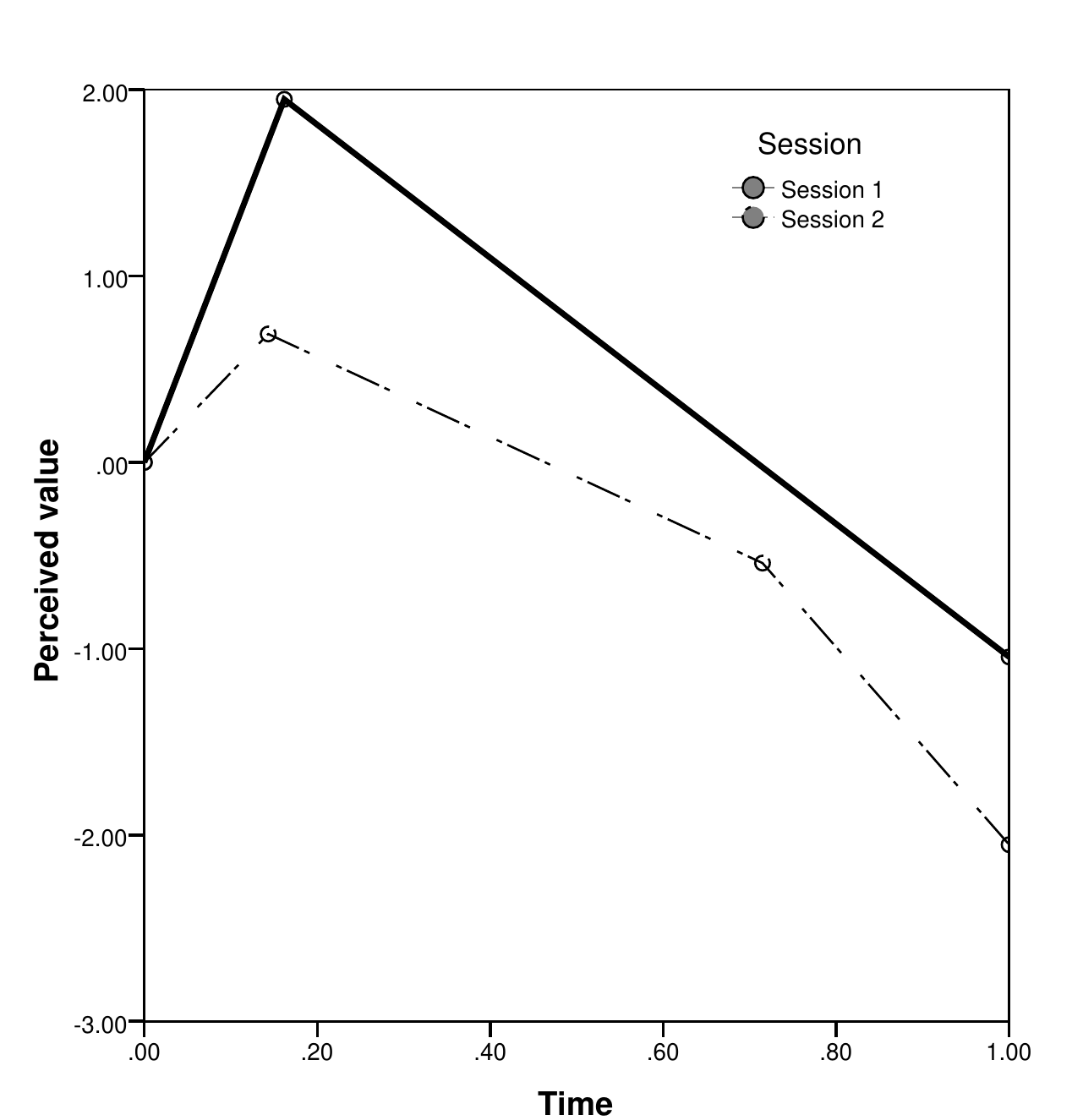} \includegraphics[width=5.7cm, trim=0.625in 4.9in 2.75in 0.8in, clip=true]{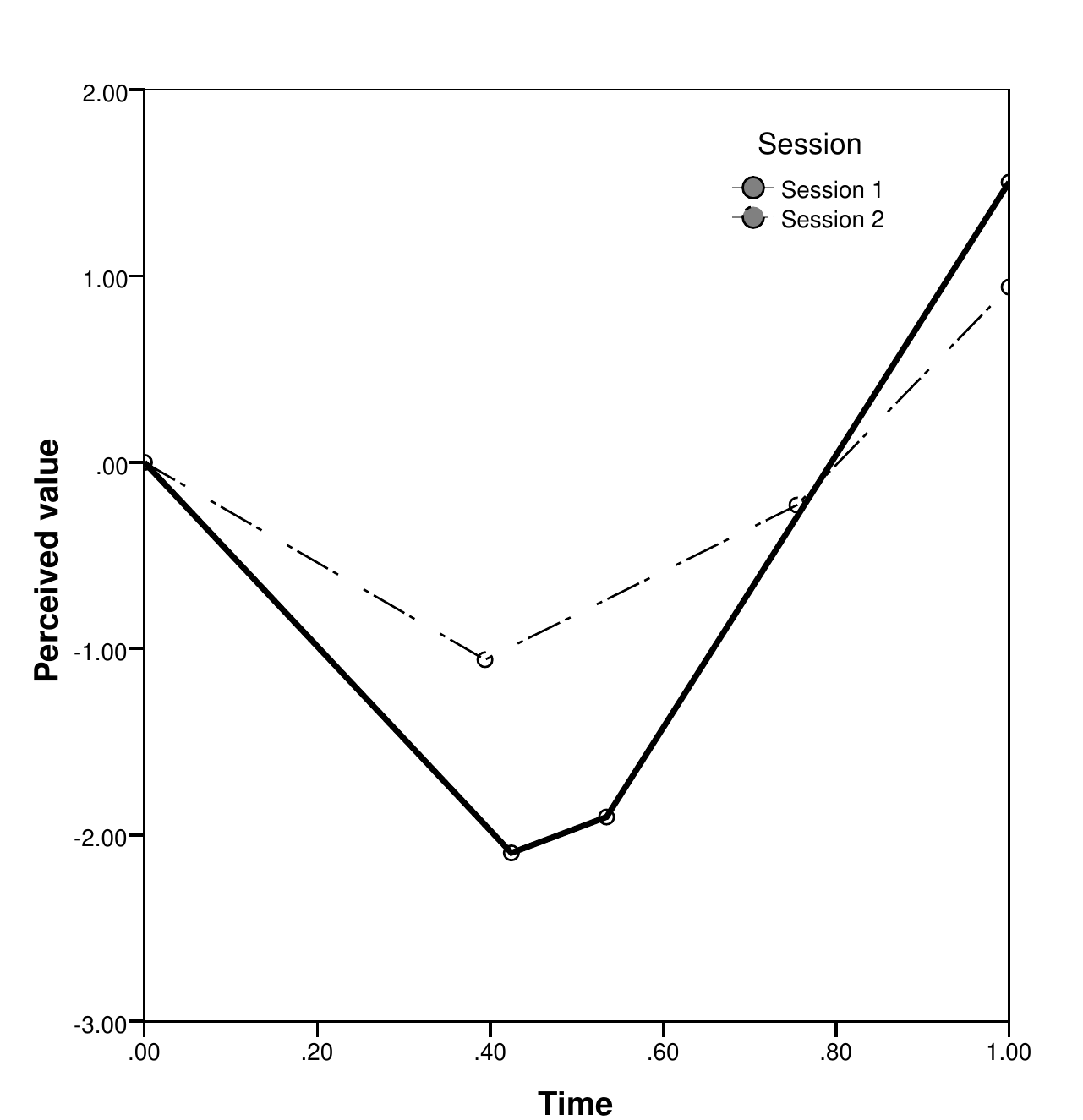}}
\centerline{\includegraphics[width=5.7cm, trim=0.625in 4.9in 2.75in 0.8in, clip=true]{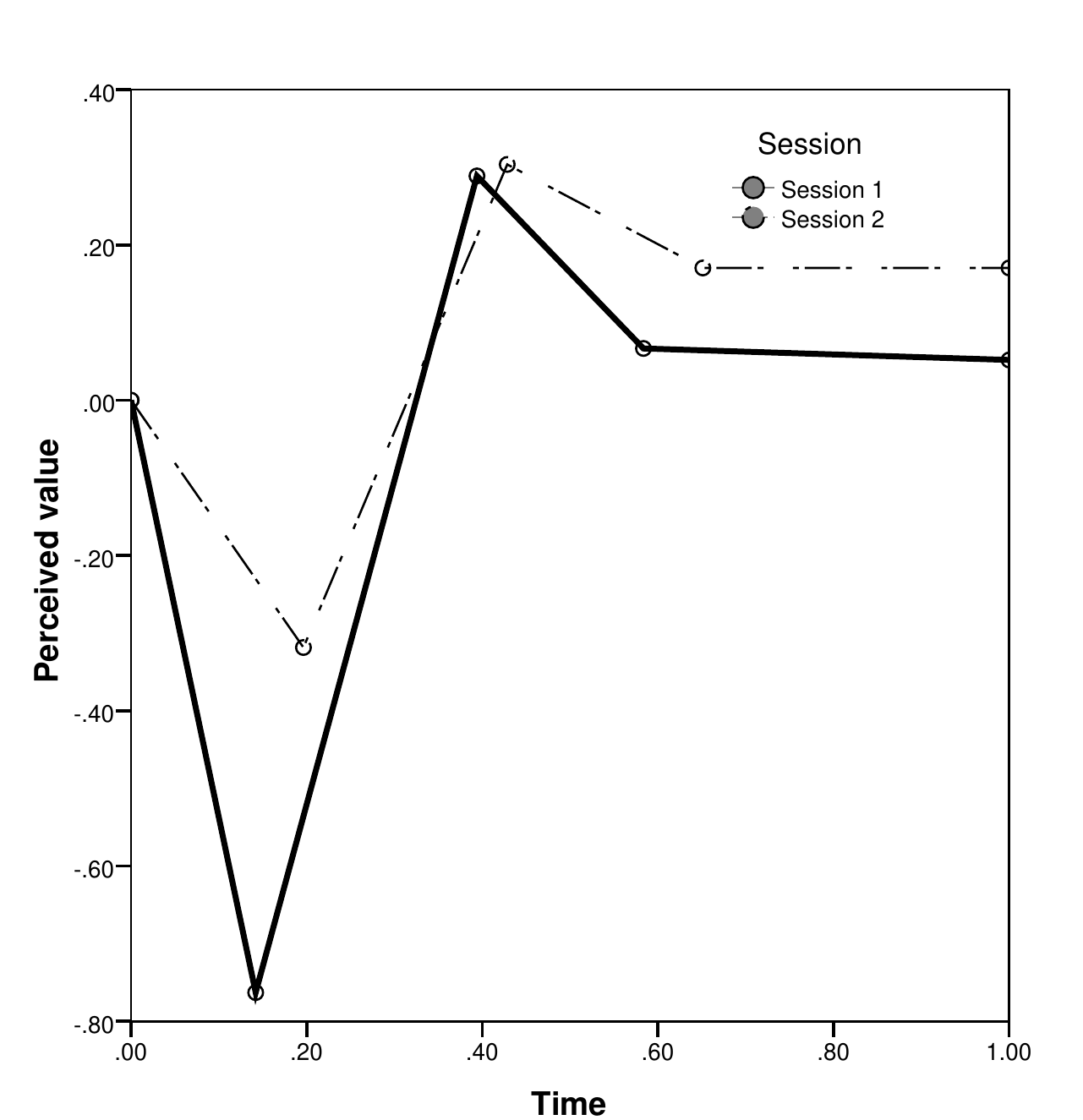} \includegraphics[width=5.7cm, trim=0.625in 4.9in 2.75in 0.8in, clip=true]{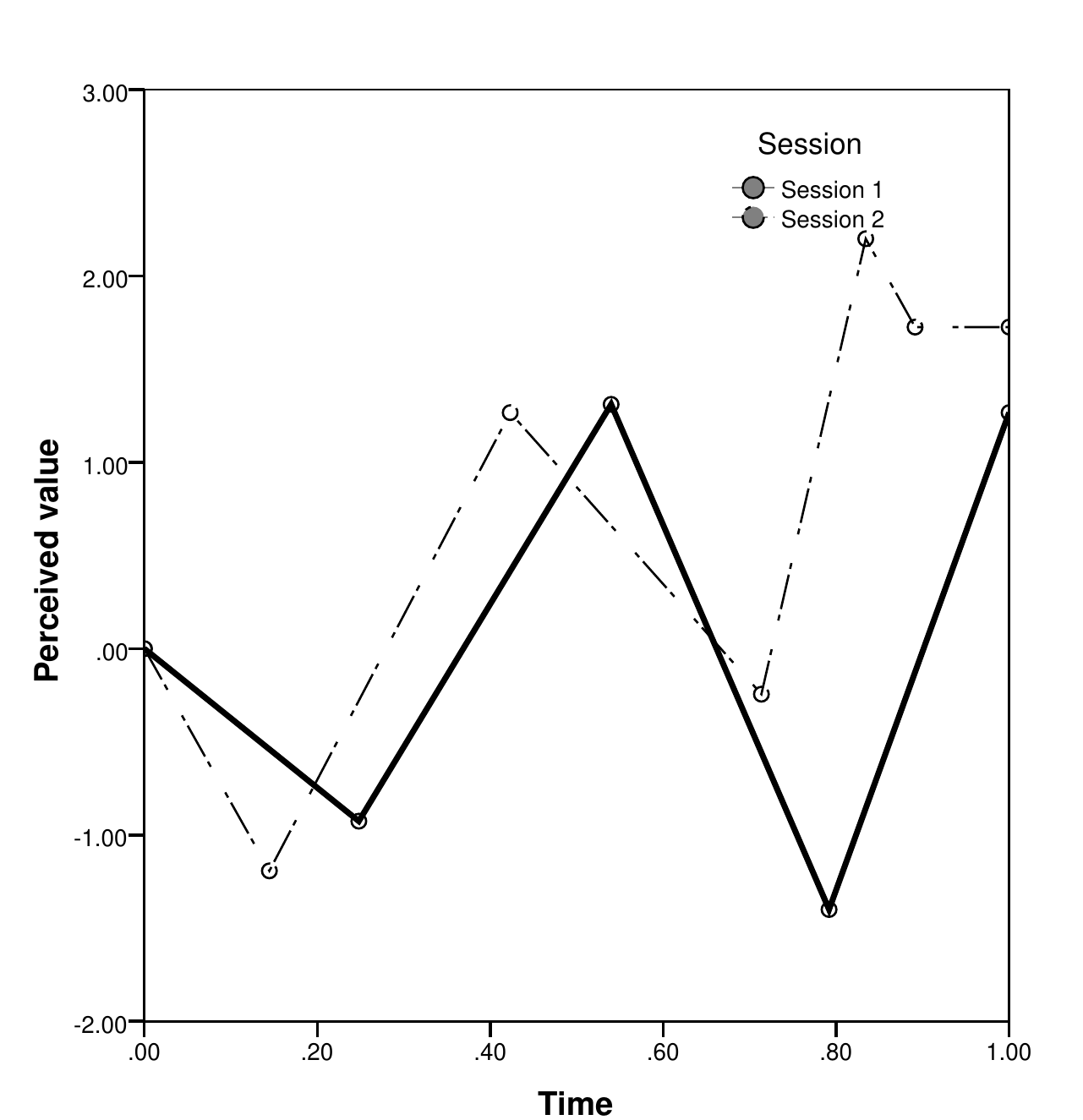}}
  \caption{Participants' sketches resulting from the Value-Account iScale tool. Part D.}
  \label{fig:SketchesDva}
\end{figure*}

\end{document}